\documentclass[fleqn,useAMS]{Papers}
\usepackage{newtxtext,newtxmath}
\usepackage[T1]{fontenc}
\usepackage{subfloat}
\usepackage{makecell}

\usepackage[square,sort,numbers]{natbib}
\usepackage{graphicx}	% Including figure files
\usepackage{amsmath}	% Advanced maths commands
\usepackage{lipsum}
\usepackage{caption}
\usepackage{comment}
\usepackage{textgreek} 
\usepackage{xcolor}  
\newcommand{\rev}[1]{{\color{black} #1}} 

\title[Statistical theory of dark matter flow]{On the statistical theory of self-gravitating collisionless dark matter flow}
\author[Z. Xu]{Zhijie (Jay) Xu,$^{1}$\thanks{E-mail: \href{mailto:zhijie.xu@pnnl.gov}{zhijie.xu@pnnl.gov}; \href{mailto:zhijiexu@hotmail.com}{zhijiexu@hotmail.com}}
\\
$^{1}$Physical and Computational Sciences Directorate, Pacific Northwest National Laboratory; Richland, WA 99354, USA\\
}

\date{Accepted XXX. Received YYY; in original form ZZZ}

\pubyear{2022}

\begin{document}
\label{firstpage}
\pagerange{\pageref{firstpage}--\pageref{lastpage}}
\maketitle

\begin{abstract}
Dark matter, if exists, accounts for five times as much as the ordinary baryonic matter. Compared to hydrodynamic turbulence, the flow of dark matter might possess the widest presence in our universe. This paper presents a statistical theory for the flow of dark matter that is compared with N-body simulations. By contrast to hydrodynamics of normal fluids, dark matter flow is self-gravitating, long-range, and collisionless with a scale dependent flow behavior. The peculiar velocity field is of constant divergence nature on small scale and irrotational on large scale. The statistical measures, i.e. correlation, structure, dispersion, and spectrum functions are modeled on both small and large scales, respectively. Kinematic relations between statistical measures are fully developed for incompressible, constant divergence, and irrotational flow. Incompressible and constant divergence flow share same kinematic relations for even order correlations. The limiting correlation of velocity $\rho_L=1/2$ on the smallest scale ($r=0$) is a unique feature of collisionless flow ($\rho_L=1$ for incompressible flow). On large scale, transverse velocity correlation has an exponential form $T_2\propto e^{-r/r_2}$ with a constant comoving scale $r_2$=21.3Mpc/h that maybe related to the horizon size at matter-radiation equality. All other correlation, structure, dispersion, and spectrum functions for velocity, density, and potential fields are derived analytically from kinematic relations for irrotational flow. On small scale, longitudinal structure function follows one-fourth law of $S^l_2\propto r^{1/4}$. All other statistical measures can be obtained from kinematic relations for constant divergence flow. Vorticity is negatively correlated for scale $r$ between 1 and 7Mpc/h. Divergence is negatively correlated for $r$>30Mpc/h that leads to a negative density correlation.
\end{abstract}

\begin{keywords}
\vspace*{-10pt}
Dark matter flow; N-body simulations; Self-gravitating; Collisionless; 
\end{keywords}

%%%%%%%%%%%%%%%%%%%%%%%%%%%%%%%%%%%%%%%%%%%%%%%%%
%%%%%%%%%%%%%%%%% BODY OF PAPER %%%%%%%%%%%%%%%%%%
\begingroup
\let\clearpage\relax
\tableofcontents
\endgroup
\vspace*{-20pt}
%\newpage

\section{ Introduction}
\label{sec:1}
The existence of dark matter is supported by many astronomical observations \citep{Rubin:1970-Rotation-of-Andromeda-Nebula-f,Rubin:1980-Rotational-Properties-of-21-Sc}. In standard $\Lambda$CDM (cold dark matter) paradigm of cosmology, the amount of dark matter is about 5 times of baryonic matter \citep{Spergel:2003-First-Year-Wilkinson-Microwave-Anisotropy,Komatsu:Seven-year-Wilkinson-Microwave-Anisotropy-Probe,Aghanim:2021-Planck-2018-results--VI--Cosmo}. Therefore, the flow of self-gravitating and collisionless dark matter possesses the widest presence in our universe. Cosmic flow velocity is a key component of self-gravitating collisionless fluid dynamics (SG-CFD) for dark matter. Many insights were obtained from the velocity field on the structure formation and evolution, and the interpretation of observational data. Due to the complicated nonlinear nature, N-body simulation is an invaluable tool to study SG-CFD, capture the very complex gravitational collapse and many effects beyond Newtonian approximations \citep{Angulo:2012-Scaling-relations-for-galaxy-c,Springel:2005-The-cosmological-simulation-co,Peebles:1989-A-Model-for-the-Formation-of-t,Efstathiou:1985-Numerical-Techniques-for-Large}. Complete review of N-body simulation methods for large scale structure and galaxy formation were discussed by \citet{Angulo:2022-Large-scale-dark-matter-simulations} and \citet{Vogelsberger:2020-Cosmological-simulations-of-galaxy}. Recent improvements were also presented with information-optimized parallel algorithm \citep{Yu:2018-CUBE-An-Information-optimized-Parallel} and fast multiple method \citep{Wang:2021-hybrid-Fast-Multipole-Method} for N-body simulations, and new approach for generating N-body initial conditions \citep{Liao:2018-An-alternative-method-to-generate-pre-initial}.

Traditionally, two distinct approaches have been applied to study the velocity field from N-body system, i.e. a halo-based approach and a statistical approach. In halo-based approach, all haloes are identified in N-body simulation with various statistics defined for particle velocity in haloes and velocity of haloes. Using this approach, velocity distributions of dark matter particles that maximize system entropy can be rigorously formulated \citep{Xu:2021-The-maximum-entropy-distributi}. %Most of our previous study focus on the former category, of which the inverse mass and energy cascade and halo mass functions can be rigorously developed \citep{Xu:2021-Inverse-mass-cascade-mass-function}. In addition, velocity distributions of dark matter particles that maximize system entropy were also formulated. Halo mass functions can be directly related to these maximum entropy distributions. 
An alternative strategy without identifying haloes, i.e. a statistical approach, works with statistical measures, such as the correlation and spectrum functions, to make statistical statement about SG-CFD. In this approach, haloes are not explicitly considered. Instead, we focus on the scale and redshift dependence of these statistical measures, which are crucial for structure formation and dynamics on both small and large scales \citep{Kitaura:2016-Bayesian-redshift-space-distor,Pueblas:2009-Generation-of-vorticity-and-ve, Xu:2022-Two-thirds-law-for-pairwise-ve}.  
However, it is not a trivial task to extract and characterize the statistics of velocity field from N-body simulations. Some fundamental problems are: 
%\begin{enumerate}[leftmargin=\parindent,align=left,labelwidth=\parindent,labelsep=0pt]
\begin{enumerate}
\item \noindent The velocity field is defined everywhere in the entire phase space, but is only sampled by N-body simulations at discrete locations of particles. This sampling has a poor quality at locations with low particle density that leads to large uncertainties on the measure of velocity power spectrum \citep{Jennings:2011-Modelling-redshift-space-disto}. 
\item \noindent The velocity field can be multi-valued and discontinuous after shell-crossing due to the collisionless nature.  
The traditional statistical approach involves computing the power spectrum of velocity (and its gradients) in Fourier space \citep{Pueblas:2009-Generation-of-vorticity-and-ve,Hahn:2015-The-properties-of-cosmic-veloc,Jelic_Cizmek:2018-The-generation-of-vorticity-in}. It uses cloud-in-cell (CIC) \citep{Hockney:1988-Computer-Simulation-Using-Part} or triangular-shaped-cloud (TSC) schemes to project density/velocity fields onto the regular mesh, which unavoidably introduces a finite sampling error \citep{Baugh:1994-A-Comparison-of-the-Evolution-,Baugh:1995-A-Comparison-of-the-Evolution-}. 
\end{enumerate}

\noindent Since both real- and Fourier-space contain the same information, there is no reason for preference of one over the other except the computational cost. Directly working on the density/velocity fields in real-space avoids an additional layer of systematic errors and information loss due to the mapping of particles onto structured grid and the conversion from Fourier-space to real space. 

The statistical theory for incompressible flow was developed for isotropic and homogeneous turbulence \citep{Taylor:1935-Statistical-theory-of-turbulan,Taylor:1938-Production-and-dissipation-of-,de_Karman:1938-On-the-statistical-theory-of-i,Batchelor:1953-The-Theory-of-Homogeneous-Turb}. However, the self-gravitating collisionless dark matter flow is intrinsically different. The cosmic peculiar velocity is not incompressible (vanishing divergence) on all scales. Instead, on small scale, the virialized haloes has a vanishing (proper) radial flow and the peculiar velocity $\textbf{u}$ satisfies a constant divergence condition or $\nabla \cdot \textbf{u}=-3Ha$, where $H$ is Hubble parameter and $a$ is scale factor \citep{Xu:2022-The-statistical-theory-of-3rd}. While on large scale, flow becomes irrotational or $\nabla \times \textbf{u}=0$. The nature of two different types of flow on small and large scales and the transition between them are the key to understand the velocity field in dark matter flow. Therefore, it is necessary to develop a complete set of kinematic relations between various statistical measures for different types of flow.

This paper focus on the two-point second order statistical measures of velocity field for SG-CFD. Using second order velocity statistics fully extracted from N-body simulation without projecting onto structured grid, we are able to:
%\begin{enumerate}[leftmargin=\parindent,align=left,labelwidth=\parindent,labelsep=0pt]
\begin{enumerate}
\item \noindent demonstrate the dark matter flow is of irrotational nature on large scale and constant divergence on small scale (Fig. \ref{fig:4}); 
\item \noindent identify an exponential form of transverse velocity correlation on large scale (Eq. \eqref{ZEqnNum971850}), such that all other relevant statistical measures for density, velocity and potential fields can all be analytically derived; 
\item \noindent identify a one-fourth (1/4) scaling law for second order velocity structure function on small scale ($S_{2}^{l} \left(r\right)\propto r^{{1/4}}$ in Eq. \eqref{ZEqnNum392323}) with all other statistical measures on small scale derived subsequently; 
\item \noindent identify a limiting velocity correlation coefficient approaching 1/2 on the smallest scale ($r\to 0$) (Fig. \ref{fig:15}). 
\end{enumerate}

This paper is organized as follows: Section \ref{sec:2} introduces the N-body simulation data used for this paper. Section \ref{sec:3} introduces the two-point second order statistical measures and develops the fundamental kinematic relations between different statistical measures. Generalization to high and arbitrary order is presented in a separate paper \citep{Xu:2022-The-statistical-theory-of-3rd}. Section \ref{sec:4} presents the statistical measures from N-body simulations, followed by the corresponding models on small and large scales in Section \ref{sec:5}.
%One-point statistics of velocities, i.e. the probability distributions, is discussed in \citep{Xu:2022-Two-thirds-law-for-pairwise-ve}, along with applications of statistical theory for dark matter particle mass and properties \citep{Xu:2022-Postulating-dark-matter-partic}, MOND (modified Newtonian dynamics) theory \citep{Xu:2022-The-origin-of-MOND-acceleratio}, and baryonic-to-halo mass relation \citep{Xu:2022-The-baryonic-to-halo-mass-rela}. 

\section{N-body simulations and numerical data}
\label{sec:2}
\begin{figure}
\includegraphics*[width=\columnwidth,angle=-90,scale=0.6]{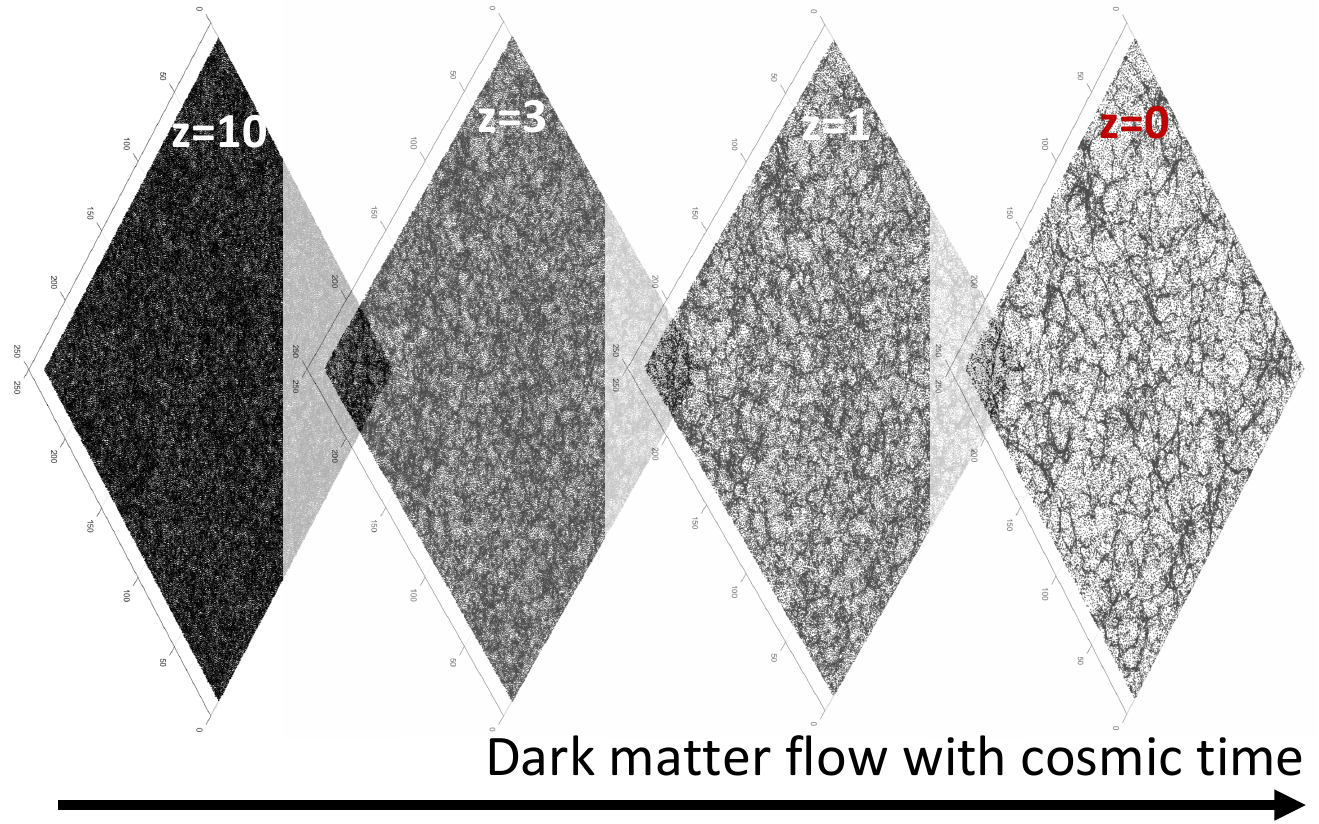}
\caption{Snapshots (239.5Mpc/h$\times$239.5Mpc/h) of N-body simulations at different redshifts for dark matter flow with cosmic time.} 
\label{fig:1-1}
\end{figure}

The numerical data were public available and generated from the N-body simulations carried out by the Virgo consortium. A comprehensive description of the data can be found in references \citep{Frenk:2000-Public-Release-of-N-body-simul,Jenkins:1998-Evolution-of-structure-in-cold}. As the first step, the current study was carried out using the simulation runs with Ω = 1 and the standard CDM power spectrum (SCDM) that started from \textit{z} = 50 to focus on the matter-dominant self-gravitating flow of collisionless dark matter. Similar analysis can be extended to other simulation with different cosmological model assumptions and parameter in the future. The same set of data have been widely used in studies from clustering statistics \citep{Jenkins:1998-Evolution-of-structure-in-cold} to formation of cluster haloes in large scale environment \citep{Colberg:1999-Linking-cluster-formation-to-l}, and test of models for halo abundances and mass functions \citep{Sheth:2001-Ellipsoidal-collapse-and-an-im}. Some key numerical parameters of N-body simulation are listed in Table \ref{tab:1}. 
Figure \ref{fig:1-1} presents several simulation snapshots at different redshifts showing the dynamics of self-gravitating dark matter leading to the structure formation and evolution.  

To analyze the nature of flow of dark matter only, we first focus on the cosmological model with dark matter only and exclude any other effects. The situation is more relevant to the early matter-dominant universe. This will also establish a benchmark for future testing and comparison. The simulation campaign from the Virgo Consortium also includes data for different cosmological models (LCDM, OCDM, and tCDM, etc.)\citep{Frenk:2000-Public-Release-of-N-body-simul}. By gradually increasing the complexity, the approach described in this paper can be applied similarly to identify effects of various cosmological parameters on the dynamics of dark matter flow. 

In addition, in this paper, we propose an approach to directly compute the real-space statistics by a pairwise averaging over all particle pairs on a given scale. This approach can maximally employ the information contained in N-body simulation and generate complete statistics on all scales without involving any projection kernels (e.g. CIC). On the other hand, it is also computationally intensive to identify all particle pairs on all scales. The N-body simulation selected has a relatively low resolution when compared to recent cosmological simulations. However, this enables a computationally affordable direct extraction of real-space two-point statistics. With increasing computing power, the same approach can be similarly applied to other simulations with higher resolution.

Finally, two relevant datasets from this N-body simulation, i.e. halo-based and correlation-based statistics of dark matter flow, can be found at Zenodo.org  \citep{Xu:2022-Dark_matter-flow-dataset-part1, Xu:2022-Dark_matter-flow-dataset-part2}, along with the accompanying presentation slides, "A comparative study of dark matter flow \& hydrodynamic turbulence and its applications" \citep{Xu:2022-Dark_matter-flow-and-hydrodynamic-turbulence-presentation}. All data files are also available on GitHub \citep{Xu:Dark_matter_flow_dataset_2022_all_files}.

\begin{table}
\caption{Numerical parameters of N-body simulation}
\begin{tabular}{p{0.25in}p{0.1in}p{0.05in}p{0.05in}p{0.05in}p{0.05in}p{0.37in}p{0.1in}p{0.4in}p{0.35in}} 
\hline 
Run & $\Omega_{0}$ & $\Lambda$ & $h$ & $\Gamma$ & $\sigma _{8}$ & \makecell{L\\(Mpc/h)} & $N$ & \makecell{$m_{p}$\\($M_{\odot}/h$)} & \makecell{$l_{soft}$\\(Kpc/h)} \\ 
\hline 
SCDM1 & 1.0 & 0.0 & 0.5 & 0.5 & 0.51 & \centering 239.5 & $256^{3}$ & 2.27$\times 10^{11}$ & \makecell{\centering 36} \\ 
\hline 
\end{tabular}
\label{tab:1}
\end{table}

\section{Fundamental kinematic relations}
\label{sec:3}
To describe the random and multiscale nature of self-gravitating collisionless dark matter flow (SG-CFD), various statistical measures can be introduced to characterize the density, velocity, and potential fields. There are restrictions on these quantities due to the symmetry implied by the assumptions of homogeneity and isotropy. Based on these assumptions, kinematic relations between different statistical measures can be rigorous developed for a given nature of flow, i.e. incompressible, constant divergence, or irrotational flow. 

In this section, we focus on the statistical measures of velocity field. Three types of statistical quantities will be introduced that are related to each other and are traditionally used to characterize the turbulent flow, i.e. the correlation functions, structure functions, and power spectrum functions. The real-space correlation functions are the most fundamental quantity and building blocks of the statistical theory for any stochastic flow. The structure functions and the power spectrum are intimately related to correlation functions to describe how the system energy is distributed and transferred across different scales. This section will introduce/review these fundamental statistical quantities and their relations with each other. This will prepare us for better understanding and interpreting the data and results from N-body simulations. 

For simplicity, we restrict our discussion to the homogeneous and isotropic flow with translational and rotational symmetry in space, respectively. The assumption of homogeneity and isotropy will greatly simplify the form of velocity correlation tensors and facilitate the development of statistical theory. The SG-CFD flow is not required to be stationary (translational symmetry in time), with flow dynamically evolving in the course of time. These statistical quantities are first introduced for generic flow. Their properties for a given type of flow (i.e. the incompressible, constant divergence, and irrotational flow) are presented subsequently. This is important as we will demonstrate that the cosmic peculiar velocity is of constant divergence and irrotational nature on small and large scales (Section \ref{sec:4}), respectively.

The starting point is to define a one-dimensional RMS (root-mean-square) velocity (or velocity dispersion) $u$
\begin{equation} 
\label{ZEqnNum586699} 
u\left(a\right)=\left(\frac{1}{3} \left\langle \boldsymbol{\mathrm{u}}\left(\boldsymbol{\mathrm{x}}\right)\cdot \boldsymbol{\mathrm{u}}\left(\boldsymbol{\mathrm{x}}\right)\right\rangle \right)^{{1/2} } ,      
\end{equation} 
where $\left\langle \right\rangle $ denotes the average over all particles in system. Here $\boldsymbol{\mathrm{x}}$ is the coordinates of a particle. The RMS velocity $u$ is a measure of mean specific kinetic energy (per unit mass). \textit{N}-body simulation used here gives $u_{0} =u\left(a=1\right)=354.61{km/s} $. The time variation of $u^{2} \left(a\right)$ can be found in Fig. \ref{fig:18} with $u^{2} \left(a\right)\propto t$ once statistically steady state is established. This leads to a constant rate of energy cascade $\varepsilon_u \propto u^2/t \approx 4.6\times 10^{-7}m^2/s^3$, a fundamental constant in the theory of mass and energy cascade \citep{Xu:2023-Universal-scaling-laws-and-density-slope}. 

\subsection{Two-point first order velocity correlation tensor }
\label{sec:3.1}
We first define a correlation tensor between velocity field $\boldsymbol{\mathrm{u}}\left(\boldsymbol{\mathrm{x}}\right)$ and a scalar field $p\left(\boldsymbol{\mathrm{x}}\right)$. Examples are the correlations between velocity field and pressure, density, or gravitational potential fields at two different locations $\boldsymbol{\mathrm{x}}$ and $\boldsymbol{\mathrm{x}}^{'} =\boldsymbol{\mathrm{x}}+\boldsymbol{\mathrm{r}}$ that are separated by a vector $\boldsymbol{\mathrm{r}}$. The most general form of this tensor in the index notation reads 
\begin{equation} 
\label{eq:2} 
Q_{i} \left(\boldsymbol{\mathrm{x}},\boldsymbol{\mathrm{r}}\right)=\left\langle u_{i} \left(\boldsymbol{\mathrm{x}}\right)p\left(\boldsymbol{\mathrm{x}}^{'} \right)\right\rangle .         
\end{equation} 
For homogeneous flow, all statistical quantities should be, by definition, independent of location $\boldsymbol{\mathrm{x}}$. The translational symmetry associated with the spatial homogeneity leads to $Q_{i} \left(\boldsymbol{\mathrm{x}},\boldsymbol{\mathrm{r}}\right)\equiv Q_{i} \left(\boldsymbol{\mathrm{r}}\right)$. The isotropy symmetry further reduces this tensor to:
\begin{equation} 
\label{ZEqnNum445111} 
Q_{i} \left(\boldsymbol{\mathrm{r}}\right)\equiv Q_{i} \left(r\right)=A_{1} \left(r\right)r_{i} ,         
\end{equation} 
where $A_{1} \left(r\right)$ is a symmetric function of \textit{r} ($A_{1} \left(r\right)$=$A_{1} \left(-r\right)$) and $r_{i}$ is the Cartesian components of vector \textbf{\textit{r}}. The divergence of two-point first order correlation tensor $Q_{i}(r)$ reads
\begin{equation} 
\label{ZEqnNum191183} 
\nabla \cdot \boldsymbol{\mathrm{Q}}\left(\boldsymbol{\mathrm{x}},\boldsymbol{\mathrm{r}}\right)=\left\langle \frac{\partial u_{i} \left(\boldsymbol{\mathrm{x}}\right)}{\partial x_{i} } p\left(\boldsymbol{\mathrm{x}}^{'} \right)\right\rangle =\frac{\partial Q_{i} \left(\boldsymbol{\mathrm{x}},\boldsymbol{\mathrm{r}}\right)}{\partial x_{i} } =-\frac{\partial Q_{i} \left(r\right)}{\partial r_{i} } ,     
\end{equation} 
where the average operator and differential operator can commute $\nabla \cdot \left\langle \bullet \right\rangle =\left\langle \nabla \cdot \bullet \right\rangle $. Because of $\boldsymbol{\mathrm{r}}=\boldsymbol{\mathrm{x}}^{'} -\boldsymbol{\mathrm{x}}$, we should have
\begin{equation}
{\partial /\partial } x_{i} =-{\partial /\partial } r_{i} \quad \textrm{and} \quad {\partial /\partial } x_{i}^{'} ={\partial /\partial } r_{i}.        
\label{eq:5}
\end{equation}
\noindent With Eqs. \eqref{ZEqnNum445111} and \eqref{eq:5}, the divergence of $Q_{i}(r)$ (Eq. \eqref{ZEqnNum191183}) reduces to
\begin{equation} 
\label{ZEqnNum725997} 
\frac{\partial Q_{i} \left(r\right)}{\partial r_{i} } =A_{1} r_{i,i} +\frac{\partial A_{1} }{\partial r} \frac{\partial r}{\partial r_{i} } r_{i} =3A_{1} +\frac{\partial A_{1} }{\partial r} r.       
\end{equation} 
Similarly, the curl of two-point first order correlation tensor is
\begin{equation}
\begin{split}
\label{ZEqnNum790489} 
\nabla \times \boldsymbol{\mathrm{Q}}\left(\boldsymbol{\mathrm{x}},\boldsymbol{\mathrm{r}}\right)&=\varepsilon _{ijk} \frac{\partial Q_{i} \left(\boldsymbol{\mathrm{x}},\boldsymbol{\mathrm{r}}\right)}{\partial x_{k} } =-\varepsilon _{ijk} \frac{\partial Q_{i} \left(r\right)}{\partial r_{k} }\\ 
&=-\varepsilon _{ijk} \left(A_{1} \delta _{ik} +\frac{r_{i} r_{k} }{r} \frac{\partial A_{1} }{\partial r} \right)=0,
\end{split}
\end{equation} 
where $\varepsilon_{ijk}$ is the Levi-Civita symbol satisfying the identity
\begin{equation}
\varepsilon_{ijk} \delta_{jk} =0 \quad \textrm{and} \quad \varepsilon_{ijk} r_{j} r_{k} =\boldsymbol{\mathrm{r}}\times \boldsymbol{\mathrm{r}}=0.  
\label{eq:8}
\end{equation}

The velocity field of self-gravitating collisionless dark matter flow (SG-CFD) can be very complex. We demonstrated that on small scale, virialized haloes are spherical and isotropic. The peculiar velocity of small haloes should have a constant divergence of $-3Ha$. On large scale, velocity is proportional to the gradient of the gravitational potential, curl free, and irrotational (Zeldovich approximation \citep{Zeldovich:1970-Gravitational-Instability---an}). Next, we will extend the discussion to three different types of flow.

\subsubsection{Results for incompressible flow}
\label{sec:3.1.1}
The incompressible flow requires $\nabla \cdot \boldsymbol{\mathrm{u}}=0$ such that the divergence of the correlation tensor vanishes everywhere (Eqs. \eqref{ZEqnNum191183} and \eqref{ZEqnNum725997}) , which leads to $A_{1} \left(r\right)=0$ or $A_{1} \left(r\right)\sim r^{-3} $. Since $A_{1} \left(r\right)$ must be symmetric of \textit{r} and regular at origin \textit{r}=0,  $A_{1} \left(r\right)=0$ must be satisfied for incompressible flow. An important conclusion is that $Q_{i}(r)$ must be zero for incompressible, homogeneous, isotropic velocity field. 

\subsubsection{Results for constant divergence flow}
\label{sec:3.1.2}
For flow with constant divergence everywhere such that $\nabla \cdot \boldsymbol{\mathrm{u}}=\theta $. This leads to a constant divergence for correlation tensor (Eq. \eqref{ZEqnNum725997}), 
\begin{equation} 
\label{ZEqnNum666615} 
\frac{\partial Q_{i} \left(r\right)}{\partial r_{i} } =3A_{1} +\frac{\partial A_{1} }{\partial r} r=-\theta \left\langle p\left(\boldsymbol{\mathrm{x}}\right)\right\rangle  
\end{equation} 
such that solution $A_{1} \left(r\right)=-{\theta \left\langle p\left(\boldsymbol{\mathrm{x}}\right)\right\rangle /3}$ is independent of $r$. 

\subsubsection{Results for irrotational flow}
\label{sec:3.1.3}
Irrotational flow requires curl free for velocity field ($\nabla \times \boldsymbol{\mathrm{u}}=0$). The curl of $Q_{i}(r)$ vanishes, i.e. $\nabla \times \boldsymbol{\mathrm{Q}}\left(\boldsymbol{\mathrm{x}},\boldsymbol{\mathrm{r}}\right)\equiv 0$ (Eq. \eqref{ZEqnNum790489}). 

\subsection{Two-point second order velocity correlation tensor}
\label{sec:3.2}
The two-point second order velocity correlation tensor quantifies the degree to which extent velocities at two locations are statistically correlated. It can be defined as (similarly, the dependence on location \textbf{\textit{x}} and direction \textbf{\textit{r}} is dropped due to homogeneity and isotropy)
\begin{equation} 
\label{ZEqnNum195930} 
Q_{ij} \left(\boldsymbol{\mathrm{r}}\right)=Q_{ij} \left(r\right)=\left\langle u_{i} \left(\boldsymbol{\mathrm{x}}\right)u_{j} \left(\boldsymbol{\mathrm{x}}^{'} \right)\right\rangle. 
\end{equation} 
In N-body simulation, this average $\left\langle \cdot \right\rangle $ is taken over \textbf{all pairs} of particles with the same separation \textit{r} (Fig. \ref{fig:1}). A more compact alternative notation is to write $Q_{ij} \left(r\right)\equiv \left\langle u_{i} u_{j}^{'} \right\rangle $, where the prime indicates that quantity is evaluated at $\boldsymbol{\mathrm{x}}^{'} =\boldsymbol{\mathrm{x}}+\boldsymbol{\mathrm{r}}$. Due to symmetry in Eq. \eqref{ZEqnNum195930}
\begin{equation} 
\label{eq:11} 
Q_{ij} \left(-r\right)=\left\langle u_{i}^{'} u_{j} \right\rangle =Q_{ji} \left(r\right)=Q_{ij} \left(r\right)=\left\langle u_{i} u_{j}^{'} \right\rangle .        
\end{equation} 
Einstein index notation is adopted hereafter that implies the summation over a specific index, 
\begin{equation} 
\label{eq:12}
Q_{ii} \left(r\right)=\left\langle u_{i} u_{i}^{'} \right\rangle=\left\langle u_{1} u_{1}^{'} \right\rangle +\left\langle u_{2} u_{2}^{'} \right\rangle +\left\langle u_{3} u_{3}^{'} \right\rangle.
\end{equation} 
The most general form of $Q_{ij}(r)$ reads 
\begin{equation} 
\label{ZEqnNum221113} 
Q_{ij} \left(\boldsymbol{\mathrm{r}}\right)=Q_{ij} \left(r\right)=A_{2} \left(r\right)r_{i} r_{j} +B_{2} \left(r\right)\delta _{ij} ,       
\end{equation} 
where $A_{2} \left(r\right)$ and $B_{2} \left(r\right)$ are scalar functions of \textit{r} only. Finally, the divergence of $Q_{ij}(r)$ can be derived from Eq. \eqref{ZEqnNum221113}
\begin{equation} 
\label{ZEqnNum321178} 
Q_{ij,i} =\frac{\partial Q_{ij} \left(r\right)}{\partial r_{i} } =\left(4A_{2} +\frac{\partial A_{2} }{\partial r} r+\frac{1}{r} \frac{\partial B_{2} }{\partial r} \right)r_{j} .       
\end{equation} 
The curl of second order correlation tensor is derived as
\begin{equation} 
\label{ZEqnNum800260} 
\begin{split}
\nabla \times Q_{ij} \left(r\right)=\varepsilon _{imk} Q_{mj,k} 
&=\left(A_{2} \varepsilon _{imj} r_{m} +\frac{1}{r} \frac{\partial B_{2} }{\partial r} \varepsilon _{ijk} r_{k} \right)\\
&=\varepsilon _{imj} r_{m} \left(A_{2} -\frac{1}{r} \frac{\partial B_{2} }{\partial r} \right).
\end{split}
\end{equation} 
 
\subsubsection{Second order velocity correlation functions}
\label{sec:3.2.1}
\begin{figure}
\includegraphics*[width=\columnwidth]{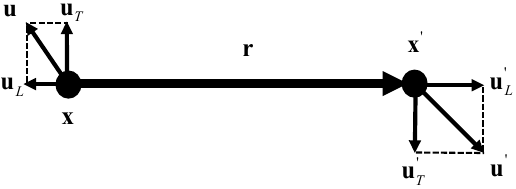}
\caption{Velocity correlation at two particles separated by vector $\textbf{r}$. Here $\boldsymbol{\mathrm{u}}_{T}$ and $\boldsymbol{\mathrm{u}}_{T}^{'} $ are two transverse velocities, while $u_{L} $ and $u_{L}^{'} $ are two longitudinal velocities at $\textbf{x}$ and $\textbf{x}^{'}$, respectively.} 
\label{fig:1}
\end{figure}

It is more convenient to introduce three scalar velocity correlation functions by index contraction of $Q_{ij} \left(r\right)$ in Eq. \eqref{ZEqnNum221113}. Namely, a total correlation function $R_{2} \left(r\right)$ can be defined as
\begin{equation} 
\label{ZEqnNum935612} 
R_{2} \left(r\right)=Q_{ij} \delta _{ij} =\left\langle \boldsymbol{\mathrm{u}}\cdot \boldsymbol{\mathrm{u}}_{}^{'} \right\rangle =\left\langle u_{i} u_{i}^{'} \right\rangle =A_{2} r^{2} +3B_{2} .      
\end{equation} 
The longitudinal correlation function $L_{2} \left(r\right)$ is defined as
\begin{equation}
\label{ZEqnNum950112} 
L_{2} \left(r\right)=Q_{ij} {r_{i} r_{j} /r^{2} } =\left\langle u_{L} u_{L}^{'} \right\rangle =A_{2} r^{2} +B_{2} ,       
\end{equation} 
where $u_{L} =\boldsymbol{\mathrm{u}}\cdot \hat{\boldsymbol{\mathrm{r}}}=u_{i} \hat{r}_{i} $ is the longitudinal velocity and $\hat{r}_{i} ={r_{i} /r} $ is the normalized Cartesian component of vector \textbf{r }between two particles. The transverse (lateral) correlation function reads:
\begin{equation} 
\label{ZEqnNum991811} 
T_{2} \left(r\right)=Q_{ij} n_{i} n_{j} ={\left\langle \boldsymbol{\mathrm{u}}_{T} \cdot \boldsymbol{\mathrm{u}}_{T}^{'} \right\rangle /2} =B_{2} \left(r\right),   
\end{equation} 
where \textbf{\textit{n}} is normal vector perpendicular to the direction of separation \textbf{r }($\boldsymbol{\mathrm{n}}\cdot \hat{\boldsymbol{\mathrm{r}}}=0$) and 
\begin{equation} 
\label{eq:19} 
\boldsymbol{\mathrm{u}}_{T} =-\left(\boldsymbol{\mathrm{u}}\times \hat{\boldsymbol{\mathrm{r}}}\times \hat{\boldsymbol{\mathrm{r}}}\right)=\boldsymbol{\mathrm{u}}-\left(\boldsymbol{\mathrm{u}}\cdot \hat{\boldsymbol{\mathrm{r}}}\right)\hat{\boldsymbol{\mathrm{r}}} 
\end{equation} 
is the transverse velocity perpendicular to vector \textbf{\textit{r} }with\textbf{ }$\times $ standing for the cross product. Figure \ref{fig:1} provides a sketch of the longitudinal and transverse velocities.

\begin{figure}
\includegraphics*[width=\columnwidth]{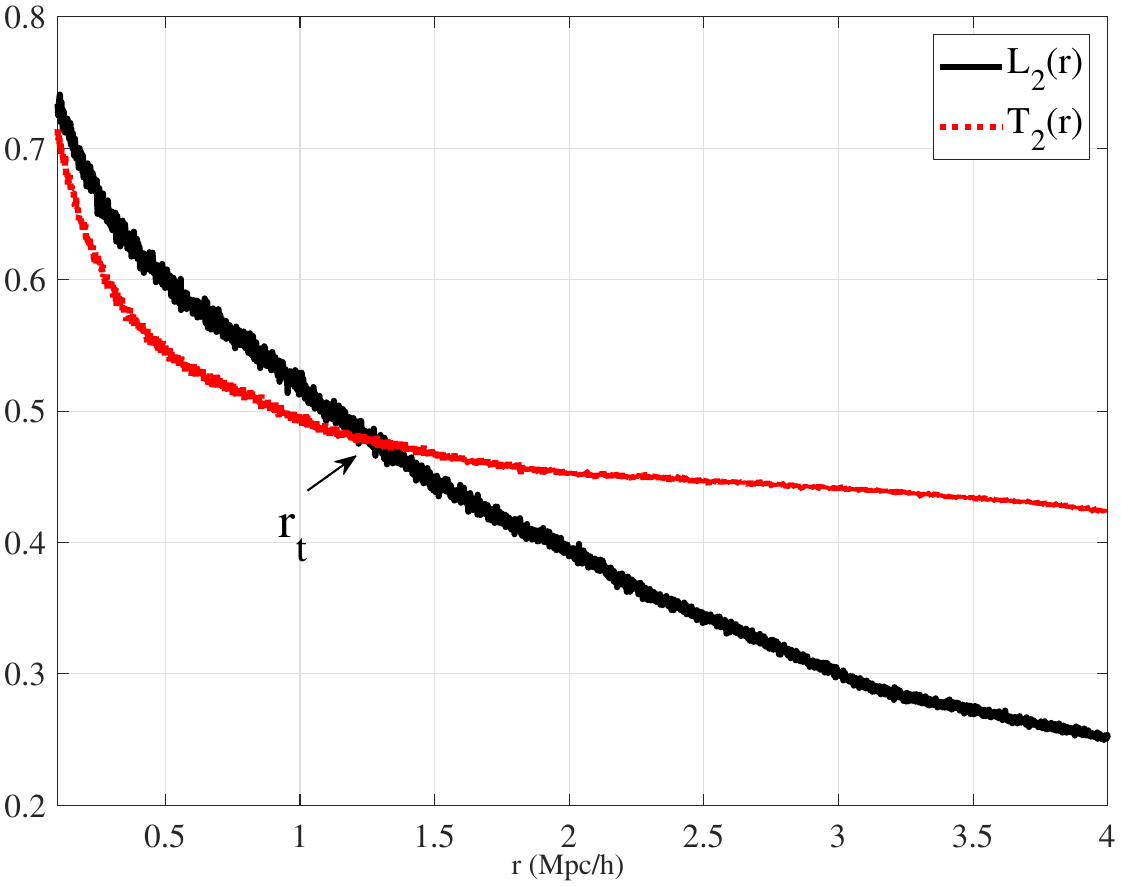}
\caption{The crossover between transverse correlation $T_{2} \left(r\right)$ and longitudinal correlation $L_{2} \left(r\right)$ at around $r_t$=1.3Mpc/h (black arrow).}
\label{fig:3b}
\end{figure}

Two correlation coefficients can be defined for longitudinal and transverse velocity,  
\begin{equation}
\rho _{L} \left(r\right)=\frac{\left\langle u_{L} u_{L}^{'} \right\rangle }{\left\langle u_{L}^{2} \right\rangle } \quad \textrm{and} \quad \rho _{T} \left(r\right)=\frac{\left\langle \boldsymbol{\mathrm{u}}_{T} \cdot \boldsymbol{\mathrm{u}}_{T}^{'} \right\rangle }{\left\langle \left|\boldsymbol{\mathrm{u}}_{T} \right|^{2} \right\rangle },  
\label{ZEqnNum890288}
\end{equation}

\noindent both of which should vanish at large distance $r\to\infty$ (velocities of two particles far from each other should be uncorrelated) and increases with $r\to0$ due to the increasing gravitational interaction (Fig. \ref{fig:15}). Similarly, we can define angles between two longitudinal and transverse velocities 
\begin{equation}
\left\langle \cos \left(\theta _{L} \right)\right\rangle =\left\langle \frac{u_{L} u_{L}^{'} }{\left|u_{L} \right|\left|u_{L}^{'} \right|} \right\rangle \textrm{,} \quad  \left\langle \cos \left(\theta _{T} \right)\right\rangle =\left\langle \frac{\boldsymbol{\mathrm{u}}_{T} \cdot \boldsymbol{\mathrm{u}}_{T}^{'} }{\left|\boldsymbol{\mathrm{u}}_{T} \right|\left|\boldsymbol{\mathrm{u}}_{T}^{'} \right|} \right\rangle.   
\label{ZEqnNum924986}
\end{equation}
\noindent The angle between velocity vector $\boldsymbol{\mathrm{u}}$ and \textbf{\textit{r}} vector and the angle between two velocities $\boldsymbol{\mathrm{u}}$ and $\boldsymbol{\mathrm{u}}^{'} $ can also be defined for all pair of particles with the same separation \textit{r},
\begin{equation}
\left\langle \cos \left(\theta _{\boldsymbol{\mathrm{ur}}} \right)\right\rangle =\left\langle \frac{\boldsymbol{\mathrm{u}}\cdot \boldsymbol{\mathrm{r}}}{\left|\boldsymbol{\mathrm{u}}\right|\left|\boldsymbol{\mathrm{r}}\right|} \right\rangle \textrm{,} \quad \left\langle \cos \left(\theta _{\boldsymbol{\mathrm{uu}}'} \right)\right\rangle =\left\langle \frac{\boldsymbol{\mathrm{u}}\cdot \boldsymbol{\mathrm{u}}^{'} }{\left|\boldsymbol{\mathrm{u}}\right|\left|\boldsymbol{\mathrm{u}}^{'} \right|} \right\rangle.   \label{ZEqnNum170842}
\end{equation}
\noindent Note that all averaging is performed over all pairs of particles with the same separation \textit{r}. These quantities will greatly improve our understanding of velocity field (Figs. \ref{fig:15}, \ref{fig:16}, \ref{fig:17a}, and \ref{fig:17b}). The total velocity correlation reads (Eqs. \eqref{ZEqnNum935612}-\eqref{ZEqnNum991811}) 
\begin{equation} 
\label{ZEqnNum610884} 
R_{2} \left(r\right)=2R\left(r\right)=L_{2} \left(r\right)+2T_{2} \left(r\right),        
\end{equation} 
where $R\left(r\right)$ is the standard velocity correlation function in turbulence literature. Three second order correlation functions from N-body simulations are presented in Figs. \ref{fig:2}, \ref{fig:3a}, and \ref{fig:3b}.

The velocity power spectrum $E_{u} \left(k\right)$ and correlation function $R\left(r\right)$ are related to each other through the Fourier transform pair
\begin{equation} 
\label{ZEqnNum609039} 
R\left(r\right)=\int _{0}^{\infty }E_{u} \left(k\right)\frac{\sin \left(kr\right)}{kr} dk,        
\end{equation} 
\begin{equation} 
\label{ZEqnNum891034} 
E_{u} \left(k\right)=\frac{2}{\pi } \int _{0}^{\infty }R\left(r\right)kr\sin \left(kr\right)dr .        
\end{equation} 

A length scale $l_{u0} $ (integral scale) is introduced as
\begin{equation} 
\label{eq:26} 
l_{u0} =\frac{1}{u^{2} } \int _{0}^{\infty }R \left(r\right)dr=\frac{\pi }{2u^{2} } \int _{0}^{\infty }E_{u}  \left(k\right)k^{-1} dk 
\end{equation} 
that defines a characteristic length scale within which, velocities are appreciably correlated. This is the length scale at which energy is injected into or taken out of the system. 

\subsubsection{Second order velocity dispersion functions and energy distribution in real space}
\label{sec:3.2.2}
The one-dimensional variance of smoothed velocity (the bulk flow velocity) with a filter of size \textit{r} can be written as follows:
\begin{equation} 
\label{ZEqnNum726048} 
\sigma _{u}^{2} \left(r\right)=\frac{1}{3} \int _{-\infty }^{\infty }E_{u} \left(k\right)W\left(kr\right)^{2} dk =\int _{r}^{\infty }E_{ur} \left(r^{'} \right)dr^{'} , 
\end{equation} 
which is the variance of smoothed velocity field with a filter of size \textit{r}, or equivalently the kinetic energy contained in all scales above \textit{r}. Here $E_{ur} \left(r\right)$ is the energy density in real space that describes the energy distribution with respect to scale \textit{r}, just like the energy spectrum $E_{u} \left(k\right)$ in Fourier space with respect to wavenumber $k$ (Fig. \ref{fig:10}). Here $W\left(x\right)$ is a window function. For a top-hat spherical filter
\begin{equation} 
\label{ZEqnNum605014} 
W\left(x\right)=\frac{3}{x^{3} } \left[\sin \left(x\right)-x\cos \left(x\right)\right]=3\frac{j_{1} \left(x\right)}{x} , \end{equation} 
where $j_{1} \left(x\right)$ is the \textit{first} order spherical Bessel function of the \textit{first} kind. The velocity dispersion on the same scale \textit{r} is
\begin{equation} 
\label{ZEqnNum904685} 
\sigma _{d}^{2} \left(r\right)=\frac{1}{3} \int _{-\infty }^{\infty }E_{u} \left(k\right)\left[1-W\left(kr\right)^{2} \right]dk ,        
\end{equation} 
which is the kinetic energy contained in scales below \textit{r}. 

The exact relation between total velocity correlation $R_{2}(r)$ and $\sigma _{u}^{2} \left(r\right)$ can be derived from Eqs. \eqref{ZEqnNum609039} and \eqref{ZEqnNum726048}
\begin{equation} 
\label{ZEqnNum910224} 
R_{2}^{} \left(2r\right)=\frac{1}{24r^{2} } \frac{\partial }{\partial r} \left(\frac{1}{r^{2} } \frac{\partial }{\partial r} \left(r^{3} \frac{\partial }{\partial r} \left(\sigma _{u}^{2} \left(r\right)r^{4} \right)\right)\right).      
\end{equation} 
The total kinetic energy are decomposed into two parts:
\begin{equation} 
\label{ZEqnNum926839} 
\sigma _{u}^{2} \left(r\right)+\sigma _{d}^{2} \left(r\right)=u_{}^{2} .          
\end{equation} 
The specific kinetic energy of entire system $u_{}^{2}$ can have two separate contributions on any scale of \textit{r}, i.e. the velocity dispersion $\sigma _{d}^{2} \left(r\right)$ for kinetic energy contained in scales below \textit{r} and velocity variance $\sigma _{u}^{2} \left(r\right)$ for kinetic energy contained in scales above \textit{r} (Fig. \ref{fig:8}). From Eq. \eqref{ZEqnNum910224}, total velocity correlation $R_{2}^{} \left(r\right)$ can be used to derive both dispersion functions. 

Just like the energy spectrum $E_{u} \left(k\right)$ in Fourier space, the energy distribution in real space with respect to scale \textit{r} can be found from dispersion functions as (from Eq. \eqref{ZEqnNum726048}),
\begin{equation} 
\label{ZEqnNum359490} 
E_{ur} \left(r\right)=-\frac{\partial \sigma _{u}^{2} \left(r\right)}{\partial r} .          
\end{equation} 
The connection with energy spectrum $E_{u} \left(k\right)$ in Fourier space can be found from Eq. \eqref{ZEqnNum726048},
\begin{equation}
E_{ur} \left(r\right)r^{2}=-\frac{4}{3} \int _{0}^{\infty }E_{u} \left(\frac{x}{r} \right)W\left(x\right)W^{'} \left(x\right)xdx,
\label{eq:33}
\end{equation}
where $x=kr$ and $W^{'}(x)$ is the derivative of $W(x)$ with respect to $x$. In principle, $E_{ur} \left(r\right)$ and $E_{u} \left(k\right)$ contain the same information on the distribution of energy.

Two velocity dispersion functions $\sigma _{d}^{2} \left(r\right)$ and $\sigma _{u}^{2} \left(r\right)$ can be derived from the velocity correlation function $R_{2}^{} \left(r\right)$, which in turn can be obtained accurately from N-body simulation (Fig. \ref{fig:13}). Specifically, for a top-hat filter, a power-law spectrum $E_{u} \left(k\right)\equiv bk^{-m} $ will lead to (from Eqs. \eqref{ZEqnNum609039} and \eqref{ZEqnNum726048}),
\begin{equation} 
\label{eq:34} 
R_{2} \left(r\right)=2b\Gamma \left(-m\right)\sin \left(\frac{m\pi }{2} \right)r^{m-1} ,        
\end{equation} 
\begin{equation} 
\label{ZEqnNum585776}
\begin{split}
\sigma _{d}^{2} \left(r\right)=&-6\cdot 2^{2+m} \left(1+m\right)\left(4+m\right)\\&\cdot \Gamma \left(-5-m\right)\sin \left(\frac{m\pi }{2} \right)br^{m-1},     
\end{split}
\end{equation} 
where two functions satisfy relation in Eq. \eqref{ZEqnNum910224}. For self-gravitating collisionless dark matter flow, 
\begin{equation}
\sigma _{d}^{2} \left(r\right)={\left(\sigma _{11}^{2} \left(r\right)+\sigma _{22}^{2} \left(r\right)+\sigma _{33}^{2} \left(r\right)\right)/3},       
\label{eq:36}
\end{equation}
\noindent where $\sigma _{ij}^{2} \left(r\right)$ is the velocity dispersion tensor on scale of \textit{r}. The off-diagonal terms are small and negligible (Fig. \ref{fig:14}).

\subsubsection{Results for incompressible flow}
\label{sec:3.2.3}
The divergence free condition for $Q_{ij}$ leads to (Eq. \eqref{ZEqnNum321178}) 
\begin{equation} 
\label{ZEqnNum823415} 
r\frac{\partial B_{2} }{\partial r} =\frac{1}{r} \frac{\partial \left(r^{2} B_{2} \right)}{\partial r} -2B_{2} =-\frac{1}{r} \frac{\partial \left(r^{4} A_{2} \right)}{\partial r} .       
\end{equation} 
Equation \eqref{ZEqnNum823415} can be rewritten as
\begin{equation} 
\label{eq:38} 
B_{2} =\frac{1}{2r} \frac{\partial \left(r^{4} A_{2} +r^{2} B_{2} \right)}{\partial r} .         
\end{equation} 
With the help of Eqs. \eqref{ZEqnNum950112}, \eqref{ZEqnNum991811} and \eqref{ZEqnNum610884}, the kinematic relations between three scalar correlations functions read
\begin{equation}
T_{2} =\frac{1}{2r} \left(r^{2} L_{2} \right)_{,r} \quad \textrm{and} \quad R_{2} =\frac{1}{r^{2} } \left(r^{3} L_{2} \right)_{,r}.      
\label{ZEqnNum314105}
\end{equation}

The two scalar functions $A_{2} \left(r\right)$ and $B_{2} \left(r\right)$, and the correlation tensor $Q_{ij} \left(r\right)$ can be finally expressed in terms of the longitudinal correlation function $L_{2} \left(r\right)$ as,
\begin{equation}
A_{2} =-\frac{1}{2r} \left(L_{2}^{} \right)_{,r} \quad \textrm{and} \quad B_{2} =\frac{1}{2r} \left(r^{2} L_{2} \right)_{,r},      
\label{eq:40}
\end{equation}
\begin{equation} 
\label{eq:41} 
Q_{ij} \left(r\right)=-\frac{1}{2r} \left[\left(L_{2}^{} \right)_{,r} r_{i} r_{j} -\left(r^{2} L_{2} \right)_{,r} \delta _{ij} \right].       
\end{equation} 
The relations between the longitudinal $L_{2} \left(r\right)$, transverse correlation functions $T_{2} \left(r\right)$, and velocity power spectrum $E_{u} \left(k\right)$ can be equivalently derived using Eqs. \eqref{ZEqnNum891034} and \eqref{ZEqnNum314105}
\begin{equation} 
\label{ZEqnNum333566} 
L_{2} \left(r\right)=\int _{0}^{\infty }E_{u} \left(k\right)\frac{2j_{1} \left(kr\right)}{kr} dk, 
\end{equation} 
\begin{equation} 
\label{eq:43} 
T_{2} \left(r\right)=\int _{0}^{\infty }E_{u} \left(k\right)\left(j_{0} \left(kr\right)-\frac{j_{1} \left(kr\right)}{kr} \right)dk,         
\end{equation} 
where $j_{n} \left(kr\right)$ is the \textit{n}th order spherical Bessel function of the \textit{first} kind. The first two spherical Bessel functions are
\begin{equation}
\begin{split}
&j_{0} \left(x\right)=\frac{\sin \left(x\right)}{x} \approx 1-\frac{x^{2} }{6},\\ 
&j_{1} \left(x\right)=\frac{\sin \left(x\right)}{x^{2} } -\frac{\cos \left(x\right)}{x} \approx \frac{x}{3} -\frac{x^{3} }{30}. 
\end{split}
\label{ZEqnNum351152}
\end{equation}
\noindent With Eqs. \eqref{ZEqnNum314105} and \eqref{ZEqnNum333566} and integration by parts, the integral length scale $l_{u0} $ for incompressible flow is
\begin{equation}
\begin{split}
&l_{u0} =\frac{1}{u^{2} } \int _{0}^{\infty }R \left(r\right)dr=\frac{1}{u^{2} } \int _{0}^{\infty }L_{2}  \left(r\right)dr,\\ 
&\int _{0}^{\infty }T_{2}  \left(r\right)dr=0.
\end{split}
\label{eq:45}
\end{equation}

\subsubsection{Results for constant divergence flow}
\label{sec:3.2.4}
For constant divergence flow with $\nabla \cdot \boldsymbol{\mathrm{u}}=\theta$, we should have
\begin{equation} 
\label{eq:46} 
Q_{ij,i} =\frac{\partial Q_{ij} \left(r\right)}{\partial r_{i} } =-\left\langle \frac{\partial u_{i} }{\partial x_{i} } u_{j}^{'} \right\rangle =-\theta \left\langle u_{j}^{'} \right\rangle =\theta \left\langle u_{j} \right\rangle .      
\end{equation} 
It can be shown that $Q_{ij,i} =0$ for velocity field with a zero mean ($\langle u_{j}^{'} \rangle =0$) such that Eq. \eqref{ZEqnNum823415} for incompressible flow is still valid for constant divergence flow. Therefore, all results for second order correlation functions in Section \ref{sec:3.2.3} are also valid for constant divergence flow. More general, the incompressible and constant divergence flow share the same kinematic relations on even order. However, kinematic relations can be different on odd order for incompressible and constant divergence flow \citep{Xu:2022-The-statistical-theory-of-3rd}.  
 
\subsubsection{Results for irrotational flow}
\label{sec:3.2.5}
The curl free condition for irrotational flow (Eq. \eqref{ZEqnNum800260}) leads to different kinematic relations between correlation functions,
\begin{equation}
R_{2} =\frac{1}{r^{2} } \left(r^{3} T_{2} \right)_{,r} \quad \textrm{and} \quad L_{2} =\left(rT_{2} \right)_{,r}. \label{ZEqnNum320035}
\end{equation}
\noindent Similarly, the velocity correlation tensor $Q_{ij} \left(r\right)$ can be written in terms of the transverse correlation function $T_{2} \left(r\right)$ only
\begin{equation} 
\label{eq:48} 
Q_{ij} \left(r\right)=\left(T_{2}^{} \right)_{,r} \frac{r_{i} r_{j} }{r} +T_{2} \delta _{ij} .         
\end{equation} 

Relations between longitudinal and transverse correlation functions and velocity power spectrum $E_{u}(k)$ can be found as (from Eq. \eqref{ZEqnNum320035}) or see \citet{Gorski:1988-On-the-Pattern-of-Perturbation}
\begin{equation} 
\label{ZEqnNum127370} 
L_{2} \left(r\right)=2\int _{0}^{\infty }E_{u} \left(k\right)\left(j_{0} \left(kr\right)-2\frac{j_{1} \left(kr\right)}{kr} \right)dk ,       
\end{equation} 
\begin{equation}
\label{ZEqnNum324737} 
T_{2} \left(r\right)=\int _{0}^{\infty }E_{u} \left(k\right)\frac{2j_{1} \left(kr\right)}{kr} dk . 
\end{equation} 

The integral length scale $l_{u0}$ for irrotational velocity reads
\begin{equation}
\begin{split}
&l_{u0} =\frac{1}{u^{2} } \int _{0}^{\infty }R \left(r\right)dr=\frac{1}{u^{2} } \int _{0}^{\infty }T_{2}  \left(r\right)dr, \\
&\int _{0}^{\infty }L_{2}  \left(r\right)dr=0.
\end{split}
\label{ZEqnNum505635}
\end{equation}
 
\subsubsection{The nature of dark matter flow on small and large scales}
\label{sec:3.2.6}
Clearly, different kinematic relations between correlation functions (Eqs. \eqref{ZEqnNum314105} and \eqref{ZEqnNum320035}) can be used to characterize the nature of flow (flow characterization). For example, we can compute the integral $R_{2i} \left(r\right)$ of total velocity correlation function $R_{2} \left(r\right)$ as
\begin{equation} 
\label{ZEqnNum956600} 
R_{2i} =\frac{1}{r^{3} } \int _{0}^{r}R_{2}  \left(y\right)y^{2} dy.         
\end{equation} 
It can be shown that $R_{2i} =L_{2} $ (from Eq. \eqref{ZEqnNum314105}) if the flow is incompressible (or constant divergence flow) and $R_{2i} =T_{2} $ (from Eq. \eqref{ZEqnNum320035}) if the flow is irrotational. This information can be used to demonstrate that the self-gravitating collisionless dark matter flow is of constant divergence on small scale and irrotational on large scale (see Fig. \ref{fig:4}). 

\subsection{Second order velocity structure functions}
\label{sec:3.3}
Structure functions are common statistical measures for turbulent flow. We will introduce different structure functions as quantitative measures for kinetic energy and enstrophy contained in scales below \textit{r}. Structure functions are relatively easy to measure from observation. 

\subsubsection{Second order longitudinal structure function and limiting correlations on small scale}
\label{sec:3.3.1}
The most common one is the second order longitudinal structure function originally defined as 
\begin{equation} 
\label{ZEqnNum250774} 
S_{2}^{lp} \left(r\right)=\left\langle \left(\Delta u_{L} \right)^{2} \right\rangle =\left\langle \left(u_{L}^{'} -u_{L}^{} \right)^{2} \right\rangle =2\left(\left\langle u_{L}^{2} \right\rangle -L_{2} \left(r\right)\right),    
\end{equation} 
where $\left\langle u_{L}^{2} \right\rangle$ is the second moment of longitudinal velocity $u_{L}$ on scale \textit{r} (Fig. \ref{fig:20}). The velocity difference on scale \textit{r} (or pairwise velocity in cosmology) is defined as $\Delta u_{L} \left(r\right)=u_{L}^{'} -u_{L}^{}$. The first order longitudinal structure function reads 
\begin{equation} 
\label{eq:54} 
S_{1}^{lp} \left(r\right)=\left\langle \Delta u_{L} \right\rangle =\left\langle u_{L}^{'} -u_{L}^{} \right\rangle , \end{equation} 
which is essentially the mean pairwise velocity. 

A modified (alternative) definition of second order longitudinal structure function is
\begin{equation} 
\label{ZEqnNum414891} 
S_{2}^{l} \left(r\right)=2\left({\mathop{\lim }\limits_{r\to 0}} \left\langle u_{L}^{} u_{L}^{'} \right\rangle -L_{2} \left(r\right)\right)=2\left(u^{2} -L_{2} \left(r\right)\right),      
\end{equation} 
where $u^{2} $ is one-dimensional velocity dispersion in Eq. \eqref{ZEqnNum586699}. Two structure functions $S_{2}^{lp} \left(r\right)\ne S_{2}^{l} \left(r\right)$ is a manifestation of the collisionless nature of dark matter flow.\\

\noindent \textbf{Remarks:} For collisional incompressible flow $\left\langle u_{L}^{2} \right\rangle =u_{}^{2} $ on all scales, where $\left\langle u_{L}^{2} \right\rangle $ is independent of the scale \textit{r}. In addition 
\begin{equation} 
\label{eq:56} 
{\mathop{\lim }\limits_{r\to 0}} L_{2} \left(r\right)={\mathop{\lim }\limits_{r\to 0}} \left\langle u_{L}^{} u_{L}^{'} \right\rangle =u_{}^{2} ={\mathop{\lim }\limits_{r\to 0}} \left\langle u_{L}^{2} \right\rangle ,        
\end{equation} 
where $u_{L}^{} $ and $u_{L}^{'} $ are fully correlated when $r\to 0$ due to the collisional interaction. Or equivalently, the correlation coefficient $\rho _{L} \left(r=0\right)=1$ (see Eq. \eqref{ZEqnNum890288}). For incompressible flow, two definitions are equivalent, i.e. $S_{2}^{lp} \left(r\right)=S_{2}^{l} \left(r\right)$ and $S_{2}^{lp} \left(0\right)=S_{2}^{l} \left(0\right)=0$. 

However, this picture is completely different for dark matter flow (SG-CFD). Due to the collisionless nature, the longitudinal velocities $u_{L}^{} $ and $u_{L}^{'} $ are not fully correlated on small scale ($r\to 0$). The maximum correlation coefficient between $u_{L}$ and $u_{L}^{'}$ is $\rho _{L} \left(r=0\right)=0.5$ that can be estimated as follows: 

Based on the halo description of self-gravitating system, any pair of particles with small separation $r\to 0$ should come from the same halo, while different pairs can be from different haloes of different size. The particle velocity dispersion in haloes can be decomposed into,
\begin{equation} 
\label{eq:57}
\begin{split}
\sigma ^{2} \left(m_{h} \right)&=\sigma _{v}^{2} \left(m_{h} \right)+\sigma _{h}^{2} \left(m_{h} \right)\\&=\left(1-\rho _{cor} \left(m_{h} \right)\right)\sigma ^{2} +\rho _{cor} \left(m_{h} \right)\sigma ^{2},    
\end{split}
\end{equation} 
i.e. the sum of halo velocity dispersion $\sigma _{h}^{2} $ and halo virial dispersion $\sigma _{v}^{2} $, where $m_h$ is the mass of the halo that the particle resides in. The correlation coefficient $\rho _{cor} $ between velocities of pair of particles in haloes of mass $m_{h} $ approximately reads,
\begin{equation} 
\label{ZEqnNum350084} 
\rho _{cor} \left(m_{h} \right)=\frac{\sigma _{h}^{2} }{\sigma ^{2} } =\frac{1}{1+{\sigma _{v}^{2} /\sigma _{h}^{2} } } .   \end{equation} 
Velocities of pair of particles with separation $r\to 0$ in small haloes should be fully correlated, i.e. $\rho _{cor} \left(m_{h} \ll m_{h}^{*} \right)\approx 1$ with $\sigma _{v}^{2} \ll \sigma _{h}^{2} $ for small haloes. While velocities of particle pairs from large haloes should be independent of each other, i.e., $\rho _{cor} \left(m_{h} \gg m_{h}^{*} \right)\approx 0$ with $\sigma _{v}^{2} \gg \sigma _{h}^{2} $. Here $m_{h}^{*} $ is a characteristic mass sale. 

The average correlation for all pairs with same $r$ from all haloes of different size can be estimated as $\sigma _{v}^{2} =\sigma _{h}^{2} $ such that $\left\langle \rho _{cor} \right\rangle =0.5$ when $r\to 0$ (see Fig. \ref{fig:15} from N-body simulation). The average correlation coefficient of longitudinal velocity for all pairs on the limiting scale $r\to 0$ is $\rho _{L} \left(r=0\right)=\left\langle \rho _{cor} \right\rangle =0.5$ such that
\begin{equation} 
\label{ZEqnNum890245} 
{\mathop{\lim }\limits_{r\to 0}} L_{2} \left(r\right)={\mathop{\lim }\limits_{r\to 0}} \left\langle u_{L}^{} u_{L}^{'} \right\rangle ={\mathop{\lim }\limits_{r\to 0}} \rho _{L} \left\langle u_{L}^{2} \right\rangle =\frac{1}{2} {\mathop{\lim }\limits_{r\to 0}} \left\langle u_{L}^{2} \right\rangle.      
\end{equation} 
Therefore, in SG-CFD, two structure functions are not equivalent, i.e., $S_{2}^{lp} \left(r\right)\ne S_{2}^{l} \left(r\right)$ (Eqs. \eqref{ZEqnNum250774} and \eqref{ZEqnNum414891}). On the smallest scale $r\to 0$, the original definition
\begin{equation}
S_{2}^{lp} \left(r=0\right)={\mathop{\lim }\limits_{r\to 0}} \left\langle u_{L}^{2} \right\rangle \quad \textrm{and} \quad S_{2}^{l} \left(r=0\right)=0,      
\label{eq:60}
\end{equation}
\noindent while on the largest scale (see Fig. \ref{fig:13}) 
\begin{equation}
\label{ZEqnNum924181} 
S_{2}^{lp} \left(r=\infty \right)=S_{2}^{l} \left(r=\infty \right)=2u_{}^{2} .        
\end{equation} 

\subsubsection{Mass-energy and momentum conservation for collisionless "annihilation" at \texorpdfstring{$r=0$}{}}
\label{sec:3.3.2}
The fact that $\rho _{L} \left(r=0\right)\ne 1$ might lead to some interesting phenomena on the smallest scale. Just like the electron-positron annihilation, let us check the simplest "one-dimension annihilation" scenario of two dark matter particles on the smallest scale $r=0$. Assuming that gravity is the only interaction and no radiation produced from "annihilation", two particles with the same mass \textit{m} only annihilate when the separation between them vanishes ($r=0$). The conservation of momentum requires,
\begin{equation}
mv_{1} +mv_{2} =m'v_{3} \quad \textrm{such that} \quad v_{3} =\frac{m}{m'}(v_{1} +v_{2}),
\label{eq:62}
\end{equation}
\noindent where $m'$ and $v_{3} $ are the rest mass and velocity of new particle after ``annihilation''.  Taking the average for all pairs of particles with $r=0$, $v_{1} =u_{L}$, and $v_{2} =u_{L}^{'}$ 
\begin{equation} 
\label{eq:63} 
\begin{split}
\left\langle v_{3}^{2} \right\rangle=\left(\frac{m}{m'} \right)^{2} \left\langle \left(v_{1} +v_{2} \right)^{2} \right\rangle&=2\left(\frac{m}{m'} \right)^{2} \left(\left\langle v_{1}^{2} \right\rangle +\left\langle v_{1} v_{2} \right\rangle \right)\\
&=2\left(\frac{m}{m'} \right)^{2} \left\langle u_{L}^{2} \right\rangle \left(1+\rho _{L0} \right),
\end{split}
\end{equation} 
where $\rho _{L0} \equiv \rho _{L} \left(r=0\right)$ is the limiting velocity correlation coefficient at $r=0$. Assuming no radiation is produced, the conservation of total mass-energy can be written as
\begin{equation} 
\label{eq:64)} 
mc^{2} +\frac{1}{2} mv_{1}^{2} +mc^{2} +\frac{1}{2} mv_{2}^{2} =m'c^{2} +\frac{1}{2} m'v_{3}^{2} ,       
\end{equation} 
where \textit{c} is the speed of light. Taking average and we have:
\begin{equation} 
\label{ZEqnNum962574}
\begin{split}
m'=m\left[1+\frac{\left\langle u_{L}^{2} \right\rangle }{2c^{2} } +\sqrt{1-\rho _{L0} \frac{\left\langle u_{L}^{2} \right\rangle }{c^{2} } +\frac{\left\langle u_{L}^{2} \right\rangle ^{2} }{4c^{4} } } \right]\\
\approx m\left[2+\left(1-\rho _{L0} \right)\frac{\left\langle u_{L}^{2} \right\rangle }{2c^{2} } \right].
\end{split}
\end{equation} 
Therefore, for $\rho_{L0} ={1/2} \ne 1$, the particle "annihilation" leads to extra mass converted from kinetic energy if gravity is the only interaction and no radiation is produced. If mass is conserved ($m^{'}=2m$), the total radiation should be produced on the order of $M \langle u_{L}^{2}\rangle$, where $M$ is the total mass of dark matter \citep{Xu:2022-Postulating-dark-matter-partic}. 
 
\subsubsection{Second order total structure function for kinetic energy}
\label{sec:3.3.3}
The second order total velocity structure functions are introduced in the vector forms, 
\begin{equation} 
\label{eq:66} 
S_{2}^{ip} \left(r\right)=\left\langle \Delta \boldsymbol{\mathrm{u}}^{2} \right\rangle =\left\langle \left(\boldsymbol{\mathrm{u}}^{'} -\boldsymbol{\mathrm{u}}\right)^{2} \right\rangle =6\left\langle u_{L}^{2} \right\rangle -2R_{2} \left(r\right),      
\end{equation} 
\begin{equation} 
\label{ZEqnNum973047} 
S_{2}^{i} \left(r\right)=6u_{}^{2} -2R_{2} \left(r\right),         
\end{equation} 
where $S_{2}^{ip} \left(r\right)$ is the original version and  $S_{2}^{i} \left(r\right)$ is the modified version. Just like the longitudinal structure functions in Section \ref{sec:3.3.1}, two definitions are equivalent for incompressible flow and $S_{2}^{i} \left(r\right)\ne S_{2}^{ip} \left(r\right)$ for SG-CFD. The relation between structure function $S_{2}^{i} \left(r\right)$ and velocity spectrum $E_{u} \left(k\right)$ can be found from Eq. \eqref{ZEqnNum609039} 
\begin{equation} 
\label{ZEqnNum594172} 
S_{2}^{i} \left(r\right)=4\int _{0}^{\infty }E_{u} \left(k\right)\left(1-j_{0} \left(kr\right)\right)dk,    
\end{equation} 
where $S_{2}^{i} \left(r=0\right)=0$, just like the modified longitudinal structure function $S_{2}^{l} \left(r=0\right)=0$. 

For a power-law velocity spectrum $E_{v} \left(k\right)\equiv bk^{-m} $, we have a power-law structure function
\begin{equation} 
\label{ZEqnNum911575} 
\begin{split}
S_{2}^{i} \left(r\right)&=4\int _{0}^{\infty }E_{u} \left(k\right)\left[1-j_{0}^{} \left(kr\right)\right]dk\\ &=-\frac{2^{2-m} \Gamma \left({3/2} \right)\Gamma \left({\left(1-m\right)/2} \right)}{\Gamma \left({\left(m+2\right)/2} \right)} br^{m-1}.
\end{split}
\end{equation} 
The exact relation between the total structure function $S_{2}^{i} \left(r\right)$ and one-dimensional velocity dispersion $\sigma _{d}^{2} \left(r\right)$ can be derived using Eqs. \eqref{ZEqnNum973047}, \eqref{ZEqnNum926839} and \eqref{ZEqnNum910224} 
\begin{equation} 
\label{ZEqnNum947585} 
S_{2}^{i} \left(2r\right)=\frac{1}{12r^{2} } \frac{\partial }{\partial r} \left(\frac{1}{r^{2} } \frac{\partial }{\partial r} \left(r^{3} \frac{\partial }{\partial r} \left(\sigma _{d}^{2} \left(r\right)r^{4} \right)\right)\right),      
\end{equation} 
where $S_{2}^{i} \left(r\right)$ is a quantity relevant to the specific kinetic energy contained below the scale \textit{r} (i.e. $\sigma _{d}^{2} \left(r\right)$ in Eq. \eqref{ZEqnNum904685}) through Eq. \eqref{ZEqnNum947585}. While the total second order correlation $R_{2}^{} \left(r\right)$ is related to the specific kinetic energy above the scale \textit{r} (i.e. $\sigma _{u}^{2} \left(r\right)$ in Eq. \eqref{ZEqnNum926839}) through Eq. \eqref{ZEqnNum910224}. 

\subsubsection{Second order structure function for enstrophy and real space enstrophy distribution}
\label{sec:3.3.4}
The enstrophy of entire system can be of interest for some applications, where the total enstrophy $E_{n} $ in the system is defined as
\begin{equation} 
\label{eq:71} 
E_{n} =\int _{0}^{\infty }E_{u} \left(k\right)k^{2} dk. 
\end{equation} 
The enstrophy is related to the integral of the gradient of velocity field with contributions from both divergence and vorticity (Eq. \eqref{ZEqnNum266353}). 

We can introduce a different structure function that is relevant to enstrophy on different scales. For a top-hat filter of size \textit{r}, we introduce the structure function $S_{2}^{x} \left(r\right)$ (see Fig. \ref{fig:11} for simulation result) 
\begin{equation} 
\label{ZEqnNum356384} 
\begin{split}
S_{2}^{x} \left(r\right)&=\frac{2}{3} r^{2} \int _{0}^{\infty }E_{u} \left(k\right)k^{2} W^{2} \left(kr\right)dk\\&=6\int _{0}^{\infty }E_{u} \left(k\right)j_{1}^{2} \left(kr\right)dk,      
\end{split}
\end{equation} 
where $W\left(kr\right)=3{j_{1} \left(kr\right)/\left(kr\right)} $ is the window function and \textit{r} is the radius of spherical filter. The structure function $S_{2}^{x} \left(r\right)$ can be related to one-dimensional enstrophy of smoothed velocity field, similar to the definition of velocity dispersion $\sigma _{u}^{2} \left(r\right)$ in Eq. \eqref{ZEqnNum726048}. The following equation shows that ${S_{2}^{x} \left(r\right)/\left(2r^{2} \right)} $ is a measure of the one-dimensional enstrophy smoothed at the scale of \textit{r}, or the total enstrophy contained in scales above \textit{r},
\begin{equation}
\label{ZEqnNum624967} 
\frac{S_{2}^{x} \left(r\right)}{2r^{2} } =\frac{1}{3} \int _{0}^{\infty }E_{u} \left(k\right)k^{2} W^{2} \left(kr\right)dk =\int _{r}^{\infty }E_{nr} \left(r^{'} \right)dr^{'}   
\end{equation} 
 
Here $E_{nr} \left(r\right)$ is the enstrophy density in real space. For collisional incompressible flow, the total enstrophy of the entire system is finite with $S_{2}^{x} \left(r\to 0\right)\propto r^{2} $. By contrast, the total enstrophy in collisionless flow diverges with $r\to 0$ such that structure function $S_{2}^{x} \left(r\right)$ should approach zero at a rate slower than $r^{2}$ (Eq. \eqref{ZEqnNum624967}). Let's apply a sharp spectral filter in the Fourier space with a cutoff wavenumber \textit{K}, at which the cutoff enstrophy equals that of the filtered velocity field, namely
\begin{equation} 
\label{eq:74} 
\int _{0}^{\infty }E_{u} \left(k\right)k^{2} W\left(kr\right)^{2} dk =\int _{0}^{K}E_{u} \left(k\right)k^{2} dk.    \end{equation} 
A relation can be established between the cutoff wavenumber \textit{K} and scale \textit{r} from Eq. \eqref{ZEqnNum624967}, 
\begin{equation}
\label{eq:75} 
\frac{3}{2} \frac{S_{2}^{x} \left(r\right)}{r^{2} } =\int _{0}^{K}E_{u} \left(k\right)k^{2} dk.  
\end{equation} 
With $K\to \infty $, $S_{2}^{x} \left(r\right)$ must approach zero with a rate slower than $r^{2} $ to enable a diverging enstrophy. 

The exact relation between two structure functions $S_{2}^{i} \left(r\right)$ and $S_{2}^{x} \left(r\right)$ can be obtained from Eqs. \eqref{ZEqnNum594172} and \eqref{ZEqnNum356384}
\begin{equation} 
\label{ZEqnNum952229} 
\frac{1}{3r^{2} } \frac{\partial }{\partial r} \left(\frac{1}{r} \frac{\partial }{\partial r} \left(S_{2}^{x} \left(r\right)r^{4} \right)\right)=\frac{\partial S_{2}^{i} \left(2r\right)}{\partial r} ,        
\end{equation} 
where both $S_{2}^{i} \left(r\right)$ and $S_{2}^{x} \left(r\right)$ can be found from the total velocity correlation function $R_{2}^{} \left(r\right)$. 

The structure function $S_{2}^{x} \left(r\right)$ has a special property that itself is a measure of the specific kinetic energy in scales below \textit{r} (see Eq. \eqref{ZEqnNum952229}, just like $\sigma _{d}^{2} \left(r\right)$), while ${S_{2}^{x} \left(r\right)/\left(2r^{2} \right)} $ is a direct measure of the enstrophy contained in scales above \textit{r} (Eq. \eqref{ZEqnNum624967} and Fig. \ref{fig:11}). Specifically, for the power-law velocity spectrum $E_{u} \left(k\right)\equiv bk^{-m} $,
\begin{equation} 
\label{ZEqnNum204217}
\begin{split}
S_{2}^{x} \left(r\right)=&-3\cdot 2^{1+m} \left(m+2\right)\left(m-1\right)\\&\cdot \Gamma \left(-3-m\right)\sin \left(\frac{m\pi }{2} \right)br^{m-1}.
\end{split}
\end{equation} 
Clearly, both Eqs. \eqref{ZEqnNum911575} and \eqref{ZEqnNum204217} satisfy the relation in Eq. \eqref{ZEqnNum952229}.  
Just like the energy distribution in real space (Eq. \eqref{ZEqnNum359490}), the enstrophy distribution in real space can be obtained from $S_{2}^{x} \left(r\right)$ as (from Eq. \eqref{ZEqnNum624967} and see Fig. \ref{fig:13}) 
\begin{equation} 
\label{ZEqnNum336678} 
E_{nr} \left(r\right)=-\frac{\partial }{\partial r} \left[{S_{2}^{x} \left(r\right)/\left(2r^{2} \right)} \right]. \end{equation} 
The relation between $E_{nr} \left(r\right)$ and velocity spectrum $E_{u} \left(k={x/r} \right)$ is
\begin{equation} 
\label{eq:79} 
E_{nr} \left(r\right)r^{4} =-\frac{2}{3} \int _{0}^{\infty }E_{u} \left(\frac{x}{r} \right)W\left(x\right)W^{'} \left(x\right)x^{3} dx .    
\end{equation} 

\subsubsection{Results for incompressible flow}
\label{sec:3.3.5}
Some kinematic relations between second order structure functions can be derived and presented here. First, the longitudinal structure function can be related to the velocity spectrum for incompressible flow by using Eqs.\eqref{ZEqnNum333566} and \eqref{ZEqnNum414891}
\begin{equation}
\label{ZEqnNum944866} 
S_{2}^{l} \left(r\right)=\frac{4}{3} \int _{0}^{\infty }E_{u} \left(k\right)\left(1-3\frac{j_{1} \left(kr\right)}{kr} \right)dk.        
\end{equation} 
The exact relations with the second order total velocity structure function can be found from Eqs. \eqref{ZEqnNum594172} and \eqref{ZEqnNum944866} 
\begin{equation} 
\label{eq:81} 
S_{2}^{i} \left(r\right)=\frac{1}{r^{2} } \frac{\partial }{\partial r} \left[r^{3} S_{2}^{l} \left(r\right)\right]. \end{equation} 
The exact relation with velocity dispersion function $\sigma _{d}^{2} \left(r\right)$ is 
\begin{equation}
\label{ZEqnNum258035} 
S_{2}^{l} \left(2r\right)=\frac{1}{12r^{5} } \frac{\partial }{\partial r} \left(r^{3} \frac{\partial }{\partial r} \left(\sigma _{d}^{2} \left(r\right)r^{4} \right)\right),       
\end{equation} 
where the longitudinal structure function $S_{2}^{l} \left(r\right)$ is also a quantitative measure of the specific kinetic energy contained in scales below \textit{r} for incompressible flow. For power-law velocity spectrum $E_{u} \left(k\right)\equiv bk^{-m} $ (from Eq. \eqref{ZEqnNum944866}) 
\begin{equation} 
\label{ZEqnNum124513} 
S_{2}^{l} \left(r\right)=-\frac{1}{3} \cdot \frac{2^{2-m} \Gamma \left({1+3/2} \right)\Gamma \left({\left(1-m\right)/2} \right)}{\Gamma \left({\left(m+4\right)/2} \right)} br^{m-1} .       
\end{equation} 
Let us assume a power-law scaling for velocity dispersion function $\sigma _{d}^{2} \left(r\right)\equiv br^{n} $ that can be inserted into Eqs. \eqref{ZEqnNum947585}, \eqref{ZEqnNum952229}, and \eqref{ZEqnNum258035} to obtain all three velocity structure functions that follow the same power-law scaling as,
\begin{equation}
\label{ZEqnNum957243} 
S_{2}^{l} \left(r\right)=\frac{\left(4+n\right)\left(6+n\right)}{12\cdot 2^{n} } br^{n} ,         
\end{equation} 
\begin{equation} 
\label{eq:85} 
S_{2}^{i} \left(r\right)=\frac{\left(3+n\right)\left(4+n\right)\left(6+n\right)}{12\cdot 2^{n} } br^{n} ,        
\end{equation} 
\begin{equation} 
\label{eq:86} 
S_{2}^{x} \left(r\right)=\frac{n\left(3+n\right)\left(6+n\right)}{4\left(2+n\right)} br^{n} .         
\end{equation} 
We will use these equations in Section \ref{sec:4} to understand the distribution of kinetic energy across different scales in N-body simulations. 
 
\subsubsection{Results for constant divergence flow}
\label{sec:3.3.6}
Just like the correlation functions in Section \ref{sec:3.2}, all results for second order structure functions of incompressible flow in Section \ref{sec:3.3.5} are still valid for constant divergence flow.

\subsubsection{Results for irrotational flow}
\label{sec:3.3.7}
Similar relations for longitudinal structure function can be derive exactly for irrotational flow. The relation to the velocity spectrum is (from Eqs. \eqref{ZEqnNum127370} and \eqref{ZEqnNum414891})
\begin{equation} 
\label{eq:87} 
S_{2}^{l} \left(r\right)=\frac{4}{3} \int _{0}^{\infty }E_{u} \left(k\right)\left(1-3j_{0} \left(kr\right)+6\frac{j_{1} \left(kr\right)}{kr} \right)dk .      
\end{equation} 
The relation with the second order total velocity structure function $S_{2}^{i} \left(r\right)$ is derived as, 
\begin{equation}
\label{ZEqnNum905257} 
\frac{\partial \left[rS_{2}^{i} \left(r\right)\right]}{\partial r} =\frac{1}{r^{2} } \frac{\partial }{\partial r} \left[r^{3} S_{2}^{l} \left(r\right)\right].        
\end{equation} 

\subsection{Correlation functions for velocity gradients}
\label{sec:3.4}
In this section, we focus on the two-point second order correlations for velocity gradients, namely the divergence field $\theta \left(\boldsymbol{\mathrm{x}}\right)=\nabla \cdot \boldsymbol{\mathrm{u}}\left(\boldsymbol{\mathrm{x}}\right)$ and vorticity field $\boldsymbol{\mathrm{\omega }}\left(\boldsymbol{\mathrm{x}}\right)=\nabla \times \boldsymbol{\mathrm{u}}\left(\boldsymbol{\mathrm{x}}\right)$. The correlation function of velocity divergence can be written as
\begin{equation} 
\label{ZEqnNum793459} 
\left\langle \theta \cdot \theta ^{'} \right\rangle =\left\langle u_{k,k} \cdot u_{q,q}^{'} \right\rangle =-\delta _{jq} \delta _{kp} Q_{kq,jp} =-Q_{pq,pq} ,       
\end{equation} 
where the second order derivative of correlation tensor $Q$ (see definition in Eq. \eqref{ZEqnNum221113}) reads
\begin{equation} 
\label{eq:90} 
\begin{split}
Q_{pq,pq} =&\frac{\partial Q_{pq} \left(r\right)}{\partial r_{p} \partial r_{q} }=\frac{\partial ^{2} A_{2} }{\partial r^{2} } r^{2} +8r\frac{\partial A_{2} }{\partial r}+12A_{2} \\&+\frac{\partial ^{2} B_{2} }{\partial r^{2} } +\frac{2}{r} \frac{\partial B_{2} }{\partial r}=\frac{1}{r^{2} } \left[\left(r^{2} L_{2} \right)_{,r} -2rT_{2} \right]_{,r}. 
\end{split}
\end{equation} 
The correlation function of vorticity field can be written as 
\begin{equation} 
\label{ZEqnNum844816} 
\begin{split}
\left\langle \boldsymbol{\mathrm{\omega }}\cdot \boldsymbol{\mathrm{\omega }}^{'} \right\rangle &=\varepsilon _{ijk} \varepsilon _{ipq} \left\langle u_{k,j} u_{q,p}^{'} \right\rangle \\&=-\left(\delta _{jp} \delta _{kq} -\delta _{jq} \delta _{kp} \right)Q_{kq,jp} . 
\end{split}
\end{equation} 
It can be easily verified that 
\begin{equation} 
\label{ZEqnNum266353} 
-\nabla ^{2} \left\langle \boldsymbol{\mathrm{u}}\cdot \boldsymbol{\mathrm{u}}^{'} \right\rangle =\left\langle \boldsymbol{\mathrm{\omega }}\cdot \boldsymbol{\mathrm{\omega }}^{'} \right\rangle +\left\langle \theta \cdot \theta ^{'} \right\rangle. 
\end{equation} 
The general expression for $\nabla ^{2} \left\langle \boldsymbol{\mathrm{u}}\cdot \boldsymbol{\mathrm{u}}^{'} \right\rangle $ is (with help of Eq. \eqref{ZEqnNum935612})
\begin{equation} 
\label{ZEqnNum226610} 
\begin{split}
\nabla ^{2} &\left\langle \boldsymbol{\mathrm{u}}\cdot \boldsymbol{\mathrm{u}}^{'} \right\rangle =\frac{1}{r^{2} } \frac{\partial }{\partial r} \left[r^{2} \frac{\partial R_{2} }{\partial r} \right]\\
&=3\frac{\partial ^{2} B_{2} }{\partial r^{2} } +\frac{6}{r} \frac{\partial B_{2} }{\partial r} +r^{2} \frac{\partial ^{2} A_{2} }{\partial r^{2} } +6r\frac{\partial A_{2} }{\partial r} +6A_{2}.   
\end{split}
\end{equation} 
Terms $-\nabla ^{2} \left\langle \boldsymbol{\mathrm{u}}\cdot \boldsymbol{\mathrm{u}}^{'} \right\rangle $ and $E_{u} \left(k\right)k^{2} $ form a Fourier transform pair 
\begin{equation} 
\label{ZEqnNum518422} 
-\nabla ^{2} \left\langle \boldsymbol{\mathrm{u}}\cdot \boldsymbol{\mathrm{u}}^{'} \right\rangle=2\int _{0}^{\infty }E_{u} \left(k\right)k^{2} \frac{\sin \left(kr\right)}{kr} dk.  
\end{equation} 
The sum of divergence and vorticity correlation functions can be found from Eqs. \eqref{ZEqnNum594172} and  \eqref{ZEqnNum518422}
\begin{equation} 
\label{ZEqnNum912296} 
\begin{split}
R_{\theta } +R_{\boldsymbol{\mathrm{\omega }}} &=\frac{1}{4r^{2} } \frac{\partial }{\partial r} \left(r^{2} \frac{\partial }{\partial r} S_{2}^{i} \left(r\right)\right)\\&=\frac{1}{96r^{2} } \frac{\partial }{\partial r} \left(\frac{1}{r^{2} } \frac{\partial }{\partial r} \left(r^{3} \frac{\partial }{\partial r} \left(S_{2}^{x} \left(r\right)r^{2} \right)\right)\right).
\end{split}
\end{equation} 
The expression for vorticity correlation reads (Eq. \eqref{ZEqnNum844816}) 
\begin{equation} 
\label{ZEqnNum441477} 
\begin{split}
R_{\boldsymbol{\mathrm{\omega }}} =\frac{\left\langle \boldsymbol{\mathrm{\omega }}\cdot \boldsymbol{\mathrm{\omega }}^{'} \right\rangle }{2} &=-\frac{\partial ^{2} B_{2} }{\partial r^{2} } -\frac{2}{r} \frac{\partial B_{2} }{\partial r}+3A_{2} +r\frac{\partial A_{2} }{\partial r}\\&=\frac{1}{r^{2} } \left[r^{2} \left(A_{2} r-\frac{\partial B_{2}^{} }{\partial r} \right)\right]_{,r}.
\end{split}
\end{equation} 
The expression for divergence correlation reads (Eq. \eqref{ZEqnNum793459})
\begin{equation} 
\label{ZEqnNum528609} 
\begin{split}
R_{\theta } =&\frac{\left\langle \theta \cdot \theta ^{'} \right\rangle }{2} =-\frac{1}{2}(\frac{\partial ^{2} B_{2} }{\partial r^{2} } +\frac{2}{r} \frac{\partial B_{2} }{\partial r} +r^{2} \frac{\partial ^{2} A_{2} }{\partial r^{2} }+8r\frac{\partial A_{2} }{\partial r}\\&+12A_{2}) =-\frac{1}{2r^{2} } \left[r^{2} \left(4A_{2} r+\frac{\partial A_{2} }{\partial r} r^{2} +\frac{\partial B_{2} }{\partial r} \right)\right]_{,r}.
\end{split}
\end{equation} 
The correlation functions for both divergence and vorticity are fully determined by two scalar functions $A_{2}$ and $B_{2}$ from the second order velocity correlation tensor $Q_{{}_{ij}} $ (Eq. \eqref{ZEqnNum221113}) for arbitrary homogeneous and isotropic flow. Both correlation functions can be identified from N-body simulation (Fig. \ref{fig:24}). 
 
\subsubsection{Results for incompressible flow}
\label{sec:3.4.1}
The divergence correlation function must be zero everywhere ($\langle \theta \cdot \theta ^{'} \rangle =0$, see Eqs. \eqref{ZEqnNum528609} and \eqref{ZEqnNum321178}). Because of the incompressibility,  $Q_{ij,ij} =0$, the vorticity correlation function can be reduced to (using Eq. \eqref{ZEqnNum314105})
\begin{equation} 
\label{eq:98} 
\left\langle \boldsymbol{\mathrm{\omega }}\cdot \boldsymbol{\mathrm{\omega }}^{'} \right\rangle =-Q_{ii,jj} =30A_{2} +18r\frac{\partial A_{2} }{\partial r} +2r^{2} \frac{\partial ^{2} A_{2} }{\partial r^{2} },     
\end{equation} 
\begin{equation} 
\label{ZEqnNum244673} 
\begin{split}
\left\langle \boldsymbol{\mathrm{\omega }}\cdot \boldsymbol{\mathrm{\omega }}^{'} \right\rangle &=-\nabla ^{2} \left\langle \boldsymbol{\mathrm{u}}\cdot \boldsymbol{\mathrm{u}}^{'} \right\rangle =-\frac{1}{r^{2} } \frac{\partial }{\partial r} \left[r^{2} \frac{\partial R_{2} }{\partial r} \right]\\&=-\frac{1}{r^{2} } \frac{\partial }{\partial r} \left[r^{2} \frac{\partial }{\partial r} \left(\frac{1}{r^{2} } \frac{\partial \left(r^{3} L_{2} \right)}{\partial r} \right)\right],
\end{split}
\end{equation} 
where the vorticity correlation is fully determined by the longitudinal correlation function $L_{2} \left(r\right)$.

The vorticity correlation can be related to the velocity power spectrum function $E_{u} \left(k\right)$ as
\begin{equation} 
\label{eq:100} 
R_{\boldsymbol{\mathrm{\omega }}} \left(r\right)=\frac{1}{2} \left\langle \boldsymbol{\mathrm{\omega }}\cdot \boldsymbol{\mathrm{\omega }}^{'} \right\rangle =\int _{0}^{\infty }E_{u} \left(k\right)k^{2} \frac{\sin \left(kr\right)}{kr} dk .       
\end{equation} 
 
\subsubsection{Results for constant divergence flow}
\label{sec:3.4.2}
All results for incompressible flow in Section \ref{sec:3.4.1} are still valid for constant divergence flow.

\subsubsection{Results for irrotational flow}
\label{sec:3.4.3}
Similarly, the vorticity correlation function should vanish for irrotational flow ($\left\langle \boldsymbol{\mathrm{\omega }}\cdot \boldsymbol{\mathrm{\omega }}^{'} \right\rangle =0$). The divergence correlation function can be reduced to
\begin{equation} 
\label{eq:101} 
\left\langle \theta \cdot \theta ^{'} \right\rangle =-\nabla ^{2} \left\langle \boldsymbol{\mathrm{u}}\cdot \boldsymbol{\mathrm{u}}^{'} \right\rangle =-\left(r^{2} \frac{\partial ^{2} A_{2} }{\partial r^{2} } +9r\frac{\partial A_{2} }{\partial r} +15A_{2} \right). 
\end{equation} 
With Eq. \eqref{ZEqnNum320035}, it is related to transverse correlation $T_2$ as
\begin{equation} 
\label{ZEqnNum432904} 
\begin{split}
\left\langle \theta \cdot \theta ^{'} \right\rangle &=-\nabla ^{2} \left\langle \boldsymbol{\mathrm{u}}\cdot \boldsymbol{\mathrm{u}}^{'} \right\rangle =-\frac{1}{r^{2} } \frac{\partial }{\partial r} \left[r^{2} \frac{\partial R_{2} }{\partial r} \right]\\&=-\frac{1}{r^{2} } \frac{\partial }{\partial r} \left[r^{2} \frac{\partial }{\partial r} \left(\frac{1}{r^{2} } \frac{\partial \left(r^{3} T_{2} \right)}{\partial r} \right)\right].
\end{split}
\end{equation} 
In addition, it can be related to velocity power spectrum as, 
\begin{equation} 
\label{ZEqnNum639543} 
R_{\theta } \left(r\right)=\frac{1}{2} \left\langle \theta \cdot \theta ^{'} \right\rangle =\int _{0}^{\infty }E_{u} \left(k\right)k^{2} \frac{\sin \left(kr\right)}{kr} dk.
\end{equation} 

\section{Statistical measures from N-body simulation}
\label{sec:4}
The simulation data described in Section \ref{sec:2} are used to compute all statistical measures described in Section \ref{sec:3}. For a given scale \textit{r}, all particle pairs with a separation \textit{r} are identified and particle position and velocity data are recorded to compute the correlation and structure functions by averaging that quantity over all pairs with the same separation \textit{r}, i.e a pairwise averaging for all scales r. This provides a complete view for the variation of these statistical measures on all scales. 

\subsection{Velocity correlation functions and flow characterization }
\label{sec:4.1}
\begin{figure}
\includegraphics*[width=\columnwidth]{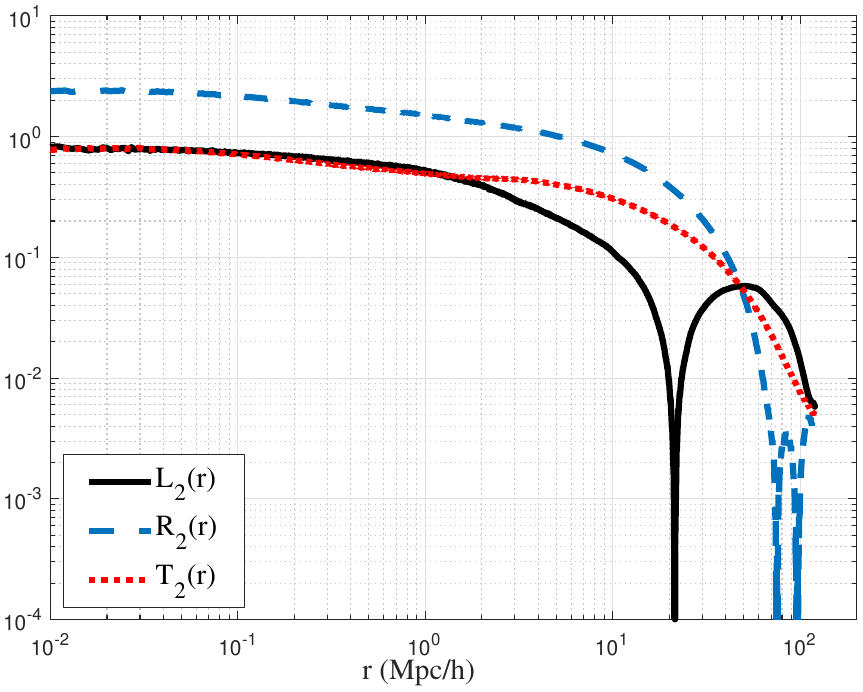}
\caption{The variation of two-point second order velocity correlation (normalized by $u_{0}^{2} $) with scale \textit{r} at z=0. The longitudinal correlation $L_{2} \left(r\right)$ becomes negative at 21.4 Mpc/h. The total velocity correlation $R_{2} \left(r\right)$ becomes negative at around 75.2 Mpc/h, and goes back to positive at 96.6Mpc/h. The transverse correlation $T_{2} \left(r\right)$ is positive for all scale \textit{r}. The cross-over of $T_{2} \left(r\right)$ and $L_{2} \left(r\right)$ is at r${}_{t}$ = 1.3 Mpc/h.}
\label{fig:2}
\end{figure}

\begin{figure}
\includegraphics*[width=\columnwidth]{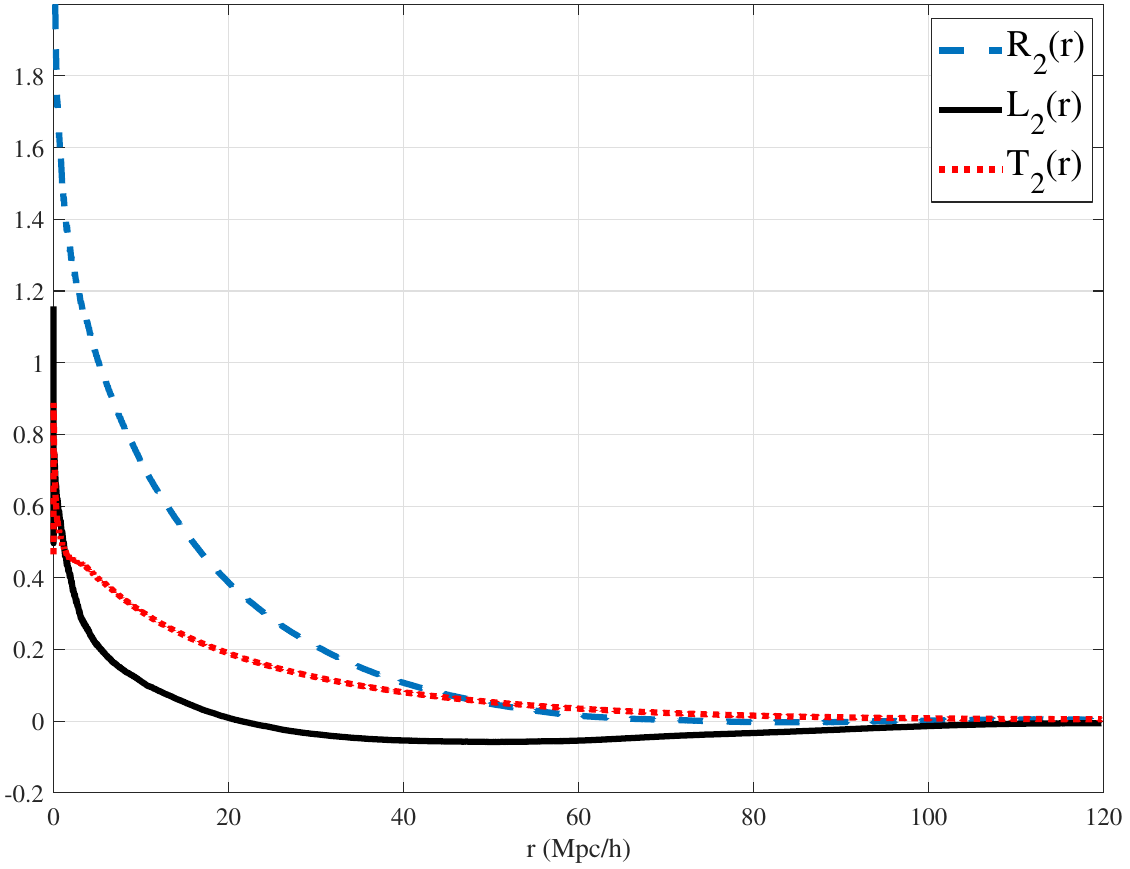}
\caption{Two-point second order velocity correlation functions (normalized by $u_{0}^{2}$)  varying with scale \textit{r} at z=0. Below $r_{t} $, most particle pairs are from the same halo. The two-point correlation functions reflect the correlation of particle velocity from the same halo and $L_{2} \left(r\right)>T_{2} \left(r\right)$ for $r<r_{t} $. While above $r_{t} $, two particles in a pair are more likely from different haloes. The correlation functions reflect the correlation of halo velocity and $T_{2} \left(r\right)>L_{2} \left(r\right)$ for $r>r_{t} $.}
\label{fig:3a}
\end{figure}

Figures \ref{fig:2} and \ref{fig:3a} present three second order velocity correlation functions varying with scale \textit{r }at z=0. The transverse correlation $T_{2} \left(r\right)$ is positive on all scales, while the longitudinal correlation $L_{2} \left(r\right)$ turns negative at $r\ge 21.4{Mpc/h} $, at which pair of particles separated by that distance will more likely move in opposite longitudinal directions, but in the same transverse directions. The total velocity correlation $R_{2} \left(r\right)$ becomes slightly negative at around r = 75.2 Mpc/h, and switch back to positive at r=96.6 Mpc/h. 

The crossover of $T_{2} \left(r\right)$ and $L_{2} \left(r\right)$ can be found at around $r_{t} $=1.3 Mpc/h, a characteristic length scale dividing two different flow regimes. Below $r_{t} $, most particle pairs are from the same halo. Two-point correlation functions reflect the correlation of particles from the same halo. The longitudinal correlation is dominant over transverse correlation. While above $r_{t} $, most pairs are from different haloes. The two-point correlation functions reflect the correlation of halo velocity and the transverse correlation is dominant.

Obviously, the comoving length scale $r_{t}$ can be related to the size of spherical halo with the characteristic mass $m_{h}^{*} \left(a\right)$. With $m_{h}^{*} \left(a\right)\propto a^{{3/2} } $ (see reference \citep{Xu:2021-Inverse-mass-cascade-mass-function}) and constant density ratio $\Delta _{c} $, we expect $r_{t} \left(a\right)\propto \left({m_{h}^{*} /\left(\Delta _{c} \rho _{0} \right)} \right)^{{1/3} } \propto a^{{1/2} }$ (Fig. \ref{fig:9}).

%\begin{subfigures}

%\end{subfigures}

\begin{figure}
\includegraphics*[width=\columnwidth]{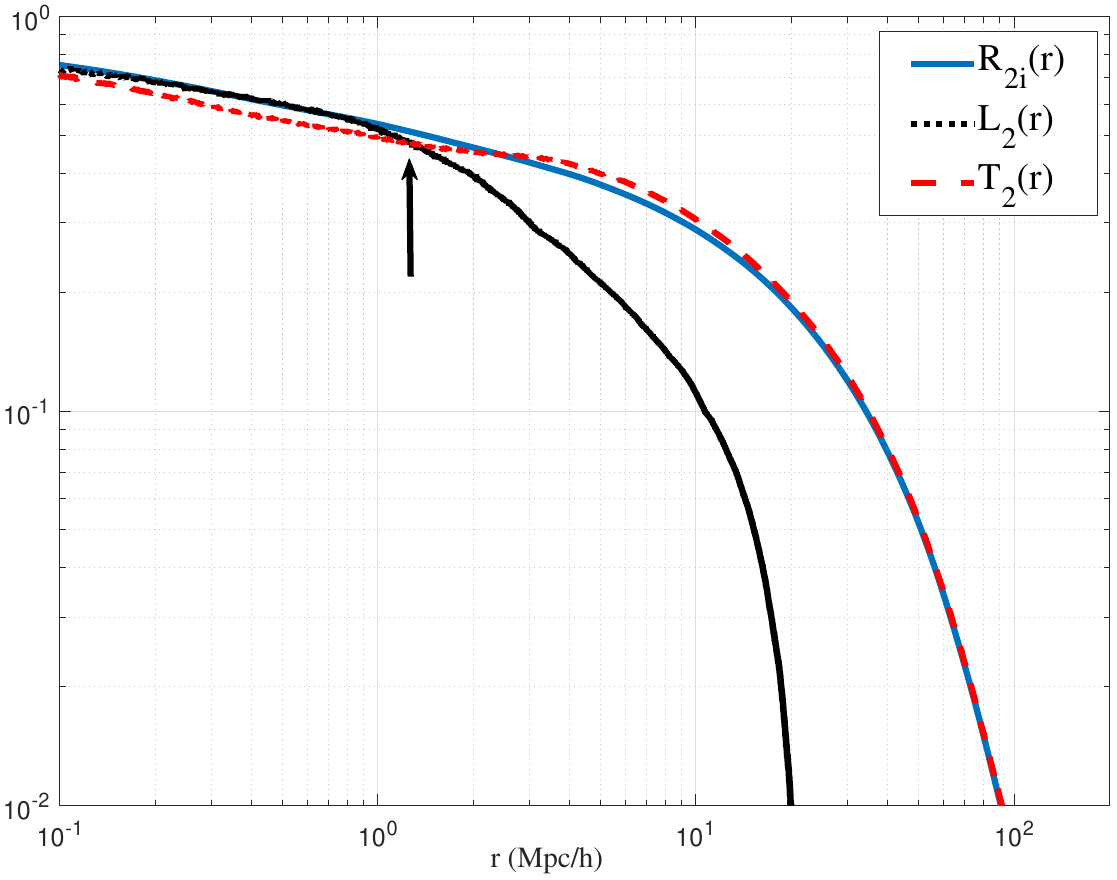}
\caption{Using correlation functions $L_{2} \left(r\right)$ and $T_{2} \left(r\right)$ to characterize types of flow. On small scale $R_{2i} \approx L_{2} $ means a constant divergence flow for $r<r_{t} $. On large sale $R_{2i} \approx T_{2} $ means an irrotational flow for$r>r_{t} $. The length scale $r_{t} \approx 1.3Mpc/h$ is defined as the scale at which $L_{2} \left(r\right)=T_{2} \left(r\right)$ (black arrow). The peculiar velocity field is of constant divergence on small scale and irrotational on large scale.}
\label{fig:4}
\end{figure}

Next, the kinematic relations developed in Section \ref{sec:3.2} (Eq. \eqref{ZEqnNum956600}) can be applied to identify the nature of flow on small and large scales. Figure \ref{fig:4} plots the variation of $R_{2i} \left(r\right)$ with scale \textit{r}, along with two correlation functions $L_{2} \left(r\right)$ and $T_{2} \left(r\right)$. In Fig. \ref{fig:4}, on small scale $r\ll r_{t} $, $R_{2i} \approx L_{2} $ means a constant divergence flow for peculiar velocity on small scale ($\nabla \cdot \boldsymbol{\mathrm{u}}=const$). While on large sale $r\gg r_{t} $, $R_{2i} \approx T_{2} $ means an irrotational velocity field ($\nabla \times \boldsymbol{\mathrm{u}}=0$). The critical length scale $r_{t} \approx 1.3Mpc/h$ is defined as the scale at which $L_{2} \left(r\right)=T_{2} \left(r\right)$ (indicated by an arrow). The irrotational flow on large scale is expected, where the Zeldovich approximation and the linear perturbation theory apply. Unlike incompressible flow that is incompressible on all scales, the peculiar velocity field of dark matter flow has a scale-dependent behavior, i.e. a constant divergence flow on small scale and an irrotational flow on large scale. 

\begin{figure}
\includegraphics*[width=\columnwidth]{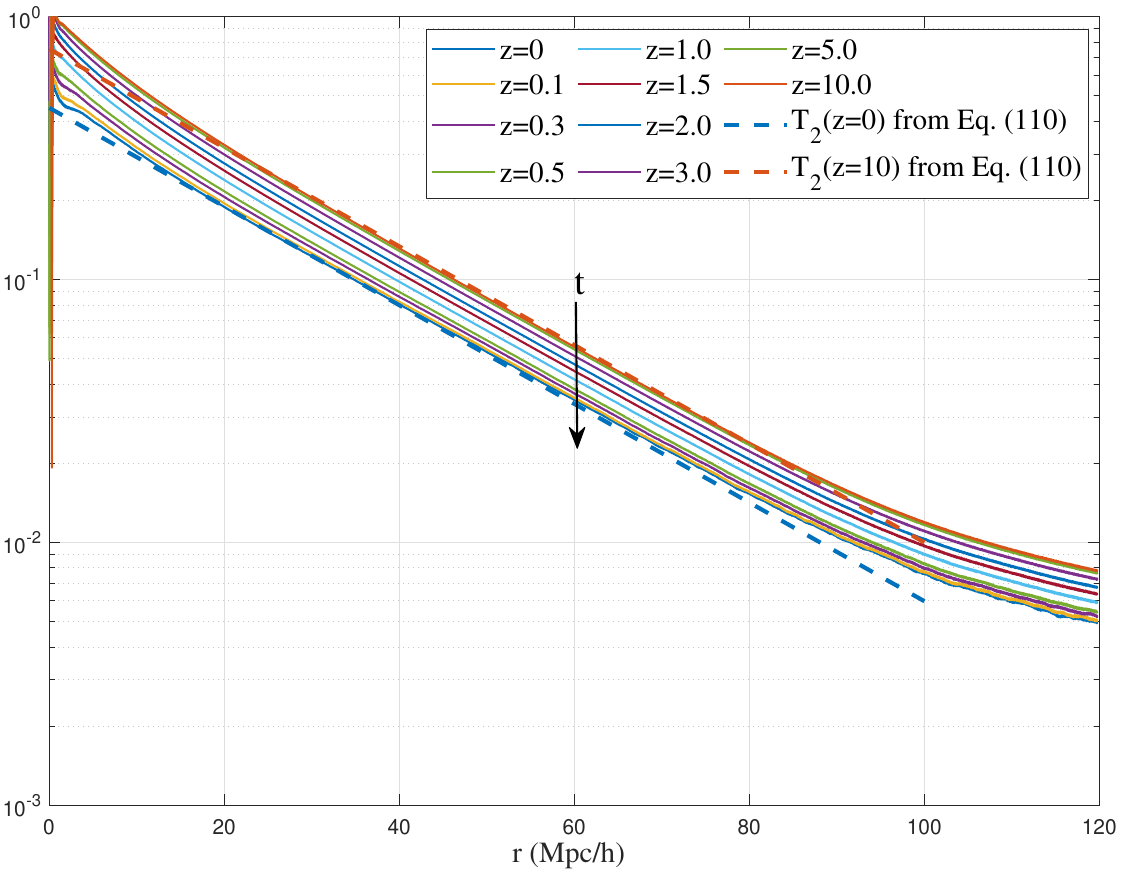}
\caption{Transverse velocity correlation function $T_{2} \left(r\right)$ varying with comoving scale \textit{r} at different redshifts z ($\ln T_{2} \left(r,z\right)$ vs. \textit{r}). The correlation function is normalized by $u^{2} \left(a\right)$. Plot suggests an exponential function for $T_{2} \left(r\right)$. Detail models are provided in Section \ref{sec:5}.}
\label{fig:5}
\end{figure}
Figure \ref{fig:5} presents the transverse velocity correlation function $T_{2} \left(r\right)$ at different redshifts \textit{z}. The correlation functions are normalized by $u^{2} \left(a\right)$, i.e. the one-dimensional velocity dispersion at different redshifts (Eq. \eqref{ZEqnNum586699}). Figure \ref{fig:5} hints an exponential form for $T_{2} \left(r\right)$ on large scale at different \textit{z}. The detail models for correlation functions are provided in Section \ref{sec:5} (Eqs. \eqref{ZEqnNum971850} and \eqref{ZEqnNum739102} for large and small scales, respectively). 

\begin{figure}
\includegraphics*[width=\columnwidth]{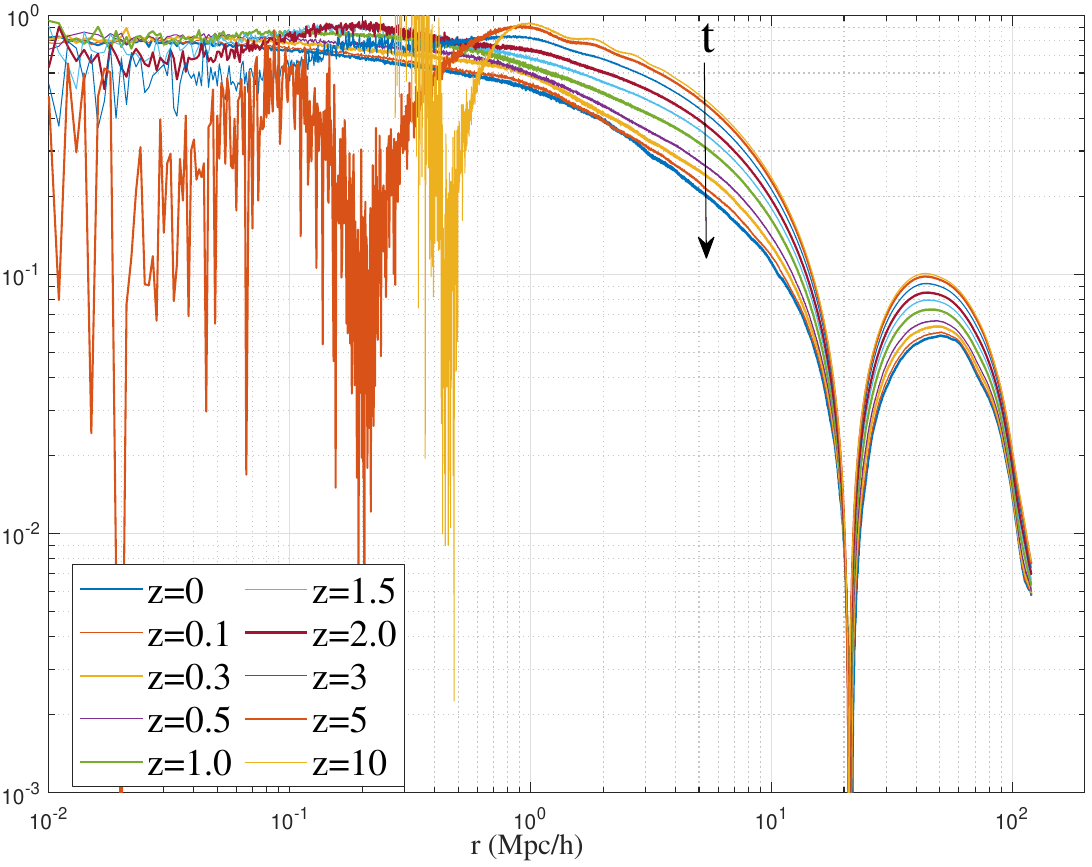}
\caption{The variation of longitudinal velocity correlation $L_{2} \left(r\right)$ (normalized by $u^{2} \left(a\right)$) with scale \textit{r} at different redshifts \textit{z}. Figure shows the evolution of longitudinal velocity correlation on small scales with more particle pairs from the same halo formed. A comoving length scale $r_{2} \approx 21.4{Mpc/h} $ can be identified that does not vary with time, where longitudinal velocity correlation becomes negative.}
\label{fig:6}
\end{figure}
Figure \ref{fig:6} presents the longitudinal velocity correlation $L_{2} \left(r\right)$ (normalized by $u^{2} \left(a\right)$). With more and more haloes formed, the number of pairs of particles from the same halo with small separation \textit{r} increases with time, along with the longitudinal correlation being gradually developed on small scales after z=3. A constant comoving length scale $r_{2} \approx 21.4{Mpc/h} $ can be identified that does not vary with time, where longitudinal velocity correlation becomes negative. Length scale $r_2$ is related to the horizon size at matter-radiation equality (see Section \ref{sec:5.1.3}). The models are presented in Section \ref{sec:5} (Eqs. \eqref{ZEqnNum344034} and \eqref{ZEqnNum178092} for large and small scales, respectively). 

\begin{figure}
\includegraphics*[width=\columnwidth]{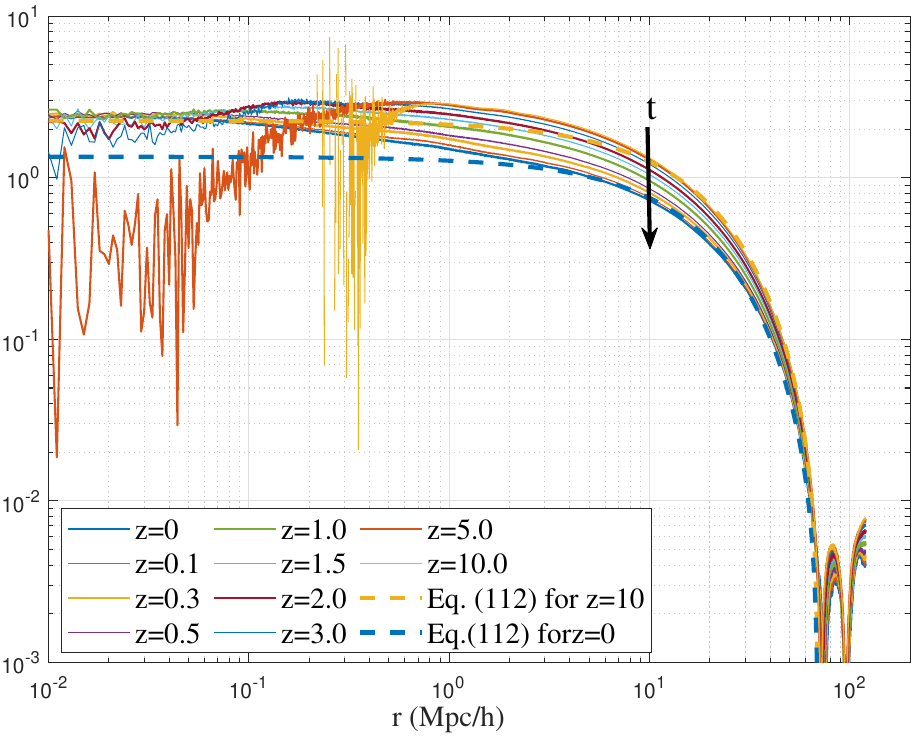}
\caption{The variation of total velocity correlation functions $R_{2} \left(r\right)$ (Eq. \eqref{ZEqnNum610884}, normalized by $u^{2} \left(a\right)$) with scale \textit{r} at different Redshift z. Models of $R_{2} \left(r\right)$ are provided in Section \ref{sec:5} and plotted in the same figure for comparison.}
\label{fig:7}
\end{figure}
Figure \ref{fig:7} presents the total velocity correlation function $R_{2} \left(r\right)$ at different redshift \textit{z} (from Eq. \eqref{ZEqnNum610884}). The detail models are presented in Section \ref{sec:5} (Eqs. \eqref{ZEqnNum235904} and \eqref{ZEqnNum955991}) for large and small scales. 
 
\subsection{Velocity dispersion functions and energy distribution}
\label{sec:4.2}
\begin{figure}
\includegraphics*[width=\columnwidth]{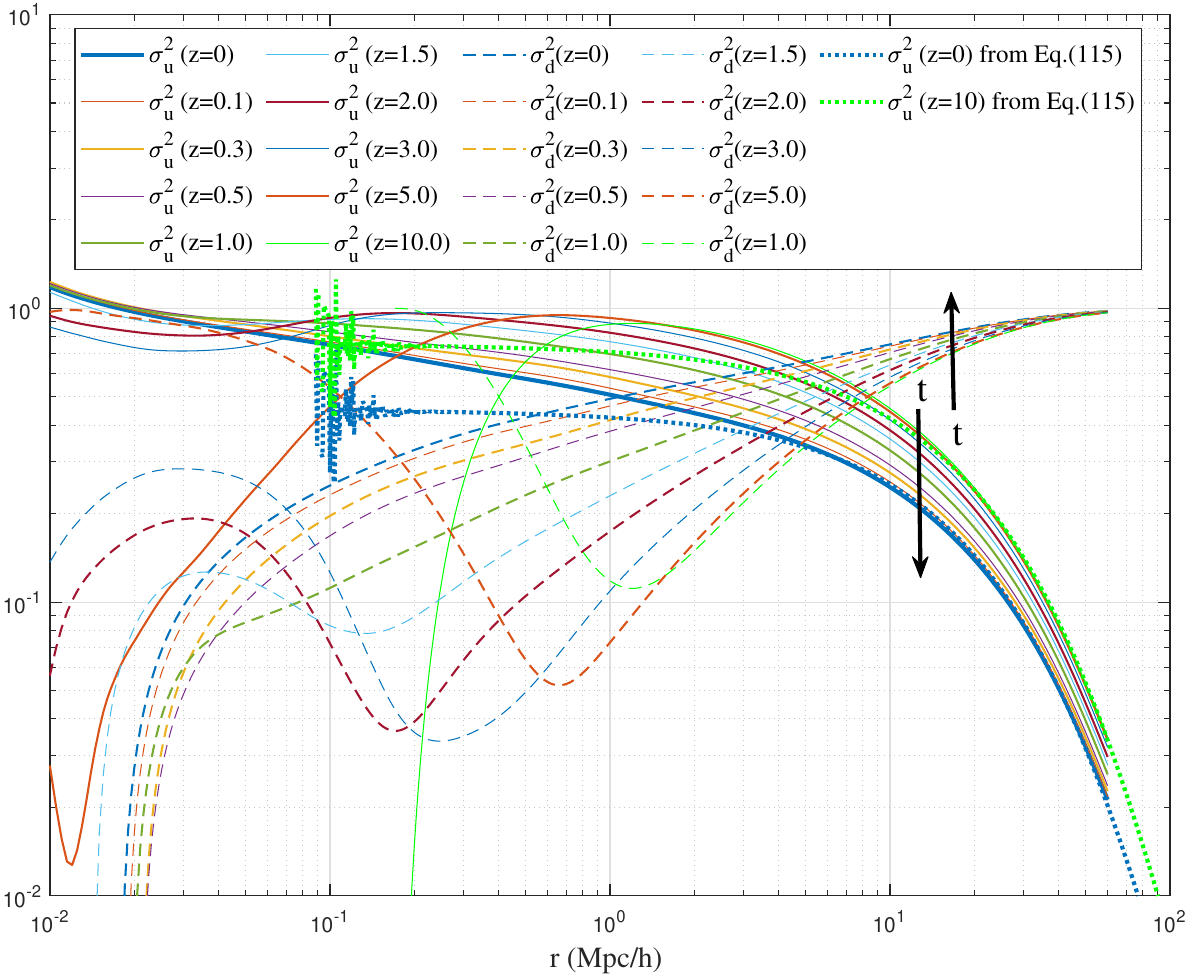}
\caption{The variation of velocity dispersion functions with scale \textit{r} at different redshift \textit{z}, normalized by velocity dispersion $u^{2} \left(a\right)$. The one dimensional dispersion of smoothed velocity $\sigma _{u}^{2} \left(r\right)$ is obtained from Eq. \eqref{ZEqnNum910224} (solid line), while velocity dispersion velocity$\sigma _{d}^{2} \left(r\right)$ is obtained from Eq. \eqref{ZEqnNum926839} (dash line). Increasing small scale power with time can be demonstrated by the increasing $\sigma _{d}^{2} \left(r\right)$. Model for $\sigma _{u}^{2} \left(r\right)$ on large scale in Eq. \eqref{ZEqnNum178726} is plotted in the same figure (dotted line) for comparison.}
\label{fig:8}
\end{figure}
\begin{figure}
\includegraphics*[width=\columnwidth]{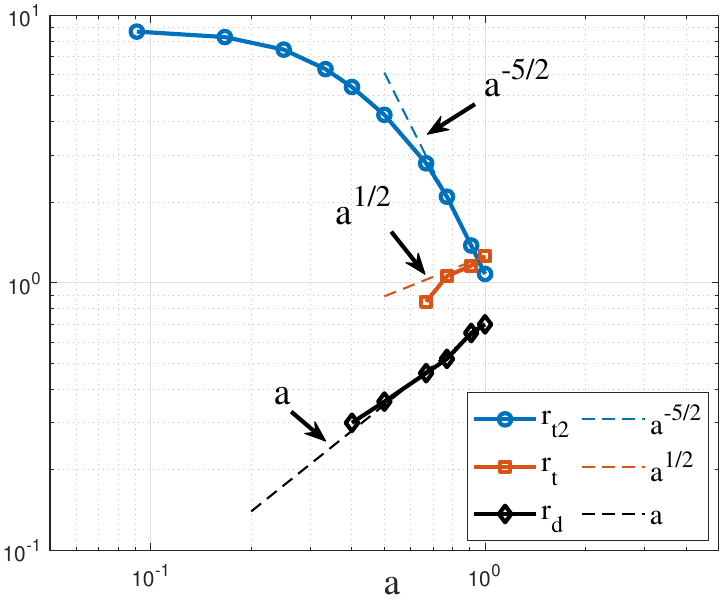}
\caption{Comoving length scales $r_{t2} $ for $\sigma _{u}^{2} \left(r_{t2} \right)=\sigma _{d}^{2} \left(r_{t2} \right)=0.5$ and $r_{t} $ for $L_{2} \left(r_{t} \right)=T_{2} \left(r_{t} \right)$ (Mpc/h) varying with \textit{a}. A scaling of $r_{t2} \propto a^{{-5/2} } $ and $r_{t} \propto a^{{1/2} } $ can be found. The scale $r_{d} \propto a$ is for the maximum velocity dispersion$\left\langle u_{L}^{2} \right\rangle $ in Eq. \eqref{ZEqnNum776534} (see Fig. \ref{fig:19}).}
\label{fig:9}
\end{figure}

Figure \ref{fig:8} presents the variation of two dispersion functions with comoving scale \textit{r} for different redshift \textit{z}. The dispersion of smoothed velocity $\sigma _{u}^{2} \left(r\right)$ is obtained from Eq. \eqref{ZEqnNum910224} (solid line), while the velocity dispersion $\sigma _{d}^{2} \left(r\right)$ is obtained from Eq. \eqref{ZEqnNum926839} (dash line). Increasing small scale power with time can be seen from the increasing $\sigma _{d}^{2} \left(r\right)$ for a given \textit{r}. The dispersion function $\sigma _{u}^{2} \left(r\right)$ on large scale can be modeled by Eq. \eqref{ZEqnNum178726} and plotted in dotted line with good agreement with N-body simulation (solid blue line). 
By setting the energy in scales above \textit{r} equals the energy contained in scales smaller than \textit{r}, i.e. $\sigma _{u}^{2} \left(r_{t2} \right)=\sigma _{d}^{2} \left(r_{t2} \right)=0.5$, a second comoving length scale $r_{t2} $ can be introduced. The kinetic energy contained in scales less than $r_{t2}$ is 50\% of the entire kinetic energy. Figure \ref{fig:9} presents the variation of $r_{t2} \propto a^{{-5/2} } $ and $r_{t} \propto a^{{1/2} }$. The length scale $r_{t2}$ decreasing with time reflects increasing fraction of kinetic energy contained in small scales with more haloes formed.

\begin{figure}
\includegraphics*[width=\columnwidth]{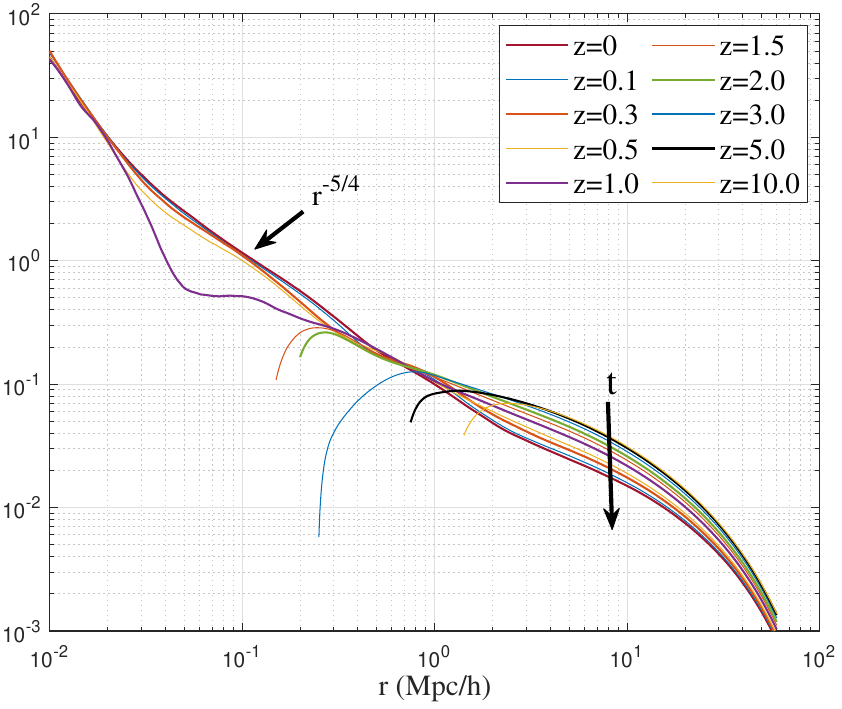}
\caption{The real space energy distribution $E_{ur} \left(r\right)$ (Eq. \eqref{ZEqnNum359490}) with unit of $h/Mpc$) on scale \textit{r} at different redshifts \textit{z} (normalized by $u^{2}\left(a\right)$). The (normalized) kinetic energy contained in large scale decreases with time, while the energy contained in small scale \textit{r} increases with time. On small scale, $E_{ur}(r) \propto r^{-5/4}$.} 
\label{fig:10}
\end{figure}
Figure \ref{fig:10} presents the energy distribution in real space from Eq. \eqref{ZEqnNum359490}. Kinetic energy contained in large scale is decreasing with time, with more energy contained in small scale. With more haloes forming, the small scale power ($\propto u^{2} \propto a^{{3/2} } $ Eq. \eqref{ZEqnNum955991}) increases faster than the large scale ($\propto a_{0} u^{2} \propto a^{1} $, Eq. \eqref{ZEqnNum235904}) with increasing fraction of kinetic energy contained in small scales.  

\subsection{Velocity structure function \texorpdfstring{$S_{2}^{x}$}{} and enstrophy distribution}
\label{sec:4.3}
\begin{figure}
\includegraphics*[width=\columnwidth]{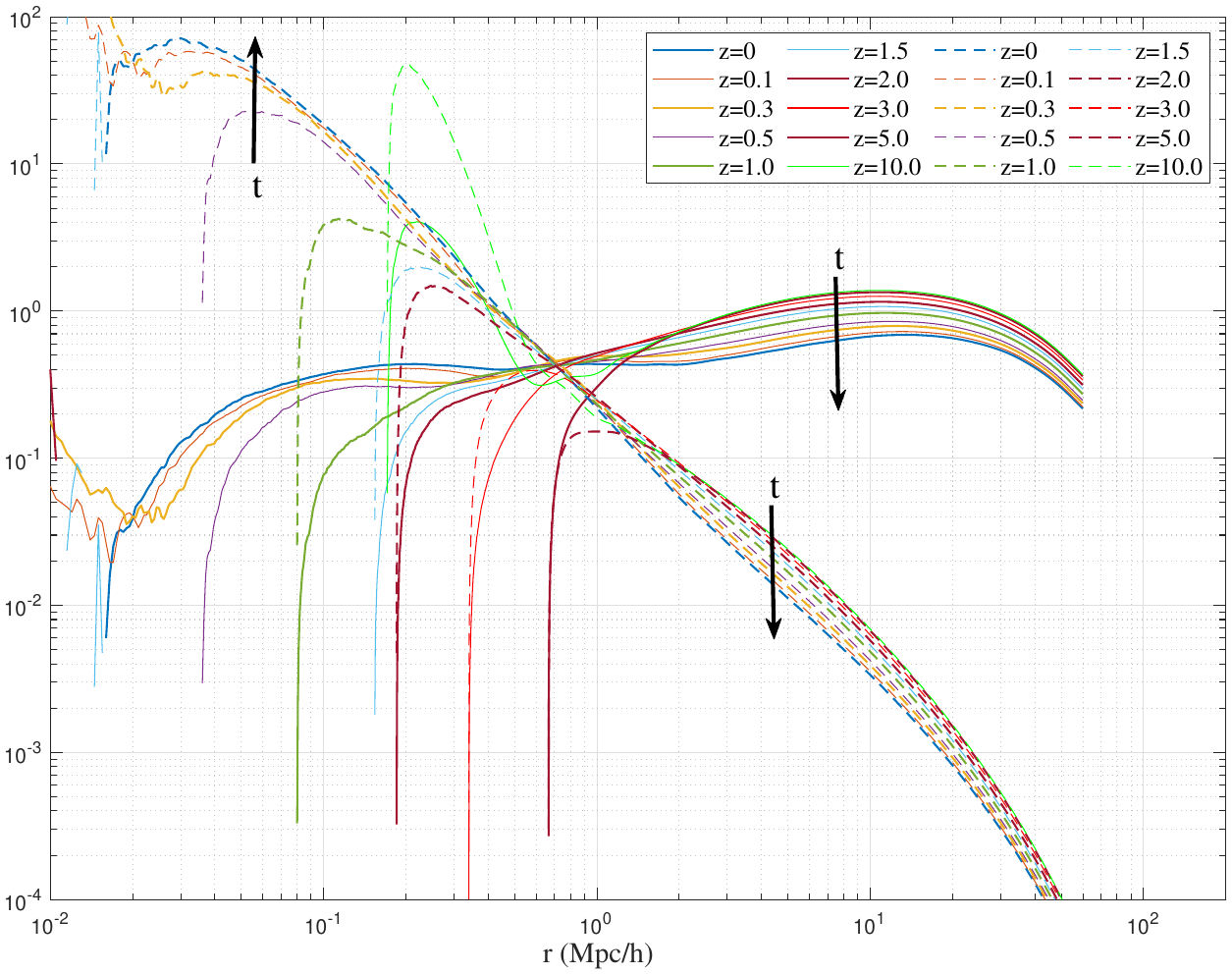}
\caption{The variation of structure function $S_{2}^{x} \left(r\right)$ (normalized by $u^{2} $ in solid lines) with scale \textit{r} for different redshifts \textit{z}. The variation of ${S_{2}^{x} \left(r\right)/\left(2r^{2} \right)} $ (Eq. \eqref{ZEqnNum624967} normalized by $u^{2} $, i.e. the enstrophy contained in scales larger than \textit{r}) is also presented in dash lines.}
\label{fig:11}
\end{figure}
Figure \ref{fig:11} presents the variation of structure function $S_{2}^{x} \left(r\right)$ (from Eq. \eqref{ZEqnNum952229}) with comoving scale \textit{r} for different redshifts \textit{z} (solid lines). The corresponding models of $S_{2}^{x} \left(r\right)$ on large and small scales are discussed in Section \ref{sec:5} (Eqs. \eqref{ZEqnNum858651} and \eqref{ZEqnNum413972}). The one-dimensional enstrophy of smoothed velocity  ${S_{2}^{x} \left(r\right)/\left(2r^{2} \right)} $ (Eq. \eqref{ZEqnNum624967}, normalized by $u^{2} $ with unit of $\left({h/Mpc} \right)^{2} $), i.e. the enstrophy contained in all scales above \textit{r}, is also plotted in dash lines. 

\begin{figure}
\includegraphics*[width=\columnwidth]{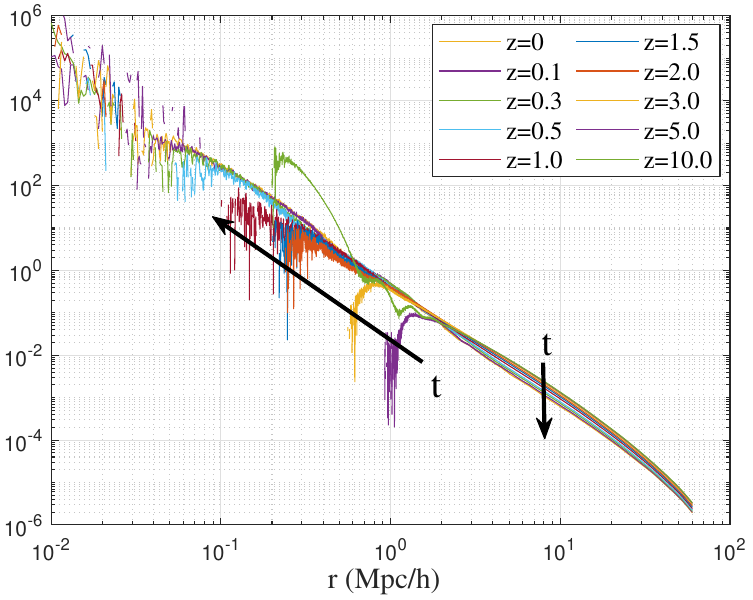}
\caption{The enstrophy distribution $E_{nr} \left(r\right)$ (Eq. \eqref{ZEqnNum624967} with unit of $(h/Mpc)^3$) in real space \textit{r} at different redshift z (normalized by $u^{2} \left(a\right)$).  In comoving system, the enstrophy contained in small scale increases with time, while the enstrophy contained in large scale decreases with time.}
\label{fig:12}
\end{figure}

Figure \ref{fig:12} presents the distribution of enstrophy $E_{nr} \left(r\right)$ (Eq. \eqref{ZEqnNum624967}) in real space \textit{r} at different redshift \textit{z}. With more haloes formed with finite angular velocity (or vorticity), the enstrophy distribution gradually extends to smaller and smaller scales. 

\subsection{Velocity structure functions \texorpdfstring{$S_2^l$}{} and one-fourth law}
\label{sec:4.4}
\begin{figure}
\includegraphics*[width=\columnwidth]{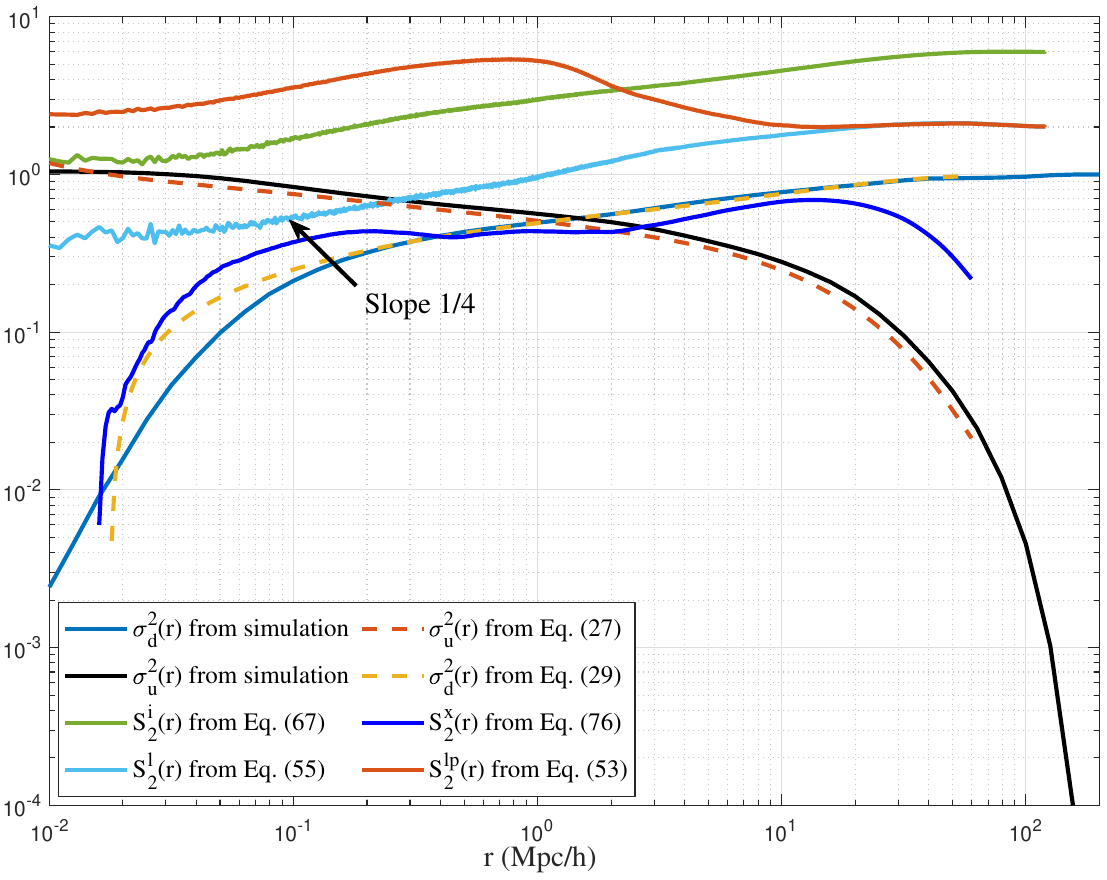}
\caption{Different velocity dispersion and structure functions at z=0, normalized by $u_{0}^{2} $. The velocity dispersion $\sigma _{d}^{2} \left(r\right)$ (Eq. \eqref{ZEqnNum904685}), dispersion of smoothed velocity $\sigma _{u}^{2} \left(r\right)$ (Eq. \eqref{ZEqnNum726048}), total velocity structure function $S_{2}^{i} \left(r\right)$ (Eq. \eqref{ZEqnNum973047}), and longitudinal structure function $S_{2}^{l} \left(r\right)$ and $S_{2}^{lp} \left(r\right)$ (Eqs. \eqref{ZEqnNum250774} and \eqref{ZEqnNum414891}), were all obtained directly from the N-body simulation data. The dispersion functions $\sigma _{u}^{2} \left(r\right)$ and $\sigma _{d}^{2} \left(r\right)$, and structure function $S_{2}^{x} \left(r\right)$ can also be obtained indirectly using Eqs. \eqref{ZEqnNum910224}, \eqref{ZEqnNum926839}, and \eqref{ZEqnNum952229} for comparison. A power-law of $S_{2}^{l} \left(r\right)\propto \sigma _{d}^{2} \left(r\right)\propto r^{{1/4} } $ can be identified. Two longitudinal structure functions $S_{2}^{lp} \left(r\right)\ne S_{2}^{l} \left(r\right)$ is clearly shown (Eqs. \eqref{ZEqnNum890245} to \eqref{ZEqnNum924181}).}
\label{fig:13}
\end{figure}
Figure \ref{fig:13} presents the variation of different velocity dispersion and structure functions with scale \textit{r} at z=0. From N-body simulation, we can directly obtain the velocity dispersion $\sigma _{d}^{2} \left(r\right)$ (Eq. \eqref{ZEqnNum904685}), dispersion of smoothed velocity $\sigma _{u}^{2} \left(r\right)$ (Eq. \eqref{ZEqnNum726048}), total velocity structure function $S_{2}^{i} \left(r\right)$ (Eq. \eqref{ZEqnNum973047}), longitudinal structure functions $S_{2}^{l} \left(r\right)$ and $S_{2}^{lp} \left(r\right)$ (Eqs. \eqref{ZEqnNum250774} and \eqref{ZEqnNum414891}), and structure function $S_{2}^{x} \left(r\right)$ from Eq. \eqref{ZEqnNum952229}. 

The dispersion $\sigma _{u}^{2} \left(r\right)$ and $\sigma _{d}^{2} \left(r\right)$ can also be obtained indirectly using the correlation function $R_{2} \left(r\right)$ (Eqs. \eqref{ZEqnNum910224} and \eqref{ZEqnNum926839}), which show good agreement on large scale with those directly obtained from simulation. On small scale, the velocity dispersion $\sigma_u^2(r)$ obtained indirectly from $R_2$ (Eq. \eqref{ZEqnNum910224}) still matches the one directly obtained from N-body simulation, while the velocity dispersion $\sigma_d^2(r)$ obtained through $\sigma_d^2(r)=u^2-\sigma_u^2(r)$ has some disagreement with that directly obtained from simulation. The potential reasons can be: 1) the sum of two dispersions $\sigma_d^2(r)+\sigma_u^2(r)$ does not exactly equal $u^2$ in N-body simulation. This might be related to the simulation resolution due to finite gravitational softening. 2) numerical artifacts introduced from the grid might have effect on the statistical isotropy on small scale. Since Eq. (30) is derived based on the homogeneity and isotropy, violation of local statistical isotropy may also contribute to this discrepancy.

A power-law of $S_{2}^{l} \left(r\right)\propto \sigma _{d}^{2} \left(r\right)\propto r^{{1/4} } $ can be identified on small scale. The dispersion function $\sigma _{d}^{2} \left(r\right)$ should share the same power-law scaling with structure functions $S_{2}^{i} \left(r\right)$ (Eq. \eqref{ZEqnNum947585}) and $S_{2}^{l} \left(r\right)$ (Eq. \eqref{ZEqnNum258035}). The reason for a power-law scaling is briefly outlined here. Let us first compute the potential energy for a sphere of radius \textit{r}  
\begin{equation} 
\label{ZEqnNum299865} 
U\left(r\right)=-\frac{\int _{0}^{r}\frac{G}{y}  M\left(y\right)\rho \left(y\right)4\pi y^{2} dy}{\int _{0}^{r}\rho \left(x\right)4\pi x^{2} dx } ,        
\end{equation} 
where $M\left(y\right)=\int _{0}^{y}\rho \left(x\right)4\pi x^{2} dx $ is the total mass in all shells with a radius smaller than $y$. 

The density $\rho \left(y\right)$ in that sphere can be related to the correlation function $\xi \left(r\right)$. By assuming a power-law of $\xi \left(r\right)$ ($\xi \left(r\right)=\left({r/6.1} \right)^{-1.75} $ in reference \citep{Zehavi:2002-Galaxy-clustering-in-early-Slo}), density $\rho \left(y\right)$ reads
\begin{equation}
\begin{split}
&\xi \left(r\right)=a^{m} \left({r/r_{\xi } } \right)^{-n}\\  
&\rho \left(y\right)=\rho _{0} \left(1+\xi \left(y\right)\right)\approx \rho _{0} a^{m} \left({y/r_{\xi } } \right)^{-n}, 
\end{split}
\label{ZEqnNum955584}
\end{equation}
\noindent where $r_{\xi } \approx 3.08Mpc/h$, $m\approx 2$, $n\approx 1.75$, and $\rho _{0} $ is the mean density at \textit{z}=0. Inserting density into Eq. \eqref{ZEqnNum299865}, 
\begin{equation} 
\label{eq:106} 
\begin{split}
U\left(r\right)&=-\frac{3-n}{5-2n} \frac{GM\left(r\right)}{r}\\&=-\frac{4\pi G\rho _{0} }{5-2n} \left(\frac{r}{r_{\xi } } \right)^{-n} =-\frac{3a^{m} H_{0}^{2} r^{2} }{2\left(5-2n\right)} \left(\frac{r}{r_{\xi } } \right)^{-n}.    
\end{split}
\end{equation} 
The specific kinetic energy of that sphere can be written as $T\left(r\right)={3/2} \sigma _{d}^{2} $, where $\sigma _{d}^{2}$ is the one-dimensional velocity dispersion of particles in that sphere. Applying the virial theorem $2T\left(r\right)+\gamma U\left(r\right)=0$ leads to equation (where $\gamma $ is virial ratio),
\begin{equation} 
\label{eq:107} 
\sigma _{d}^{2} \left(r\right)=\frac{\gamma \left(H_{0} r\right)^{2} }{2r^{5} } \frac{\int _{0}^{r}\left(1+\xi \left(y\right)\right)\left(1+\bar{\xi }\left(y\right)\right)y^{4} dy }{\left(1+\bar{\xi }\left(r\right)\right)} .      
\end{equation} 
For power-law density correlation in Eq. \eqref{ZEqnNum955584} 
\begin{equation} 
\label{ZEqnNum206214} 
\sigma _{d}^{2} \left(r\right)=\frac{\gamma a^{m} \left(H_{0} r\right)^{2} }{2\left(5-2n\right)} \left(\frac{r}{r_{\xi } } \right)^{-n} .         
\end{equation} 
The structure function  can be computed from Eq. \eqref{ZEqnNum258035} (peculiar velocity is of constant divergence on small scale, where Eq. \eqref{ZEqnNum258035} can be applied) as
\begin{equation} 
\label{ZEqnNum489540} 
S_{2}^{l} \left(r\right)=\frac{\gamma \left(6-n\right)\left(8-n\right)}{24\left(5-2n\right)2^{2-n} } a^{m} \left(H_{0} r\right)^{2} \left(\frac{r}{r_{\xi } } \right)^{-n} .       
\end{equation} 
Please note that the constant divergence flow shares the same kinematic relations as that of incompressible flow on second order (see Section \ref{sec:3.3.6}). With $n=7/4$, we can estimate $\sigma _{d}^{2} \left(r\right)\approx 0.19\gamma a^{m} r^{{1/4} } u_{0}^{2} $ and $S_{2}^{l} \left(r\right)\approx 0.3533\gamma a^{m} r^{{1/4} } u_{0}^{2} $, both of which follows a power-law scaling $\propto r^{{1/4} } $. The detail models for structure functions  are provided in Section \ref{sec:5} (Eqs. \eqref{ZEqnNum509836} and \eqref{ZEqnNum392323}).

\begin{figure}
\includegraphics*[width=\columnwidth]{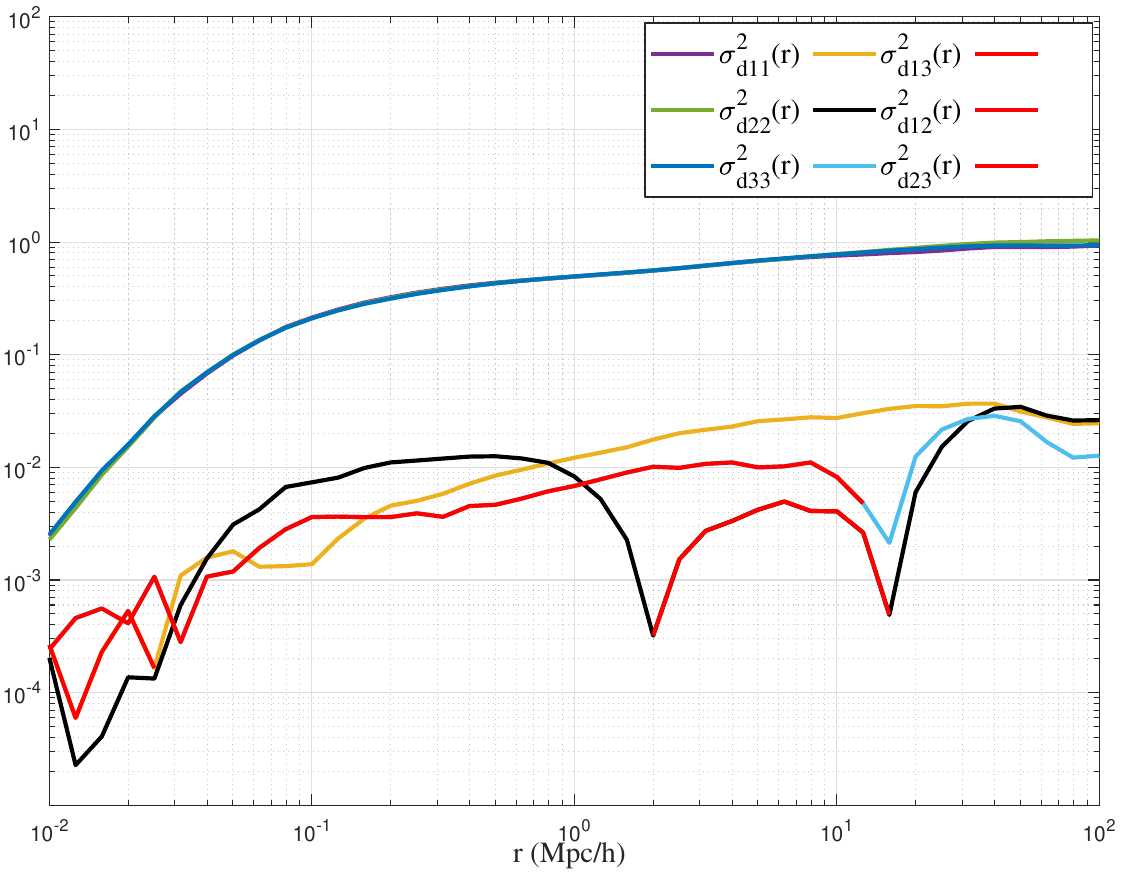}
\caption{Plot of six components of velocity dispersion tensor varying with scale \textit{r}. (red indicates that the value is negative). The off-diagonal terms are much smaller compared to the diagonal terms. The velocity dispersion tensor is almost isotropic with diagonal terms $\sigma _{d11}^{2} \approx \sigma _{d22}^{2} \approx \sigma _{d33}^{2} $ on all scales.}
\label{fig:14}
\end{figure}
Finally, the velocity dispersion tensor $\sigma _{dij}^{2} \left(r\right)$ was found from simulation by computing the average velocity dispersion within a sphere of radius \textit{r} centered around randomly selected particles. Figure \ref{fig:14} plots the six components of the velocity dispersion tensor varying with scale \textit{r}, where red color indicates the value is negative. The off-diagonal terms are negligible. Velocity dispersion tensor is isotropic with diagonal terms $\sigma _{d11}^{2} \approx \sigma _{d22}^{2} \approx \sigma _{d33}^{2} $ on all scales.  

\subsection{Longitudinal and transverse velocities on different scales}
\label{sec:4.5}
\begin{figure}
\includegraphics*[width=\columnwidth]{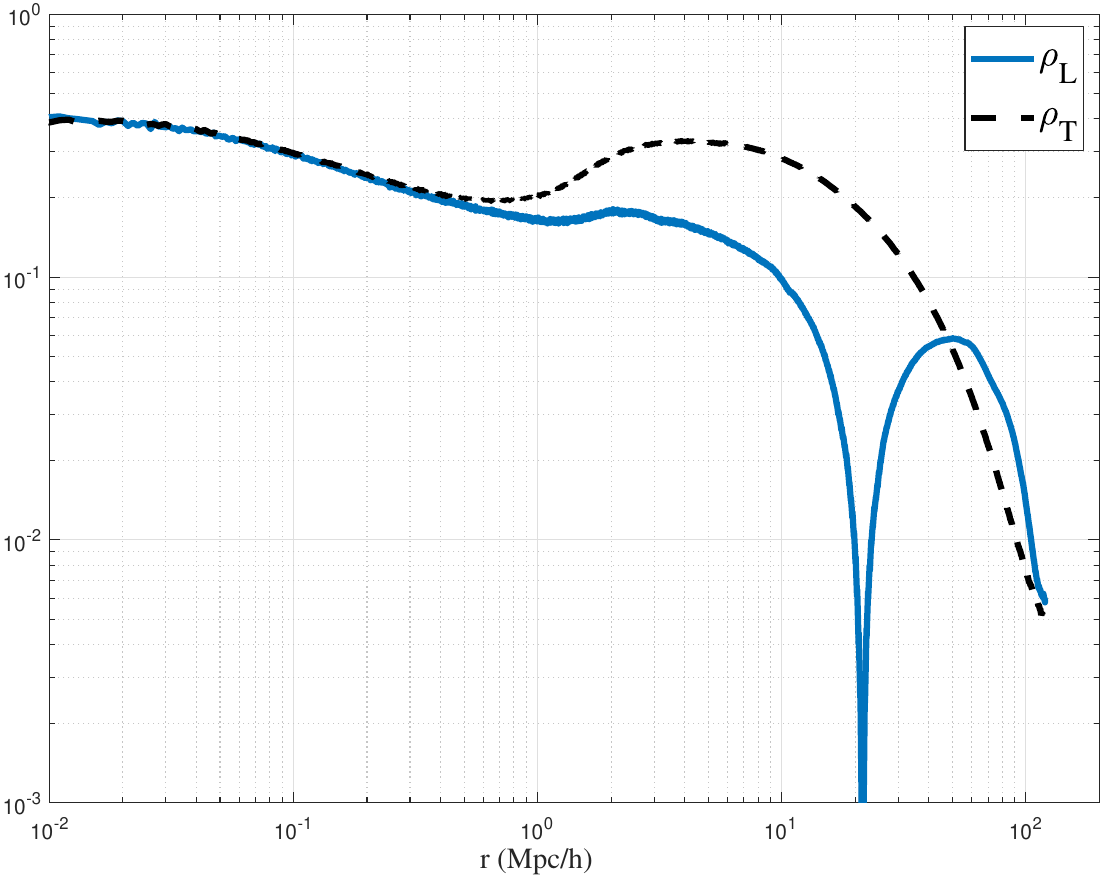}
\caption{The correlation coefficients $\rho _{L} \left(r\right)$ for longitudinal velocity and $\rho _{T} \left(r\right)$for transverse velocity (Eq. \eqref{ZEqnNum890288}) varying with scale \textit{r} at \textit{z}=0. The transverse correlation $\rho _{T} \left(r\right)>0$ for all scales, while the longitudinal correlation becomes negative for scales $r>r_{2} $. Both correlations approach 0.5 for small \textit{r}. The longitudinal velocities $u_{L}^{} $ and $u_{L}^{'} $ are strongly correlated on small scale, and weakly and inversely correlated on large scale $r>r_{2} $. The transverse velocities $u_{T}^{} $ and $u_{T}^{'} $ are positively correlated on all scales.}
\label{fig:15}
\end{figure}
Figure \ref{fig:15} presents the correlation coefficients of longitudinal and transverse velocities ($\rho _{L} \left(r\right)$ and $\rho _{T} \left(r\right)$ in Eq. \eqref{ZEqnNum890288}) varying with scale \textit{r}. The transverse correlation is positive $\rho _{T} \left(r\right)>0$ for all scales., while the longitudinal correlation becomes negative for scales $r>r_{2} \approx 21.4{Mpc/h} $. It can be shown that $u_{L}^{} $ (or $u_{T}^{} $) and $u_{L}^{'} $ (or $u_{T}^{'} $) are strongly correlated (both $u_{L}^{} $ and $u_{L}^{'} $ point to the same direction) on small scales with both $\rho _{L} $ and $\rho _{T} $ approaching 0.5 for small \textit{r}. This is a distinct feature of SG-CFD (Eq. \eqref{ZEqnNum350084}) discussed in Section \ref{sec:3.3.1}. By contrast, velocity is fully correlated $\rho _{L} =\rho _{T} =1$ for incompressible collisional flow. On large scale with $r>r_{2} $, the longitudinal velocities $u_{L}^{} $ and $u_{L}^{'} $ are inversely correlated. The increase in the correlation coefficient between scales 1 and 4 Mpc/h may be due to the sharp drop in the velocity dispersion $\left\langle \left|\boldsymbol{\mathrm{u}}_{T} \right|^{2} \right\rangle $ and $\left\langle u_{L}^{2} \right\rangle $ (Fig. \ref{fig:20}).
\begin{figure}
\includegraphics*[width=\columnwidth]{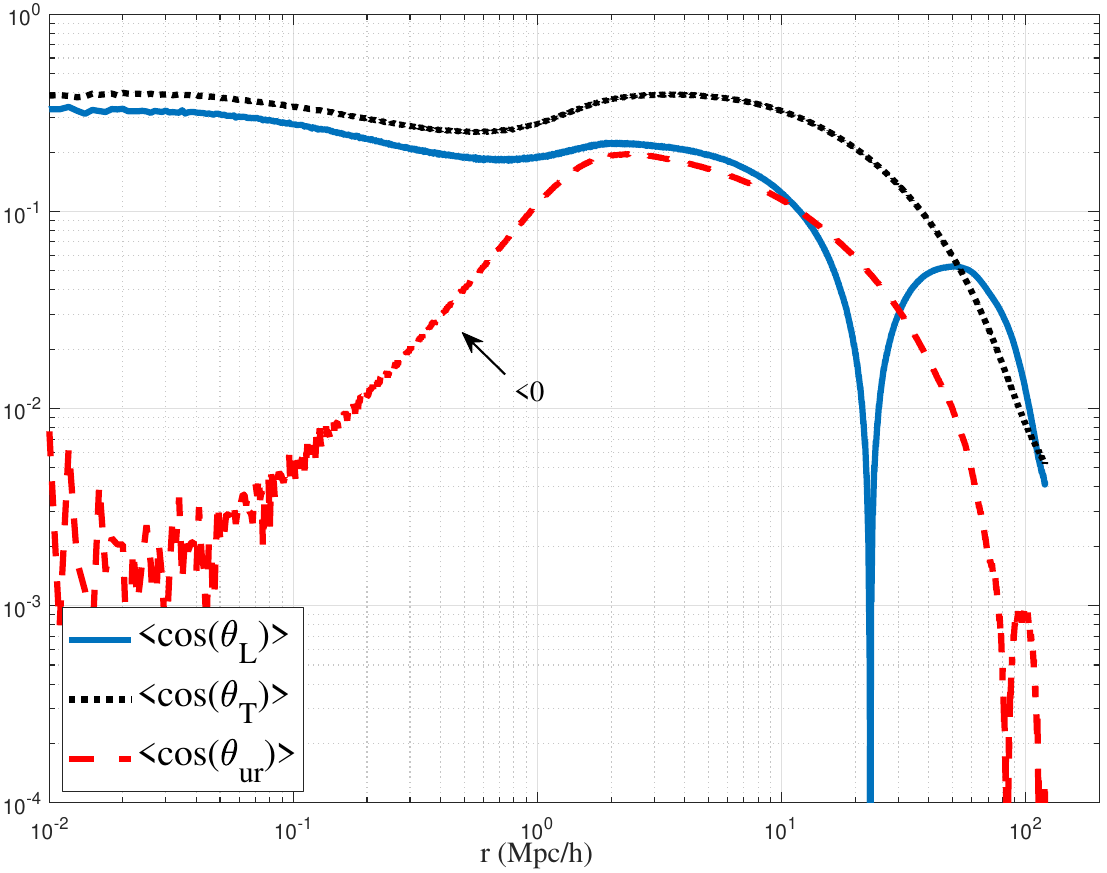}
\caption{The variation of angles $\left\langle \cos \theta _{L} \right\rangle $,  $\left\langle \cos \theta _{T} \right\rangle $ and $\left\langle \cos \theta _{ur} \right\rangle $ (Eqs. \eqref{ZEqnNum924986} and \eqref{ZEqnNum170842}) with scale \textit{r} at \textit{z}=0. Note that statistical independence between \textbf{u} and \textbf{r} fields in incompressible homogeneous turbulence is not the case for SG-CFD, because of $\left\langle u_{L} \right\rangle =\left\langle u_{i} \hat{r}_{i} \right\rangle \ne \left\langle u_{i} \right\rangle \left\langle \hat{r}_{i} \right\rangle =0$. Both fields are statistically uncorrelated only on very small or large scales where $\left\langle u_{L} \right\rangle \to 0$.}
\label{fig:16}
\end{figure}

Similarly, Fig. \ref{fig:16} plots the variation of $\left\langle \cos \theta _{L} \right\rangle $,  $\left\langle \cos \theta _{T} \right\rangle $ and $\left\langle \cos \theta _{ur} \right\rangle $ (defined in Eqs. \eqref{ZEqnNum924986} and \eqref{ZEqnNum170842}) with scale \textit{r} at \textit{z}=0. Note that velocity \textbf{u} and vector \textbf{r} fields are independent in incompressible homogeneous turbulence. This is not true for SG-CFD because of $\left\langle u_{L} \right\rangle =\left\langle \boldsymbol{\mathrm{u}}\cdot \boldsymbol{\mathrm{r}}\right\rangle \ne 0$. Both fields are only statistically uncorrelated on small or large scales, but correlated on intermediate scales. 

\begin{figure}
\includegraphics*[width=\columnwidth]{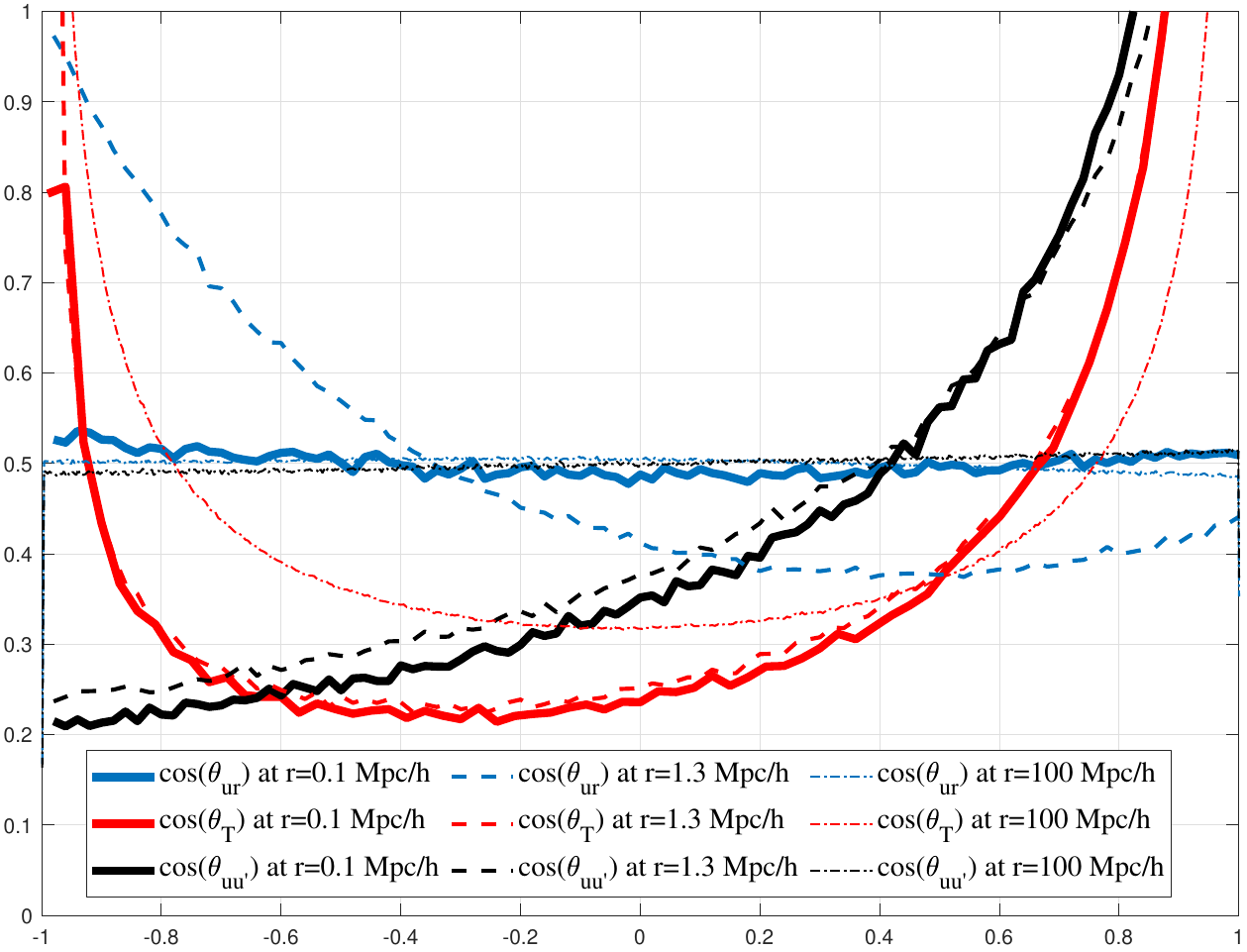}
\caption{The probability distributions of $\cos \theta _{\boldsymbol{\mathrm{ur}}} $, $\cos \theta _{T} $ and $\cos \theta _{\boldsymbol{\mathrm{uu'}}} $ on small, middle, and large scales \textit{r} at redshift z=0. The cosine of angle between \textbf{u} and \textbf{r}, i.e. $\cos \theta _{\boldsymbol{\mathrm{ur}}}$, has a uniform distribution on both small and large scales where all pairs are either from same or different haloes. \textbf{u} and \textbf{r} are uncorrelated on small and large scales.} 
\label{fig:17a}
\end{figure}

\begin{figure}
\includegraphics*[width=\columnwidth]{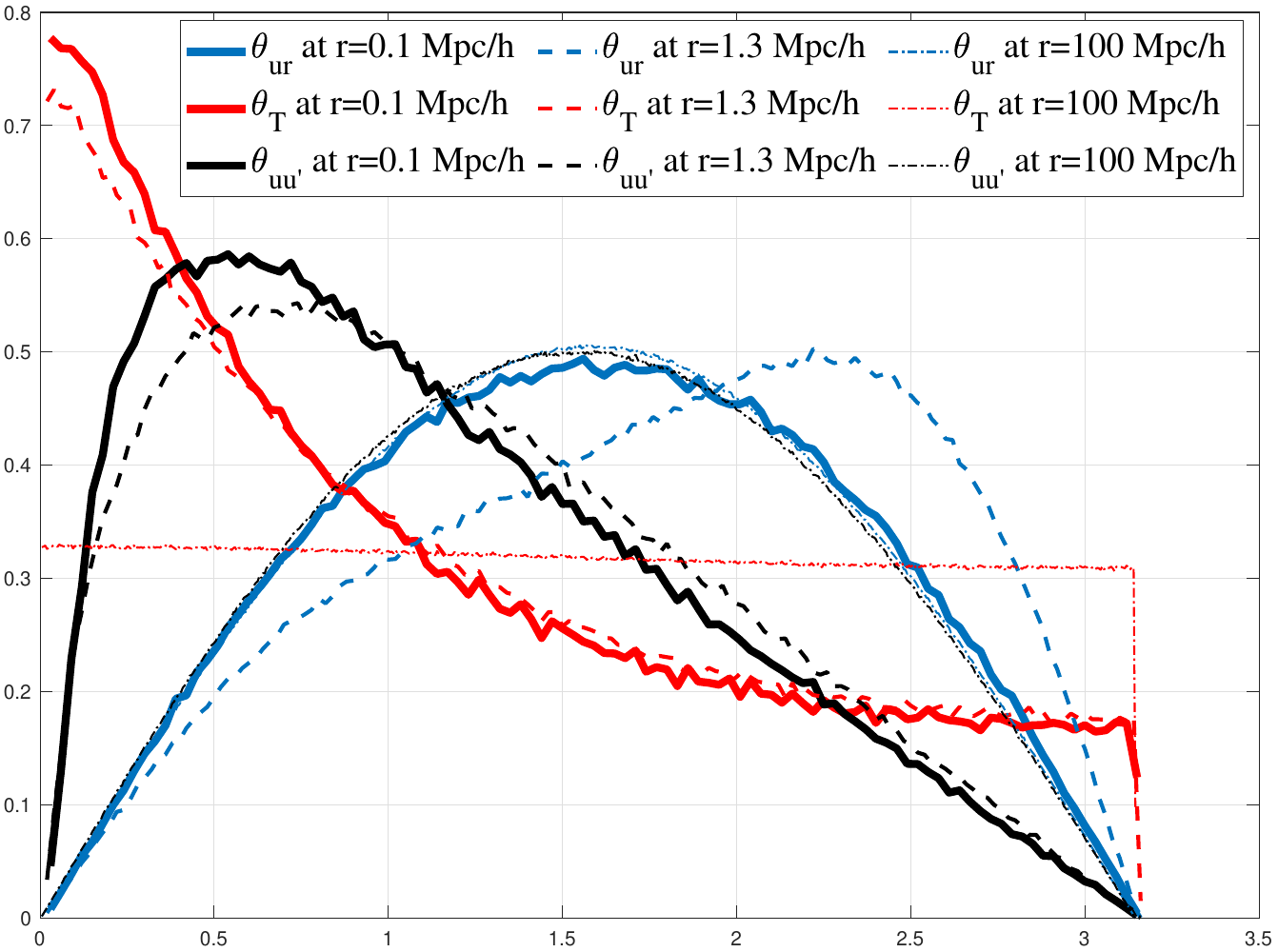}
\caption{The probability distributions of angles $\theta _{\boldsymbol{\mathrm{ur}}} $, $\theta _{T} $, and $\theta _{\boldsymbol{\mathrm{uu'}}} $) on small, middle, and large scales \textit{r} at redshift z=0. On intermediate scale, the skewed distribution of angle $\theta _{\boldsymbol{\mathrm{ur}}} $ indicates that \textbf{u} and \textbf{r }fields are correlated. The angle $\theta _{T} $ between two transverse velocities $\boldsymbol{\mathrm{u}}_{T} $ and $\boldsymbol{\mathrm{u}}_{T}^{'} $ skews toward $\theta _{T} <{\pi /2} $ on small scale due to strong gravity and shifts to uniform distribution on large scale. The angle $\theta _{\boldsymbol{\mathrm{uu}}^{'} } $ also skews toward $<{\pi /2} $ on small scale and $\cos \theta _{\boldsymbol{\mathrm{uu'}}} $ has a uniform distribution on large sale where both $\boldsymbol{\mathrm{u}}$ and $\boldsymbol{\mathrm{u}}_{}^{'} $ are uncorrelated.}
\label{fig:17b}
\end{figure}

To further investigate the statistical properties of \textbf{u} and \textbf{r} fields, we identify all pairs of particles with a given separation \textit{r} and compute the histogram of three angles $\theta _{\boldsymbol{\mathrm{ur}}} $, $\theta _{T} $, and $\theta _{\boldsymbol{\mathrm{uu'}}} $ in Eqs. \eqref{ZEqnNum924986} and  \eqref{ZEqnNum170842}. Figures \ref{fig:17a} and \ref{fig:17b} present the probability distribution of the cosine of three angles (i.e. $\cos \theta _{\boldsymbol{\mathrm{ur}}}$, $\cos \theta _{T}$ and $\cos \theta _{\boldsymbol{\mathrm{uu'}}}$) and the distribution of angles (i.e. $\theta _{\boldsymbol{\mathrm{ur}}} $, $\theta _{T} $, and $\theta _{\boldsymbol{\mathrm{uu'}}} $) on three different scales of \textit{r} at z=0. 

The cosine of angle between \textbf{u} and \textbf{r, }i.e.\textbf{ }$\cos \theta _{\boldsymbol{\mathrm{ur}}} $, has a uniform distribution on both small and large scales where all pairs are either from the same or from different haloes (r=0.1Mpc/h and r=100Mpc/h). Namely, \textbf{u} and \textbf{r} fields are uncorrelated on small and large scales. For a particle with a given \textbf{u}, the unit vector $\hat{\boldsymbol{\mathrm{r}}}$ is uniformed distributed on a unit sphere (distribution is isotropic with equal probability to find another particle separated by \textit{r} along any direction). However, on the intermediate scale where pairs of particles can be either from same haloes or different haloes ($r_{t} =1.3{Mpc/h}$), the skewed distribution of angle $\theta _{\boldsymbol{\mathrm{ur}}}$ indicates that for a particle with given \textbf{u}, probability to find another particles is not isotropic and prefers an angle $\theta_{\boldsymbol{\mathrm{ur}}} >{\pi/2}$. 

The distribution of angle $\theta _{T} $ skews toward $\theta _{T} <{\pi /2} $ on small scale where two transverse velocities are strongly correlated.  Angle $\theta _{T} $ shifts to uniform distribution between [0 $\piup$] on large scale where two transverse velocities $\boldsymbol{\mathrm{u}}_{T} $ and $\boldsymbol{\mathrm{u}}_{T}^{'} $ are uncorrelated. 

The angle $\theta _{\boldsymbol{\mathrm{uu}}^{'} } $ also skews toward $\theta_{uu^{'}} <{\pi/2}$ on small scale due to strong gravity and $\cos \theta _{\boldsymbol{\mathrm{uu'}}} $ has a uniform distribution on large scale where $\boldsymbol{\mathrm{u}}$ and $\boldsymbol{\mathrm{u}}^{'}$ are uncorrelated. On large scale, the possibility to find the velocity $\boldsymbol{\mathrm{u}}^{'} $ is same along any direction for a given velocity \textbf{u}.

\section{Models for various statistical measures}
\label{sec:5}
In this section, we present models for second order statistical measures from N-body simulation. The kinematic relations in Section \ref{sec:3} and simulation results in Section \ref{sec:4} can be combined to provide simple models for density, velocity, and potential fields. Since the self-gravitating collisionless dark matter flow has two distinct regimes: 1) irrotational flow on large scale, and 2) constant divergence flow on small scale, models for each regime are discussed separately followed by combined models for the entire range.  

\subsection{Statistical measures on large scale (linear regime)}
\label{sec:5.1}
\subsubsection{Velocity correlation/dispersion/structure functions}
\label{sec:5.1.1}
The N-body simulation (Fig. \ref{fig:5}) strongly suggests an exponential form for transverse velocity correlation function on large scale
\begin{equation} 
\label{ZEqnNum971850} 
T_{2} \left(r,a\right)=a_{0} u^{2} \exp \left(-\frac{r}{r_{2} } \right),         
\end{equation} 
which should be an intrinsic property of large-scale dynamics. Here pre-factor $a_{0} \left(a\right)$ is function of time and $r_{2}$=21.3Mpc/h is a characteristic length scale. This exponential form is not a coincidence and must be deeply rooted in the dynamics and kinematics on large scale, which is demonstrated in a separate paper on the dynamic relations between correlation functions of different order \citep{Xu:2022-The-statistical-theory-of-3rd}. 

\begin{figure}
\includegraphics*[width=\columnwidth]{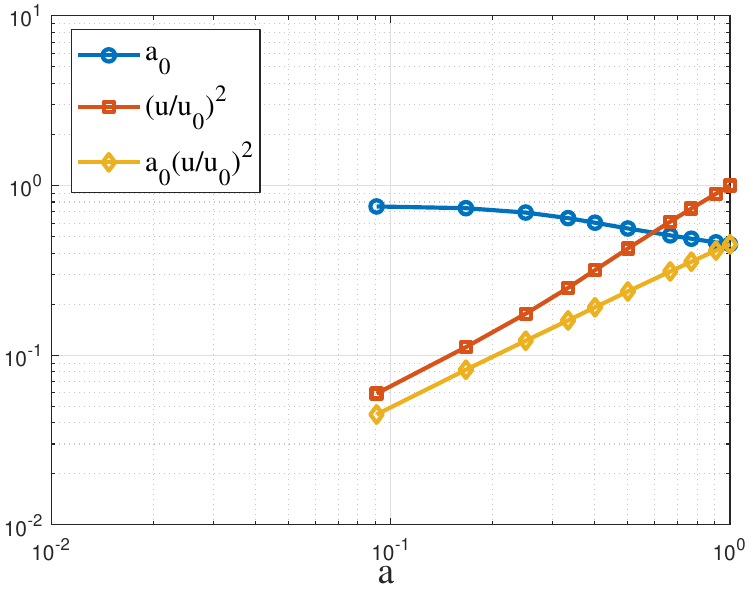}
\caption{The variation of pre-factor $a_{0} \left(a\right)$ and velocity dispersion $u^{2} $ with scale factor \textit{a}. Clearly, $a_{0} u^{2} \propto a$ such that $a_{0} \left({u/u_{0} } \right)^{2} =0.45a$, where $u_{0}^{2} =u^{2} \left(a=1\right)$ is the velocity dispersion at $z=0$.}
\label{fig:18}
\end{figure}

In the present context, what is important is that this provides a very simple and good description of velocity correlation on large scales. Figure \ref{fig:18} plots the variation of pre-factor $a_{0} \left(a\right)$ (obtained by fitting Eq. \eqref{ZEqnNum971850} to data in Fig. \ref{fig:5}) and velocity dispersion $u^{2} $ (in Eq. \eqref{ZEqnNum586699}) with scale factor \textit{a}. Clearly, $a_{0} u^{2} \propto a$ such that $a_{0} \left({u/u_{0} } \right)^{2} =0.45a$, where $u_{0}^{2} =u^{2} \left(a=1\right)$ is the dispersion at present epoch. On large scale, kinetic energy increases $\propto a_{0} u^{2} \propto a$.

With kinematic relations developed in Section \ref{sec:3}, other correlation functions of the same order can be derived easily. For irrotational flow, the relation between three scalar correlation functions (Eq. \eqref{ZEqnNum320035}) enables us to derive the longitudinal correlation function,
\begin{equation} 
\label{ZEqnNum344034} 
L_{2} \left(r,a\right)=a_{0} u^{2} \exp \left(-\frac{r}{r_{2} } \right)\left(1-\frac{r}{r_{2} } \right) 
\end{equation} 
with $L_{2} =0$ at $r=r_{2} $. The plot of $L_{2} $ at different redshift \textit{z} in Fig. \ref{fig:6} clearly indicates that $r_{2}$=21.3 Mpc/h should be relatively independent of time or redshift. 

The total correlation function (see Fig. \ref{fig:7}) reads 
\begin{equation} 
\label{ZEqnNum235904} 
R_{2} \left(r,a\right)=\left\langle \boldsymbol{\mathrm{u}}\cdot \boldsymbol{\mathrm{u}}^{'} \right\rangle =2R\left(r\right)=a_{0} u^{2} \exp \left(-\frac{r}{r_{2} } \right)\left(3-\frac{r}{r_{2} } \right),     
\end{equation} 
with  $R_{2} =0$ at $r=3r_{2} $. The \textit{m}th order moment of $R_{2} \left(r\right)$ is  
\begin{equation} 
\label{ZEqnNum485183} 
\int _{0}^{\infty }R_{2} \left(r\right) r^{m} dr=(2-m)\Gamma \left(1+m\right)r_{2}^{1+m} a_{0} u^{2} ,       
\end{equation} 
where $a_{0} =0.45a$ and $r_{2} =23.13{Mpc/h} $. In principle, the constant comoving lengths scale $r_{2}$ should be related to the horizon size at matter-radiation equality, as shown in Eq. \eqref{eq:134}. 

The correlation length of velocity field (Eq. \eqref{ZEqnNum505635}), 
\begin{equation} 
\label{eq:114} 
l_{u0} =\frac{1}{u^{2} } \int _{0}^{\infty }R \left(r\right)dr=\frac{1}{2u^{2} } \int _{0}^{\infty }R_{2}  \left(r\right)dr=2a_{0} r_{2} .    
\end{equation} 
The velocity dispersion $\sigma _{u}^{2} \left(r\right)$ on scale \textit{r} reads (from Eq. \eqref{ZEqnNum910224})
\begin{equation}
\label{ZEqnNum178726} 
\begin{split}
\sigma _{u}^{2} \left(r\right)&=\frac{3}{2} a_{0} u^{2} \left(\frac{r_{2} }{r} \right)^{2} \left\{-15\left(\frac{r_{2} }{r} \right)^{4} +3\left(\frac{r_{2} }{r} \right)^{2}+\exp \left(-\frac{2r}{r_{2} } \right)\right.\\&\left.\cdot \left[15\left(\frac{r_{2} }{r} \right)^{4} +30\left(\frac{r_{2} }{r} \right)^{3} +27\left(\frac{r_{2} }{r} \right)^{2} +14\left(\frac{r_{2} }{r} \right)+4\right]\right\},
\end{split}
\end{equation} 
where $\sigma _{u}^{2} \left(r\right)\propto r^{-4} $ for $r\to \infty $ and $\sigma _{u}^{2} \left(r\right)=a_{0} u^{2} \propto a$ for $r\to 0$. Comparison with N-body simulation is shown in Fig. \ref{fig:8}. 

The second order longitudinal structure function $S_{2}^{l} \left(r\right)$ on large scale reads (from Eq. \eqref{ZEqnNum905257})
\begin{equation} 
\label{ZEqnNum509836} 
S_{2}^{l} \left(r\right)=2u^{2} \left[1+a_{0} \left(\frac{r}{r_{2} } -1\right)\exp \left(-\frac{r}{r_{2} } \right)\right],  \end{equation} 
while the second order structure function $S_{2}^{x} \left(r\right)$ can be found from Eqs. \eqref{ZEqnNum973047} and \eqref{ZEqnNum952229}
\begin{equation} 
\label{ZEqnNum858651} 
\begin{split}
&S_{2}^{x}\left(r\right)=\frac{3}{2} a_{0} u^{2} \left\{\left[1+3\left(\frac{r_{2} }{r} \right)^{2} \right]\left(\frac{r_{2} }{r} \right)^{2}\right.\\
&\left.-\exp \left(\frac{-2r}{r_{2} } \right)\left[1+\left(\frac{r_{2} }{r} \right)^{2} \right]\left[3\left(\frac{r_{2} }{r} \right)^{2} +6\left(\frac{r_{2} }{r} \right)+4\right]\right\}.
\end{split}
\end{equation} 
The correlation function of the vorticity field vanishes on large scale $R_{\boldsymbol{\mathrm{\omega }}} =0$ because of the irrotational nature, while the correlation of divergence field can be derived from Eq. \eqref{ZEqnNum432904}
\begin{equation} 
\label{ZEqnNum973449}
\begin{split}
R_{\theta } \left(r,a\right)&=\frac{1}{2} \left\langle \left(\nabla \cdot \boldsymbol{\mathrm{u}}\right)\left(\nabla \cdot \boldsymbol{\mathrm{u}}^{'} \right)\right\rangle\\ &=\frac{a_{0} u^{2} }{2rr_{2} } \exp \left(-\frac{r}{r_{2} } \right)\left[\left(\frac{r}{r_{2} } \right)^{2} -7\left(\frac{r}{r_{2} } \right)+8\right],
\end{split}
\end{equation} 
where vanishing divergence correlation $R_{\theta } =0$ at two locations of ${\left(7\pm \sqrt{17} \right)r_{2} /2} $ (33.3 and 128.7Mpc/h). 
 
\subsubsection{Density/potential correlation functions}
\label{sec:5.1.2}
We first introduce a log-density field $\eta(x)=\log (1+\delta)\approx \delta$ and $\eta(x)\approx\delta(x)$ for small overdensity $\delta $. For weakly nonlinear regime, the well-established Zeldovich approximation \citep{Zeldovich:1970-Gravitational-Instability---an} and the linear perturbation theory relate the density $\delta \left(\boldsymbol{\mathrm{x}}\right)$ to potential $\phi \left(\boldsymbol{\mathrm{x}}\right)$ and velocity fields, 
\begin{equation}
\begin{split}
&\delta \approx \eta =-\frac{\nabla \cdot \boldsymbol{\mathrm{u}}}{aHf\left(\Omega _{m} \right)}, \quad \boldsymbol{\mathrm{u}}=-\frac{Hf\left(\Omega _{m} \right)\nabla \phi }{4\pi G\rho a},\\
&\textrm{and} \quad \delta \approx \eta =\frac{\nabla ^{2} \phi }{4\pi G\rho a^{2} }, 
\end{split}
\label{ZEqnNum513737}
\end{equation}
where $f\left(\Omega _{m} \right) \approx \Omega_m^{0.6}$ is a function of matter content $\Omega _{m} $ and $f\left(\Omega _{m} =1\right)=1$ for matter dominant model. With the help of Eqs. \eqref{ZEqnNum432904} and \eqref{ZEqnNum513737}, the density correlation on large scale should read
\begin{equation} 
\label{eq:120} 
\begin{split}
\xi \left(r,a\right)&=\left\langle \delta \left(\boldsymbol{\mathrm{x}}\right)\cdot \delta \left(\boldsymbol{\mathrm{x}}^{'} \right)\right\rangle =\frac{\left\langle \theta \left(\boldsymbol{\mathrm{x}}\right)\cdot \theta \left(\boldsymbol{\mathrm{x}}^{'} \right)\right\rangle }{\left(aHf\left(\Omega _{m} \right)\right)^{2} }\\
&=-\frac{1}{\left(aHf\left(\Omega _{m} \right)\right)^{2} } \left[\frac{1}{r^{2} } \frac{\partial }{\partial r} \left(r^{2} \frac{\partial R_{2} }{\partial r} \right)\right].
\end{split}
\end{equation} 
Density correlation can be modeled as (with $R_{2}$ in Eq. \eqref{ZEqnNum235904})
\begin{equation} 
\label{ZEqnNum762470} 
\xi \left(r,a\right)=\frac{{a_{0}u^{2} }/{(rr_{2}) }}{\left(aHf\left(\Omega _{m} \right)\right)^{2} } \cdot  \exp \left(-\frac{r}{r_{2} } \right)\left[\left(\frac{r}{r_{2} } \right)^{2} -7\left(\frac{r}{r_{2} } \right)+8\right].
\end{equation} 
The derivative of density correlation identifies the peak of correlation
\begin{equation} 
\label{ZEqnNum250582} 
\begin{split}
\frac{d\xi }{dr} &=-\frac{{a_{0} u^{2} }/{(r^{2} r_{2}) }}{\left(aHf\left(\Omega _{m} \right)\right)^{2} } \\
&\cdot  \exp \left(-\frac{r}{r_{2} } \right)\left[\left(\frac{r}{r_{2} } \right)^{3} -8\left(\frac{r}{r_{2} } \right)^{2} +8\left(\frac{r}{r_{2} } \right)+8\right].
\end{split}
\end{equation} 
Two peaks of correlations can be found at $2r_{2}$ and $(3+\sqrt{13})r_{2} \approx 6.6r_{2}$, respectively. 
\begin{comment} 
In principle, $r_{2}$ may be related to the size of sound horizon $r_{s}$ \citep{Eisenstein:2005-Detection-of-the-baryon-acoust}, the peak in the correlation function.
\end{comment}
The \textit{m}th order moment of the density correlations reads  
\begin{equation} 
\label{eq:123} 
\int _{0}^{\infty }\xi \left(r\right) r^{m} dr=\frac{a_{0} u^{2}r_{2}^{m-1}}{\left(aHf\left(\Omega _{m} \right)\right)^{2} } \left(m-4\right)\left(m-2\right)\Gamma \left(m\right).
\end{equation} 

The averaged correlation $\bar{\xi }\left(r,a\right)$ on large scale should read
\begin{equation}
\label{ZEqnNum197404} 
\begin{split}
&\bar{\xi }\left(r,a\right)=\frac{3}{r^{3} } \int _{0}^{r}\xi \left(y,a\right) y^{2} dy=-\frac{3}{\left(aHf\left(\Omega _{m} \right)\right)^{2} r} \frac{\partial R_{2} }{\partial r}\\&=\frac{a_{0} u^{2} }{\left(aHf\left(\Omega _{m} \right)\right)^{2} } \frac{3}{rr_{2} } \exp \left(-\frac{r}{r_{2} } \right)\left(4-\frac{r}{r_{2} } \right). 
\end{split}
\end{equation} 
Equation \eqref{ZEqnNum197404} can be combined with pair conservation equation \citep{Peebles:1980-The-Large-Scale-Structure-of-t} to derive the mean pairwise velocity on large scale \citep{Xu:2022-Two-thirds-law-for-pairwise-ve}, i.e., $\langle\Delta u_L\rangle \propto -Har\bar{\xi}(r,a)$.

\begin{figure}
\includegraphics*[width=\columnwidth]{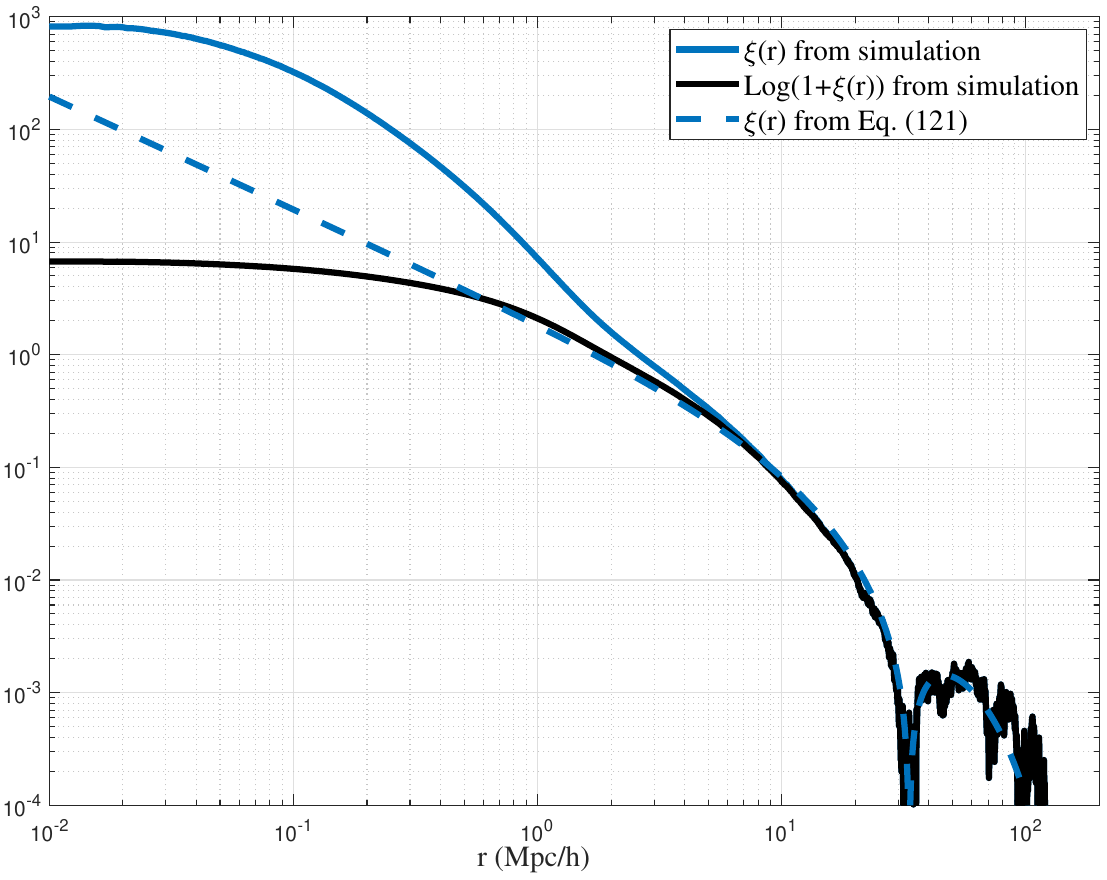}
\caption{Comparison of density correlation $\xi \left(r\right)$ at \textit{z}=0 with model (Eq. \eqref{ZEqnNum762470}) derived from velocity correlation functions. The model agrees well with simulation on large scale. The model agrees with $\ln \left[1+\xi \left(r\right)\right]$ even better. The proposed model can be applied to a wider range for the correlation of log-density field $\eta(x)=\log(1+\delta)$.}  
\label{fig:19}
\end{figure}
Figure \ref{fig:19} presents the comparison of density correlation model in  Eq. \eqref{ZEqnNum762470} with $\xi \left(r\right)$ from simulation at \textit{z}=0. The model of $\xi \left(r\right)$ derived from velocity correlation function agrees well with simulation on large scale. The model agrees with $\ln \left[1+\xi \left(r\right)\right]$ even better in a wider range of scale \textit{r}. 

Finally, the correlation of the potential field can be obtained from Eq. \eqref{ZEqnNum513737}. For matter dominant model, $H^{2} ={8\pi G\rho /3} $, the gradient of potential has a correlation
\begin{equation}
\label{eq:125} 
\begin{split}
R_{\nabla \phi }&=\frac{1}{2} \left\langle \nabla \phi \left(\boldsymbol{\mathrm{x}}\right)\cdot \nabla \phi \left(\boldsymbol{\mathrm{x}}^{'} \right)\right\rangle =\frac{9}{4} \left(\frac{aH}{f\left(\Omega _{m} \right)} \right)^{2} R\left(r\right)\\
&=\frac{9}{8} \left(\frac{aH}{f\left(\Omega _{m} \right)} \right)^{2} a_{0} u^{2} \exp \left(-\frac{r}{r_{2} } \right)\left(3-\frac{r}{r_{2} } \right).
\end{split}
\end{equation} 
Relations between $\phi$ and its gradient are (Eq. \eqref{ZEqnNum226610})
\begin{equation} 
\label{eq:126} 
R_{\nabla \phi } =-\frac{1}{r^{2} } \frac{\partial }{\partial r} \left[r^{2} \frac{\partial R_{\phi } }{\partial r} \right],
\end{equation} 
\begin{equation} 
\label{eq:127} 
R_{\nabla ^{2} \phi } =-\frac{1}{r^{2} } \frac{\partial }{\partial r} \left[r^{2} \frac{\partial R_{\nabla \phi } }{\partial r} \right]=\frac{\partial ^{4} R_{\phi } }{\partial r^{4} } +\frac{4}{r} \frac{\partial ^{3} R_{\phi } }{\partial r^{3} } .
\end{equation} 
The correlation function of potential field reads 
\begin{equation} 
\label{ZEqnNum590702} 
\begin{split}
R_{\phi }&=\frac{1}{2} \left\langle \phi \left(\boldsymbol{\mathrm{x}}\right)\cdot \phi \left(\boldsymbol{\mathrm{x}}^{'} \right)\right\rangle\\ &=\frac{9}{8} \left(\frac{aH}{f\left(\Omega _{m} \right)} \right)^{2} a_{0} u^{2} r_{2}^{2} \exp \left(-\frac{r}{r_{2} } \right)\left[\left(\frac{r}{r_{2} } \right)+1\right],
\end{split}
\end{equation} 
where $R_{\phi } \propto a^{0} $ such that the potential fluctuation $\left\langle \phi ^{2} \right\rangle $ is independent of time on large scale, i.e., a constant potential perturbation in the linear regime.

\subsubsection{Velocity/density/potential spectrum in Fourier space}
\label{sec:5.1.3}
With all correlation functions developed, the corresponding spectrum functions in Fourier space can be easily obtained. The spectrum of velocity (from Eq. \eqref{ZEqnNum891034}) reads
\begin{equation} 
\label{ZEqnNum976884} 
E_{u} \left(k\right)=a_{0} u^{2} \frac{8}{\pi r_{2} } \frac{k^{-2} }{\left(1+{1/\left(kr_{2}^{} \right)^{2} } \right)^{3} } ,        
\end{equation} 
where $E_{u} \left(k\right)\propto k^{4} $ for $kr_{2}^{} \ll 1$, which is a requirement of the conservation of angular momentum. Note that $E_{u} \left(k\right)\propto k^{-2} $ for  $kr_{2}^{} \gg 1$  (a signature of the Burger's equation in the weakly nonlinear regime). The maximum $E_{u} \left(k\right)$ is
\begin{equation}
E_{u} \left(k_{\max } \right)=\frac{256}{125\pi } r_{2} a_{0} u^{2} \quad \textrm{with} \quad k_{\max } r_{2} =\sqrt{2}.
\label{eq:130}
\end{equation}
The spectrum of divergence is obtained from Eq. \eqref{ZEqnNum973449}
\begin{equation} 
\label{eq:131} 
E_{\theta } \left(k\right)=a_{0} u^{2} \frac{8}{\pi r_{2} } \frac{1}{\left(1+{1/\left(kr_{2}^{} \right)^{2} } \right)^{3} } . \end{equation} 
The density spectrum $E_{\delta } \left(k\right)$ can be obtained from the density correlation in Eq. \eqref{ZEqnNum762470}
\begin{equation} 
\label{eq:132} 
\begin{split}
E_{\delta } \left(k\right)&=\frac{2}{\pi } \int _{0}^{\infty }\xi \left(r,a\right)kr\sin \left(kr\right)dr\\
&=\frac{16a_{0} u^{2} }{\left(aHf\left(\Omega _{m} \right)\right)^{2} \pi r_{2} } \frac{1}{\left(1+{1/\left(kr_{2}^{} \right)^{2} } \right)^{3} }.
\end{split}
\end{equation} 
The usual matter power spectrum $P_{\delta } \left(k,a\right)$ is related to the density spectrum $E_{\delta } \left(k,a\right)$ as 
\begin{equation} 
\label{eq:133}
\begin{split}
P_{\delta } \left(k,a\right)&={2\pi ^{2} E_{\delta } \left(k,a\right)/k^{2} }\\
&=\frac{32\pi a_{0} u^{2} r_{2} }{\left(aHf\left(\Omega _{m} \right)\right)^{2} } \frac{1}{\left(kr_{2}^{} \right)^{2} \left(1+{1/\left(kr_{2}^{} \right)^{2} } \right)^{3} }.
\end{split}
\end{equation} 
With $k_{\max} r_{2} =\sqrt{2} $, the maximum matter power spectrum is
\begin{equation} 
\label{eq:134} 
P_{\delta } \left(k_{\max } ,a\right)=\frac{128\pi a_{0} u^{2} r_{2} }{27\left(aHf\left(\Omega _{m} \right)\right)^{2} } .  \end{equation} 
The spectrum of potential field $E_{\phi } \left(k\right)$ can be obtained from Fourier transform of Eq. \eqref{ZEqnNum590702}
\begin{equation}
\label{ZEqnNum759927} 
E_{\phi } \left(k\right)=\frac{18}{\pi r_{2} } \left(\frac{aH}{f\left(\Omega _{m} \right)} \right)^{2} \frac{a_{0} u^{2} k^{-4} }{\left(1+{1/\left(kr_{2}^{} \right)^{2} } \right)^{3} } .       
\end{equation} 

\begin{figure}
\includegraphics*[width=\columnwidth]{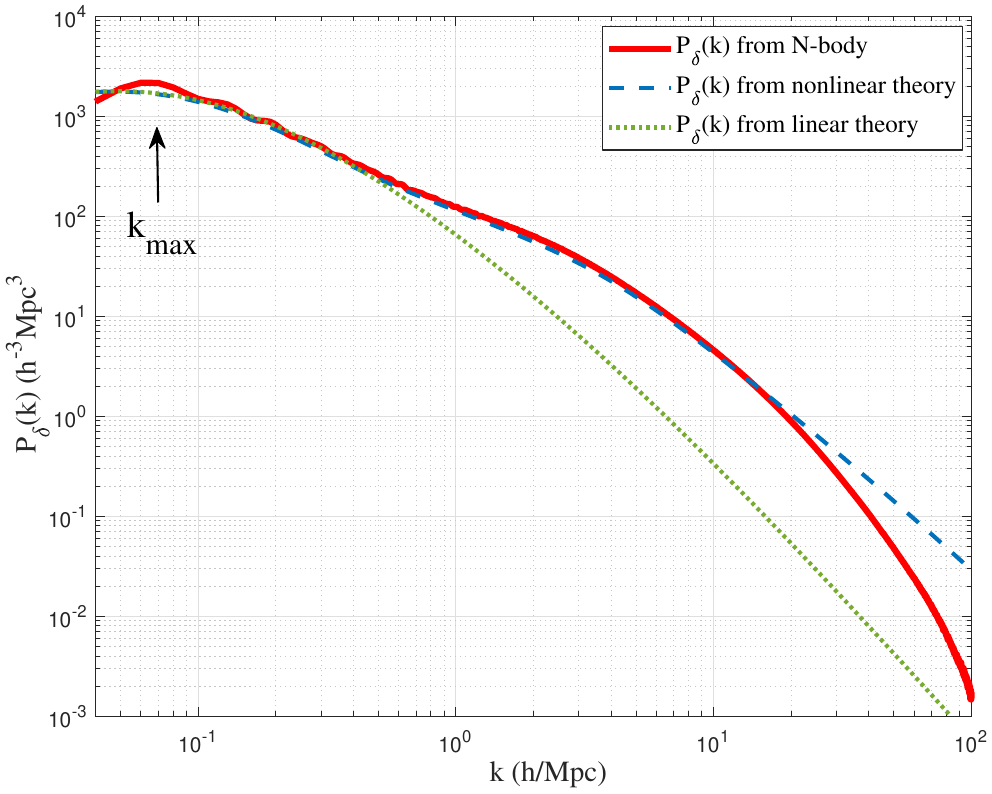}
\caption{The variation of density power spectra $P_{\delta}(k)$ with wavenumber $k$ at \textit{z}=0. The linear and nonlinear theory prediction are also presented for comparison. The pivot wavenumber $k_{\max}$ (or scale $r_2$) is dependent on the matter content $\Omega_mh^2$. The corresponding maximum power is shown in Eq. \eqref{eq:134}.}
\label{fig:21-2}
\end{figure}

In this section, the comoving length scale $r_2$ corresponds to a wavenumber $k_{\max}$ with the maximum power in density spectrum (see Fig. \ref{fig:21-2}). The pivot wavenumber $k_{\max}$ represents the scale of particle horizon at the equality epoch \citep{Eisenstein:1997-Baryonic-features-in-the-matter} that is only dependent on the matter content $\Omega_mh^2$ ($k_{\max}\propto \Omega_mh^2$). Therefore, the comoving scale $r_2$ should be relevant to the horizon size at matter-radiation equality. The sound horizon at the drag epoch, i.e., the comoving distance that a sound wave can travel between the big bang and the redshift of drag epoch, is also related to wavenumber $k_{\max}$ and dependent on both matter content and the fraction of baryon component \citep{Eisenstein:1997-Baryonic-features-in-the-matter}. Therefore, the length scale $r_2$ should be related (not equal) to that sound horizon. Figure \ref{fig:21-2} plots the density spectrum $P_{\delta}(k)$ that is obtained by the Fourier transform of the density correlation $\xi(r)$ in Fig. \ref{fig:19}.  The density correlation $\xi(r)$ is obtained directly from N-body simulation. The corresponding predictions from the linear and nonlinear theory are also presented for comparison \citep{Jenkins:1998-Evolution-of-structure-in-cold}. Good agreement with the nonlinear model in most range of $k$ validates the numerical implementation in this work.

\subsubsection{Modeling longitudinal structure function on large scale}
\label{sec:5.1.4}
The model for structure function $S_{2}^{l} \left(r\right)$ is presented in Eq. \eqref{ZEqnNum509836}. With longitudinal correlation function $L_{2} \left(r\right)$ derived on large scale (Eq. \eqref{ZEqnNum344034}), the other second order structure function $S_{2}^{lp} \left(r\right)=\left\langle \left(\Delta u_{L} \right)^{2} \right\rangle =2\left(\left\langle u_{L}^{2} \right\rangle -L_{2} \left(r\right)\right)$ (the dispersion of pairwise velocity $\Delta u_{L} $ in Eq. \eqref{ZEqnNum250774}) requires a detail model of longitudinal velocity dispersion function $\left\langle u_{L}^{2} \right\rangle $.

\begin{figure}
\includegraphics*[width=\columnwidth]{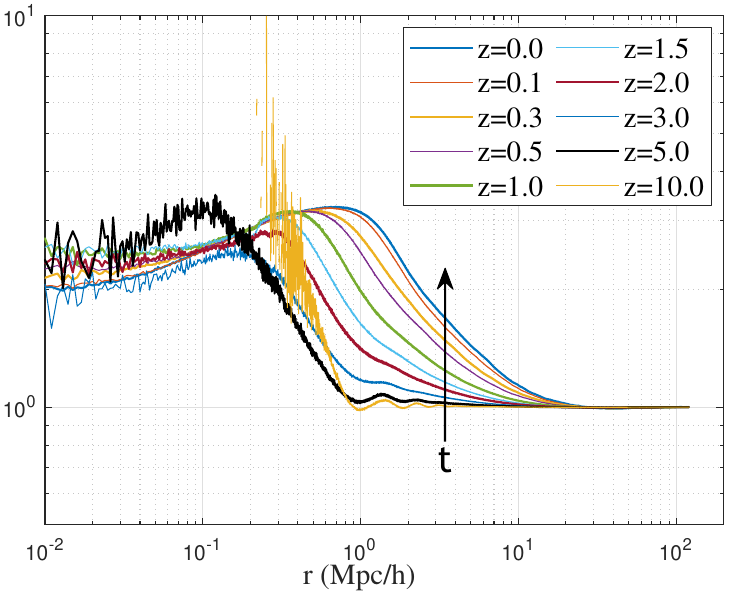}
\caption{The variation of longitudinal velocity dispersion $\left\langle u_{L}^{2} \right\rangle $ with scale \textit{r} at different redshifts \textit{z} (normalized by dispersion $u^{2} $). The scale $r_{d} $ at which $\left\langle u_{L}^{2} \right\rangle $ is at its maximum increases with time, where $r_{d} \approx 0.7a{Mpc/h} $ reflects the increasing halo size with time (see Fig. \ref{fig:9}). $\left\langle u_{L}^{2} \right\rangle =u^{2} $ on all scales at early \textit{z} such that $S_{2}^{lp} \left(r\right)=S_{2}^{l} \left(r\right)$. With halo gradually formed, $\left\langle u_{L}^{2} \right\rangle $ converges to $2u^{2}$ on small scale. Model for $\left\langle u_{L}^{2} \right\rangle $ is presented in Eq. \eqref{ZEqnNum776534}.}
\label{fig:20}
\end{figure}
Figure \ref{fig:20} presents the variation of $\left\langle u_{L}^{2} \right\rangle $ with scale \textit{r} at different redshift \textit{z}. Initially $\left\langle u_{L}^{2} \right\rangle =u^{2}$ on all scales at early \textit{z} such that $S_{2}^{lp} \left(r\right)=S_{2}^{l} \left(r\right)$. With halo structure gradually formed, $\left\langle u_{L}^{2} \right\rangle $ quickly converges to $2u^{2} $ on small scale where pair of particles are from the same halo, as discussed in Section \ref{sec:3.3.1} (Eq. \eqref{ZEqnNum890245}). The comoving length scale $r_{d} $ at which $\left\langle u_{L}^{2} \right\rangle $ is at its maximum increases with time, where $r_{d} \approx 0.70a{Mpc/h} $ (Fig. \ref{fig:9}) might be related to the size of haloes that increases with time.

\begin{figure}
\includegraphics*[width=\columnwidth]{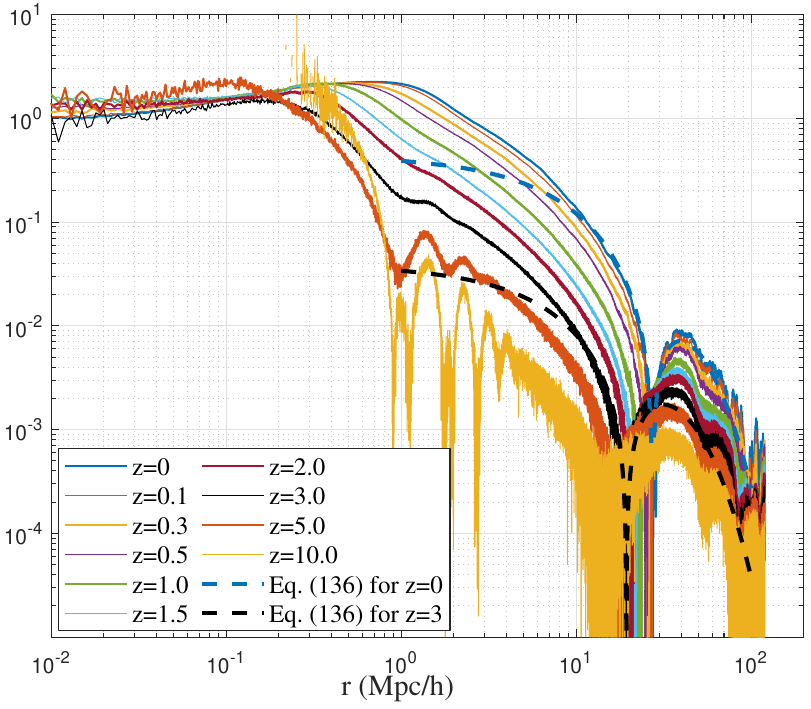}
\caption{The variation of longitudinal velocity dispersion ${\left\langle u_{L}^{2} \right\rangle /u^{2} } -1$ with scale \textit{r} at different redshift \textit{z}. Equation \eqref{ZEqnNum776534} provides a good model for $\left\langle u_{L}^{2} \right\rangle $ on the large scale above $10{Mpc/h} $ for all redshift.}
\label{fig:21}
\end{figure}
Inspired by the functional form of correlation functions in Section \ref{sec:5.1.1}, a good equation to fit $\left\langle u_{L}^{2} \right\rangle $ on large scale can be found as,
\begin{equation} 
\label{ZEqnNum776534} 
\left\langle u_{L}^{2} \right\rangle =u^{2} \left[1+a_{d} \exp \left(-\frac{r}{r_{d1} } \right)\left(1-\frac{r}{r_{d2} } \right)\right],       
\end{equation} 
where $a_{d} =0.44a^{{7/4}}$, $r_{d1} =11.953$Mpc/h and $r_{d2} =27.4a^{{1/4}}$Mpc/h. The dispersion of pairwise velocity $\left\langle \left(\Delta u_{L} \right)^{2} \right\rangle $ on large scale can be easily obtained with Eqs. \eqref{ZEqnNum776534} and \eqref{ZEqnNum250774}. Its behavior on small scale follows a two-thirds law that is discussed in a separate paper \citep{Xu:2022-Postulating-dark-matter-partic}, which can be relevant to dark matter particle mass and properties. Figure \ref{fig:21} presents the variation of $\left\langle u_{L}^{2} \right\rangle -1$ with scale \textit{r} at different redshift \textit{z}, compared with model in Eq. \eqref{ZEqnNum776534}. 

\subsection{Statistical measures on small scale (nonlinear regime)}
\label{sec:5.2}
\subsubsection{Structure functions on small scale (nonlinear regime)}
\label{sec:5.2.1}
\begin{figure}
\includegraphics*[width=\columnwidth]{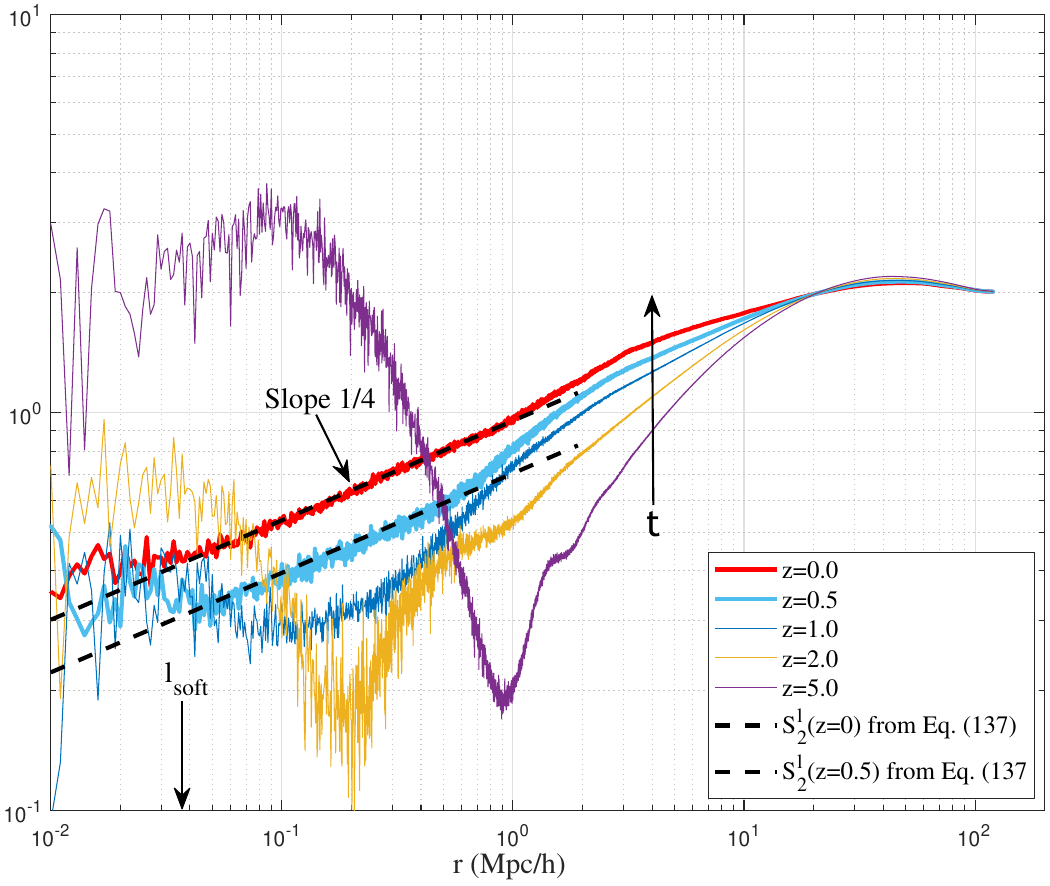}
\caption{The variation of longitudinal structure function $S_{2}^{l} \left(r\right)$ with scale \textit{r} at different redshifts \textit{z}. The structure function is normalized by $u^{2} \left(a\right)$. With more haloes formed, a power-law scaling of $S_{2}^{l} \left(r\right)\propto r^{{1/4} } $ can be established on small scale after z=1.0 and is plotted for \textit{z}=0.5 and \textit{z}=0.0 in the same figure for comparison.}
\label{fig:22}
\end{figure}

The small-scale structure functions follow a power-law scaling, as suggested from the N-body simulation data and virial theorem discussed in Section \ref{sec:4.4}. The longitudinal structure function $S_{2}^{l} \left(r\right)=2\left(u_{}^{2} -L_{2} \left(r\right)\right)$ (Eq. \eqref{ZEqnNum414891}) follows a power-law on small scale as $S_{2}^{l} \left(r\right)\propto r^{{1/4} } $. Figure \ref{fig:22} plots the variation of $S_{2}^{l} \left(r\right)$ with scale \textit{r} at different redshift \textit{z}. A power-law scaling $S_{2}^{l} \left(r\right)\propto r^{{1/4}}$ is fully developed on small scale after \textit{z}=1.0 with more haloes formed in the system. \rev{It is worth noting the effect of gravitational softening length $l_{soft}$ (black arrow in figure) on the statistics at small scales. The deviation from the 1/4 scaling law was found at scales around or below $l_{soft}$. To further confirm the statistics on small scales, results from more recent cosmological simulations with smaller softening length should be explored.}

From data in Fig. \ref{fig:22}, we model the longitudinal structure function in a general power-law form as
\begin{equation} 
\label{ZEqnNum392323} 
S_{2}^{l} =2u^{2} \left({r/r_{1} } \right)^{n} ,           
\end{equation} 
with $n\approx {1/4} $ and $r_{1} \left(a\right)\approx r_{1}^{*} a^{-3} $ is a length scale dependent on the scale factor \textit{a} with $r_{1}^{*} \approx 19.4{Mpc/h} $. The one-fourth power-law scaling of $S_{2}^{l} $ was discussed in Section \ref{sec:4.4} (Eq. \eqref{ZEqnNum489540}) that might be related to the power-law scaling of density correlation $\xi \left(r\right)$ via virial theorem. 

\subsubsection{Correlation/structure/dispersion functions on small scale}
\label{sec:5.2.2}
With structure function  from Eq. \eqref{ZEqnNum392323} and kinematic relation in Eq. \eqref{ZEqnNum258035}, the longitudinal velocity correlation $L_{2} \left(r\right)$ has the form of,
\begin{equation} 
\label{ZEqnNum178092} 
L_{2} \left(r\right)=u^{2} \left[1-\left(\frac{r}{r_{1} } \right)^{n} \right].          
\end{equation} 
Similarly, kinematic relations developed in Section \ref{sec:3} can be used to derive the total velocity correlations (Eq. \eqref{ZEqnNum314105}) and transverse correlation from Eq. \eqref{ZEqnNum610884} 
\begin{equation} 
\label{ZEqnNum955991} 
R_{2} =u^{2} \left[3-\left(3+n\right)\left(\frac{r}{r_{1} } \right)^{n} \right],           
\end{equation} 
\begin{equation} 
\label{ZEqnNum739102} 
T_{2} =u^{2} \left[1-\frac{2+n}{2} \left(\frac{r}{r_{1} } \right)^{n} \right].         
\end{equation} 
On small scale, velocity dispersion function reads (Eq. \eqref{ZEqnNum910224})
\begin{equation} 
\label{eq:141} 
\sigma _{d}^{2} \left(r\right)=\frac{24\cdot 2^{n} }{\left(4+n\right)\left(6+n\right)} u^{2} \left(\frac{r}{r_{1} } \right)^{n} \approx 1.0745u^{2} \left(\frac{r}{r_{1} } \right)^{n}.  
\end{equation} 
Corresponding structure functions are (Eqs. \eqref{ZEqnNum414891} and \eqref{ZEqnNum973047})
\begin{equation}
S_{2}^{l} \left(r\right)=2u^{2} \left(\frac{r}{r_{1} } \right)^{n}, \quad S_{2}^{i} \left(r\right)=2\left(3+n\right)u^{2} \left(\frac{r}{r_{1} } \right)^{n}.    
\label{eq:142}
\end{equation}
The structure function for enstrophy is (from Eq. \eqref{ZEqnNum952229})
\begin{equation}
\label{ZEqnNum413972} 
S_{2}^{x} \left(r\right)=\frac{6n\left(3+n\right)\cdot 2^{n} }{\left(4+n\right)\left(2+n\right)} u^{2} \left(\frac{r}{r_{1} } \right)^{n} =0.6063u^{2} \left(\frac{r}{r_{1} } \right)^{n} .      
\end{equation} 
On small scale, the peculiar velocity has a constant divergence such that $R_{\theta } =0$. The vorticity correlation can be obtained from Eq. \eqref{ZEqnNum244673}
\begin{equation} 
\label{ZEqnNum352266} 
R_{\boldsymbol{\mathrm{\omega }}} =\frac{1}{2} \left\langle \boldsymbol{\mathrm{\omega }}\left(\boldsymbol{\mathrm{x}}\right)\cdot \boldsymbol{\mathrm{\omega }}\left(\boldsymbol{\mathrm{x}}^{'} \right)\right\rangle =\frac{n\left(1+n\right)\left(3+n\right)}{2r^{2} } u^{2} \left(\frac{r}{r_{1} } \right)^{n} .      
\end{equation} 
The corresponding velocity and vorticity spectrum functions are (from Eqs. \eqref{ZEqnNum891034} and \eqref{ZEqnNum639543})
\begin{equation}
E_{u} \left(k\right)=Cu^{2} r_{1}^{-n} k^{-\left(1+n\right)}, \quad E_{\boldsymbol{\mathrm{\omega }}} \left(k\right)=Cu^{2} r_{1}^{-n} k^{\left(1-n\right)}.
\label{eq:145}
\end{equation}
\noindent where with parameters $n={1/4}$, $r_{1} =19.4a^{-3} {Mpc/h}$, the constant 
\begin{equation} 
\label{eq:146} 
C=-\frac{2\left(3+n\right)\Gamma \left({\left(n+3\right)/2} \right)}{2^{1-n} \Gamma \left({3/2} \right)\Gamma \left({-n/2} \right)} =0.4485.
\end{equation} 
The other two relevant power spectrum for velocity and vorticity fields are $P_{u}(k)\propto {E_{u}(k)/k^{2}}$ and $P_{\omega}(k)\propto {E_{\omega}(k)/k^{2}}$.

\subsection{Statistical measures on entire range of scales}
\label{sec:5.3}
With correlation functions determined on both small and large scales, we are now ready to model correlation functions for the entire range of flow. The goal is to provide a smooth and differentiable velocity correlations for the entire range such that the correlations of vorticity and divergence can be obtained as derivatives of velocity correlations (Eqs. \eqref{ZEqnNum441477} and \eqref{ZEqnNum528609}). The modeling results in this section are good for redshift z=0, while the same method can be extended to model correlation functions at any redshift. 

We first introduce a general method to smoothly connect two known functions $f_{1} \left(r\right)$ and $f_{2} \left(r\right)$
\begin{equation} 
\label{ZEqnNum294673} 
f_{\left(fit\right)} \left(r\right)=f_{1} \left(r\right)\left(1-s\left(r\right)\right)^{n1} +f_{2} \left(r\right)s\left(r\right)^{n2} ,       
\end{equation} 
where $s\left(r\right)$ is the interpolation function to smoothly connect $f_{1} \left(r\right)$ and $f_{2} \left(r\right)$ with
\begin{equation} 
\label{ZEqnNum986217} 
s\left(r\right)=\frac{1}{1+x_{b} e^{-{\left(r-x_{c} \right)/x_{a} } } } . \end{equation} 
With the property that $s\left(r\gg x_{c} \right)=1$ and $s\left(r\ll x_{c} \right)=0$, $f_{\left(fit\right)} \left(r\gg x_{c} \right)=f_{2} \left(r\right)$ and $f_{\left(fit\right)} \left(r\ll x_{c} \right)=f_{1} \left(r\right)$, where $x_{a} $, $x_{b} $, $x_{c} $, $n_{1} $, and $n_{2} $ are parameters for each smooth connection. 

The velocity correlation function $R_{2} \left(r\right)$ will be modeled first as an example. Here two functions corresponding to $R_{2} \left(r\right)$ on small and large scales are (Eqs. \eqref{ZEqnNum955991} and \eqref{ZEqnNum235904})
\begin{equation} 
\label{ZEqnNum360823} 
f_{1} \left(r\right)=R_{2s} \left(r\right)=3-\left(3+n\right)\left(\frac{r}{r_{1} } \right)^{n}  
\end{equation} 
and
\begin{equation} 
\label{ZEqnNum438462} 
f_{2} \left(r\right)=R_{2l} \left(r\right)=a_{0} \exp \left(-\frac{r}{r_{2} } \right)\left(3-\frac{r}{r_{2} } \right).       
\end{equation} 
The fitted correlation function for entire range can be expressed as
\begin{equation} 
\label{ZEqnNum701881} 
R_{2\left(fit\right)} \left(r\right)=R_{2s} \left(1-s\left(r\right)\right)^{n1} +R_{2l} \left(s\left(r\right)\right)^{n2} .
\end{equation} 
Here five fitting parameters can be obtained by minimizing the discrepancy between $R_{2\left(fit\right)} \left(r\right)$ and the correlation function from N-body simulation. The longitudinal and transverse correlations can be modeled in a similar way. The five fitting parameters for three correlation functions are listed in the first three rows in Table \ref{tab:3}. 

For a better fitting of longitudinal and transverse correlations, an alternative method is proposed. Let us first introduce a new function
\begin{equation} 
\label{ZEqnNum496950} 
R_{t} \left(r\right)=3T_{2} \left(r\right)-R_{2} \left(r\right)={\left(R_{2} \left(r\right)-3L_{2} \left(r\right)\right)/2} . \end{equation} 
Once we can fit the function $R_{t} \left(r\right)$ and with fitted $R_{2} \left(r\right)$ from Eq. \eqref{ZEqnNum701881}, $L_{2} \left(r\right)$ and $T_{2} \left(r\right)$ can be obtained from Eq. \eqref{ZEqnNum496950}
\begin{equation}
{T_{2} =\left(R_{t} +R_{2} \right)/3} \quad \textrm{and} \quad {L_{2} =\left(R_{2} -2R_{t} \right)/3}.       
\label{eq:153}
\end{equation}
Here we have three segments of function $R_{t} \left(r\right)$  
\begin{equation} 
\label{ZEqnNum277395} 
\begin{split}
&R_{ts} \left(r\right)=-\frac{n}{2} \left(\frac{r}{r_{1} } \right)^{n}, \quad R_{tm} \left(r\right)=\frac{r}{r_{3} } -0.1,\\
&\textrm{and} \quad R_{tl} \left(r\right)=a_{0} \exp \left(-\frac{r}{r_{2} } \right)\left(\frac{r}{r_{2} } \right),
\end{split}
\end{equation} 
where$R_{ts} \left(r\right)$ is function  on small scale (from Eqs.  \eqref{ZEqnNum178092} and \eqref{ZEqnNum955991}) and $R_{tl} \left(r\right)$ is function  on large scale (from Eqs. \eqref{ZEqnNum344034} and \eqref{ZEqnNum235904}). The middle segment of function $R_{t} \left(r\right)$, i.e., $R_{tm} \left(r\right)$, is directly obtained from N-body simulation. The key parameters for correlations in entire range of $r$ (Eqs. \eqref{ZEqnNum360823}, \eqref{ZEqnNum438462} and \eqref{ZEqnNum277395}) are listed in Table \ref{tab:2}.

\begin{table}
\caption{Key parameters for correlation functions at z=0}
\begin{tabular}{p{0.1in}p{0.1in}p{0.4in}p{0.45in}p{0.45in}p{0.45in}p{0.4in}}
\hline 
$n$ & $a_{0}$ &\makecell{$u_{0}$\\(km/s)} &\makecell {$r_{1}$\\(Mpc/h)} &\makecell{ $r_{2}$\\ (Mpc/h)} &\makecell{ $r_{3}$\\ (Mpc/h)} &\makecell{ $r_{t}$\\ (Mpc/h)} \\ \hline 
\centering 0.25 &\centering 0.45 &\centering 354.61 &\centering 19.4 &\centering 23.1321 &\centering 12.5 & 1.3\\ \hline 
\end{tabular}
\label{tab:2}
\end{table}
A total of three smooth connections must be constructed to fit function $R_{t} \left(r\right)$. First, $R_{t1} \left(r\right)$ will be constructed by connecting two functions $R_{ts} \Leftrightarrow R_{tm} $. Second, $R_{t2} \left(r\right)$ will be constructed by connecting the other two functions $R_{tm} \Leftrightarrow R_{tl} $. Finally, $R_{t(fit)} $ can be constructed by connecting two functions $R_{t1} \Leftrightarrow R_{t2} $. All fitting parameters are also presented in the last three rows in Table \ref{tab:3}. 

\begin{figure}
\includegraphics*[width=\columnwidth]{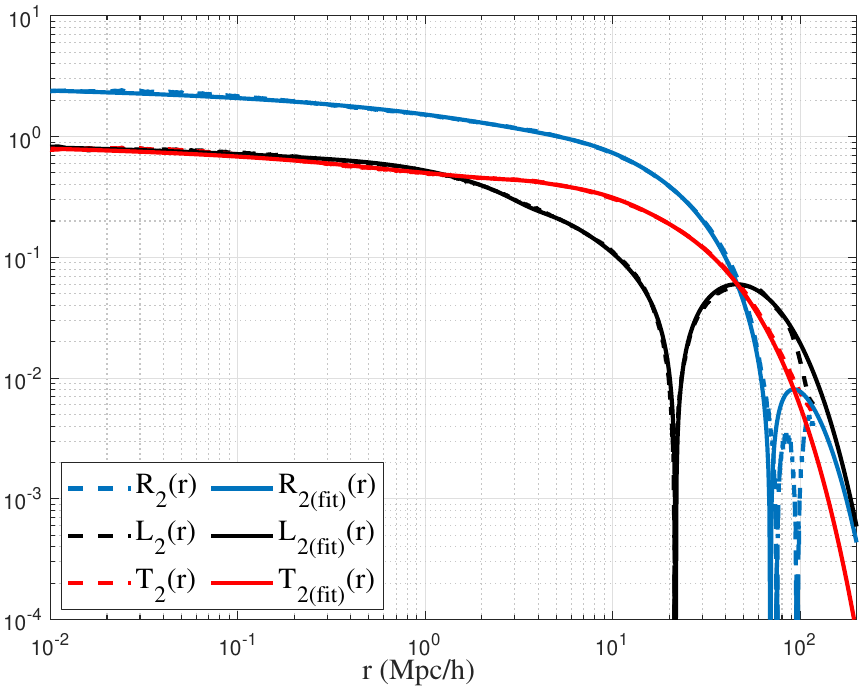}
\caption{The second order velocity correlation functions (solid lines) with fitting parameters in Tables \ref{tab:2} and \ref{tab:3}. Good agreement can be found when compared to original correlation functions (dash lines) from N-body simulation. All correlations are normalized by the velocity dispersion $u_{0}^{2} $ at \textit{z}=0.}
\label{fig:23}
\end{figure}

\begin{figure}
\includegraphics*[width=\columnwidth]{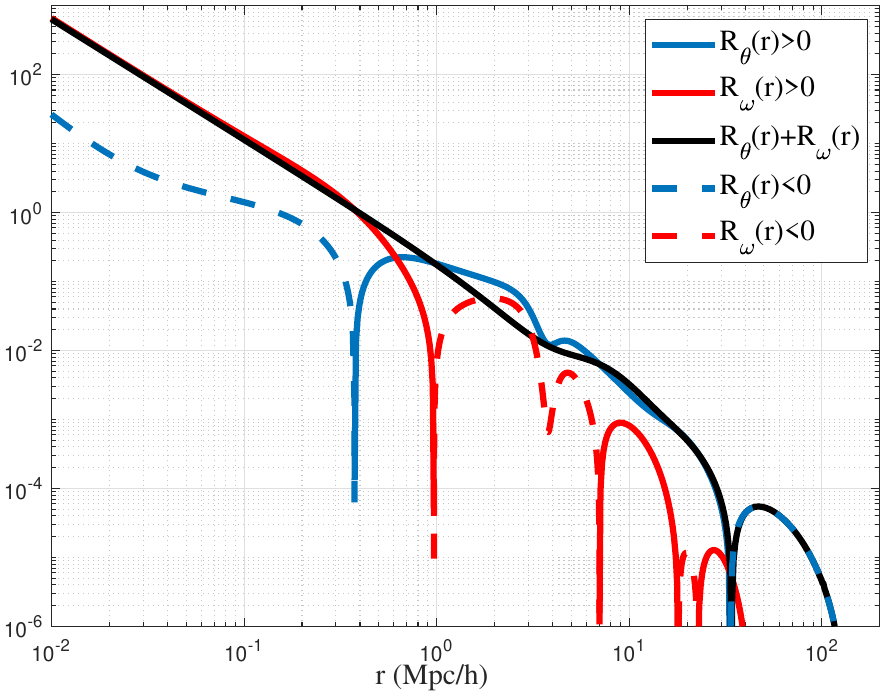}
\caption{Correlation functions of velocity divergence ($R_{\theta } $) and vorticity ($R_{\boldsymbol{\mathrm{\omega }}} $) in the unit of ${u_{0}^{2}/\left({Mpc/h} \right)^{2} } $. The divergence correlation$R_{\theta } $ is dominant on large scale (irrotational flow) and $R_{\boldsymbol{\mathrm{\omega }}} $ is dominant on small scale (constant divergence flow). The divergence $\theta $ is slightly negatively correlated on scale $r>30{Mpc/h} $, while the vorticity $\boldsymbol{\mathrm{\omega }}$ is negatively correlated for scale \textit{r} between 1Mpc/h and 7Mpc/h (particle pairs from different haloes may have opposite vorticity) and positively correlated on small scales $r<1{Mpc/h}$ (particle pairs from the same halo are more likely rotating in the same direction).}
\label{fig:24}
\end{figure}

\begin{table}
\caption{Fitting parameters for modeling correlation functions}
\begin{tabular}{p{1.1in}p{0.37in}p{0.15in}p{0.37in}p{0.15in}p{0.18in}} 
\hline 
 &\makecell{$x_{a}$\\(Mpc/h)} &\makecell{$x_{b}$} &\makecell{ $x_{c}$\\(Mpc/h)} &\makecell{$n_{1}$} & \makecell{$n_{2}$} \\ 
 \hline 
$R_{\left(fit\right)} =\left(R_{s} \Leftrightarrow R_{l} \right)$  &\centering 2.70 & 1.35 &\centering  1.83 & 0.59 & 1.09 \\ 
\hline 
$L_{2\left(fit\right)} =\left(L_{2s} \Leftrightarrow L_{2l} \right)$ &\centering 3.62 & 2.02 &\centering  6.36 & 0.78 & 3.52 \\ \hline 
$T_{2\left(fit\right)} =\left(T_{2s} \Leftrightarrow T_{2l} \right)$ &\centering  1.28 & 2.64 &\centering  0.37 & 0.30 & 1.57 \\ \hline 
$R_{t1} =\left(R_{ts} \Leftrightarrow R_{tm} \right)$ & \centering 0.21 & 0.19 &\centering 0.16 & 0.5 & 7.2 \\ \hline 
$R_{t2} =\left(R_{tm} \Leftrightarrow R_{tl} \right)$ & \centering 6.05 & 2.05 &\centering 11.75 & 1.39 & -0.6 \\ \hline 
$R_{t(fit)} =\left(R_{t1} \Leftrightarrow R_{t2} \right)$ &\centering 0.5 & 0.4 &\centering 3.7 & 0.9 & 1.0 \\ \hline 
\end{tabular}
\label{tab:3}
\end{table}
With all fitting parameters listed in Tables \ref{tab:2} and \ref{tab:3}, Fig. \ref{fig:23} plots three fitted correlations that agrees very well with the original correlations from N-body simulations. With good models of velocity correlations, the divergence and vorticity correlations and $R_{\boldsymbol{\mathrm{\omega }}} \left(r\right)$ can be obtained using Eqs. \eqref{ZEqnNum441477} and \eqref{ZEqnNum528609} in Section \ref{sec:3} with functions $A_{2} ={\left(L_{2} -T_{2} \right)/2} $ and $B_{2} =T_{2} $ (from Eqs. \eqref{ZEqnNum935612} to \eqref{ZEqnNum991811}). Two correlations of velocity gradient are presented in Fig. \ref{fig:24}, where divergence $\theta $ is negatively correlated on scale $r>30{Mpc/h} $, while vorticity $\boldsymbol{\mathrm{\omega }}$ is negatively correlated for scale \textit{r} between 1 Mpc/h and 7 Mpc/h (pair of particles from different haloes) and positively correlated on small scales $r<1{Mpc/h}$.

\section{Conclusions}
Understanding the velocity field is important for the dynamics of self-gravitating collisionless dark matter flow (SG-CFD). To describe the random and multiscale nature of velocity field, different statistical measures are introduced that involve the correlation, structure, dispersion, and spectrum functions. Velocity field in SG-CFD exhibits a scale-dependent flow behavior, i.e., constant divergence on small scale and irrotational nature on large scale (Fig. \ref{fig:4}). 

To understand the nature of flow across entire range of scales, fundamental kinematic relations between different statistical measures need to be developed for different types of flow. Current discussion is restricted to the homogeneous and isotropic flow with translational and rotational symmetry in space. This greatly simplifies the development of kinematic relations for incompressible, constant divergence, and irrotational flow. The incompressible and constant divergence flow share the same kinematic relations for even order correlations, while they can be different for odd order correlations. One example is that the first order velocity correlation tensor $Q_{i} \left(r\right)\equiv 0$ (Eq. \eqref{ZEqnNum725997}) for incompressible flow, while $Q_{i} \left(r\right)\equiv const$ is required for constant divergence flow (Eq. \eqref{ZEqnNum666615}). Both types of flow have the same kinematic relations for second order correlations. 

For second order velocity correlation tensor $Q_{ij}(r)$, three scalar correlation functions (total $R_{2} $, longitudinal $L_{2} $, and transverse $T_{2} $ in Figs. \ref{fig:5}, \ref{fig:6} and \ref{fig:7}) can be obtained by contraction of indices (Eqs. \eqref{ZEqnNum935612} to \eqref{ZEqnNum991811}), with kinematic relations in Eqs. \eqref{ZEqnNum314105} and \eqref{ZEqnNum320035} for three different types of flow. These relations are then used to characterize the nature of SG-CFD flow. Simulation results confirms the constant divergence flow on small and irrotational flow on large scales  (Eq. \eqref{ZEqnNum956600} and Fig. \ref{fig:4}). The velocity dispersion functions $\sigma _{u}^{2} \left(r\right)$ and $\sigma _{d}^{2} \left(r\right)$ (Eqs. \eqref{ZEqnNum726048} and \eqref{ZEqnNum904685} and Fig. \ref{fig:8}) represent the kinetic energy contained below or above scale \textit{r} and are related to the correlation $R_{2}$ (Eq. \eqref{ZEqnNum910224}). The function $E_{ur}$ describes the real-space energy distribution on different scales (Eq. \eqref{ZEqnNum359490} and Fig. \ref{fig:10}).

Two definitions of longitudinal structure functions $S_{2}^{lp} $ and $S_{2}^{l} $ are introduced (Eqs. \eqref{ZEqnNum250774}, \eqref{ZEqnNum414891} and Figs. \ref{fig:13}, \ref{fig:22}). Both definitions are equivalent for incompressible flow and are different for dark matter flow due to the collisionless nature. The limiting correlation of longitudinal velocity on the smallest scale $\rho _{L} \left(r=0\right)=0.5$ is a distinct feature of collisionless flow (Eq. \eqref{ZEqnNum350084} and Fig. \ref{fig:15}), while $\rho _{L} \left(r=0\right)=1$ for incompressible flow. Assuming gravity is the only interaction and no radiation produced from particle "annihilation" on the smallest scale, this feature leads to an increase in particle mass converted from kinetic energy during annihilation (Eq. \eqref{ZEqnNum962574}). 

The other two structure functions ($S_{2}^{i} $ and $S_{2}^{x} $ in Figs. \ref{fig:11} and \ref{fig:13}) are used to represent the kinetic energy below a given scale and the total enstrophy above a given scale. They are related to each other via Eq. \eqref{ZEqnNum952229} and to dispersion functions via Eq. \eqref{ZEqnNum947585}. The real-space enstrophy distribution on different scales ($E_{nr}$) can be obtained from $S_{2}^{x} $ (Eq. \eqref{ZEqnNum336678} and Fig. \ref{fig:12}). The correlation functions of vorticity and divergence are also developed for different types of flow (Eqs. \eqref{ZEqnNum912296} to \eqref{ZEqnNum528609}).

On large scale, the transverse velocity correlation follows a simple exponential form (Eq. \eqref{ZEqnNum971850} and Fig. \ref{fig:5}) with a single co-moving constant length scale $r_{2}$ that might be related to the horizon size at matter-radiation equality. All other second order velocity correlation/structure/dispersion functions can be analytically derived using kinematic relations for irrotational flow (Eqs. \eqref{ZEqnNum344034} to \eqref{ZEqnNum973449}). The density and potential correlations on large scale can be derived accordingly (Eqs. \eqref{ZEqnNum762470} to \eqref{ZEqnNum590702} and Fig. \ref{fig:19}), along with the spectrum functions of velocity, density, and potential in Eqs. \eqref{ZEqnNum976884} to \eqref{ZEqnNum759927}. 

On small scale, the longitudinal structure function exhibits a one-fourth power-law $S_{2}^{l} \left(r\right)\propto r^{{1/4} } $ (Eq. \eqref{ZEqnNum392323} and Fig. \ref{fig:22}), which can be derived using density correlation (Eqs. \eqref{ZEqnNum206214} and \eqref{ZEqnNum489540}) and virial theorem. Similarly, all other correlation and structure functions are obtained analytically using the kinematic relations for constant divergence flow on small scale (Eq. \eqref{ZEqnNum178092} to \eqref{ZEqnNum352266}). The spectrum function of velocity follows $E_{u} \left(k\right)\propto k^{-{5/4} } $ (or equivalently $P_{u} \left(k\right)\propto k^{-{13/4} } $) and vorticity spectrum follows $E_{\boldsymbol{\mathrm{\omega }}} \left(k\right)\propto k^{{3/4} } $ (or equivalently $P_{\boldsymbol{\mathrm{\omega }}} \left(k\right)\propto k^{{-5/4} } $).

Finally, the correlation functions for entire range of scales can be constructed by smoothly connecting correlations on small and large scales with a given interpolation function (Eq. \eqref{ZEqnNum986217} and Fig. \ref{fig:23}). The divergence and vorticity correlations on entire range of scales can be obtained and presented in Fig. \ref{fig:24}. The vorticity $\omega$ is negatively correlated on intermediate scale \textit{r} between 1Mpc/h and 7Mpc/h, while divergence $\theta$ is negatively correlated on large sale $\mathrm{>}$30Mpc/h that leads to a negative correlation in matter density (Fig. \ref{fig:19}). 

In this paper, the kinematics provides relations between statistical measures of the same order. Dynamic equations from evolution of velocity field are needed to derive the dynamic relations between statistical measures on different orders. Future work involves the general kinematic relations of any order and dynamic relations on different scales \citep{Xu:2022-The-statistical-theory-of-3rd}.

%\begin{acknowledgments}
%\end{acknowledgments}

\section*{Acknowledgements}
This research was supported by the Laboratory Directed Research and Development at Pacific Northwest National Laboratory (PNNL). PNNL is a multiprogram national laboratory operated for the U.S. Department of Energy (DOE) by Battelle Memorial Institute under Contract no. DE-AC05-76RL01830.
%\appendix

\section*{Data Availability}
Two datasets underlying this article, i.e. a halo-based and correlation-based statistics of dark matter flow, are available on Zenodo \citep{Xu:2022-Dark_matter-flow-dataset-part1,Xu:2022-Dark_matter-flow-dataset-part2}, along with the accompanying presentation slides "A comparative study of dark matter flow \& hydrodynamic turbulence and its applications" \citep{Xu:2022-Dark_matter-flow-and-hydrodynamic-turbulence-presentation}. All data files are also available on GitHub \citep{Xu:Dark_matter_flow_dataset_2022_all_files}.

%\nocite{*}
\bibliographystyle{Papers}
\bibliography{Papers}% Produces the bibliography via BibTeX.
\label{lastpage}
\end{document}